\documentclass[USenglish,UKenglish]{article}
\usepackage[margin=2.5cm]{geometry}
\usepackage[utf8]{inputenc}
\usepackage[T1]{fontenc}
\usepackage{amssymb, amsmath}
\usepackage{graphicx}
\usepackage{url, xcolor, hyperref}
\usepackage{babel}
\usepackage{isodate}
\usepackage{booktabs}
\usepackage{multirow}
\usepackage{longtable}
\hypersetup{colorlinks, linkcolor={red!70!black}, citecolor={blue!70!black},
urlcolor={red!70!black} }
\usepackage{siunitx}
\usepackage[normalem]{ulem}
\usepackage{caption}

\newrobustcmd*{\bftabnum}{%
  \bfseries
  \sisetup{output-decimal-marker={\textmd{.}}}%
}

\usepackage{settings_custom}
\usepackage{algorithm}
\usepackage{algpseudocode}
\usepackage{xcolor}
\usepackage{comment}
\usepackage{array}
\usepackage{tikz}
\usepackage{authblk}
\usepackage{stackengine}
\usepackage[backend=biber, sorting=nyt, doi=false,
    url=false,
    isbn=false,    
    eprint=false  ]{biblatex}

\usepackage{glossaries}

\makeglossaries
\usepackage{xspace}  

\newacronym{BDCM}{BDCM}{backtracking dynamical cavity method}
\newacronym{HPR}{HPR}{history passing reinforcement}
\newacronym{RRG}{RRG}{random regular graph}
\newacronym{RS}{RS}{replica symmetric}
\newacronym{BP}{BP}{belief propagation}
\newacronym{SA}{SA}{simulated annealing}
\newacronym{1RSB}{1RSB}{one-step replica symmetry breaking}
\newacronym{d1RSB}{d1RSB}{dynamical one-step replica symmetry breaking}
\newacronym{s1RSB}{s1RSB}{static one-step replica symmetry breaking}
\newacronym{BPR}{BPR}{belief propagation reinforcement}

\newcommand{\bpr}{\gls{BPR}\xspace}
\newcommand{\bdcm}{\gls{BDCM}\xspace}
\newcommand{\rs}{\gls{RS}\xspace}
\renewcommand{\bp}{\gls{BP}\xspace}
\newcommand{\oRSB}{\gls{1RSB}\xspace}
\newcommand{\doRSB}{\gls{d1RSB}\xspace}
\newcommand{\soRSB}{\gls{s1RSB}\xspace}
\newcommand{\hpr}{\gls{HPR}\xspace}
\newcommand{\rrg}{\gls{RRG}\xspace}
\newcommand{\rrgs}{\glspl{RRG}\xspace}
\newcommand{\sa}{\gls{SA}\xspace}

\usetikzlibrary{arrows.meta, positioning, fit, shapes}
\newcommand\bsbar[1]{\stackunder[1pt]{$\scriptstyle#1$}{\rule{1.2ex}{.07ex}}}
\newcommand\bbar[1]{\stackunder[1pt]{$#1$}{\rule{1.2ex}{.07ex}}}
\newcommand{\dd}{\mathop{}\!\mathrm{d}}

\title{Minority Takeover in Majority Dynamics: Searching for Rare Initializations via the History Passing Algorithm}

\author[a]{Marek Jankola}
\author[b]{Freya Behrens}
\author[b]{C\'edric Koller}
\author[b]{Lenka Zdeborov\'a}

\affil[a]{Institute of Theoretical Physics, Charles University}
\affil[b]{Statistical Physics of Computation Laboratory, \'Ecole Polytechnique F\'ed\'erale de Lausanne (EPFL)}

\date{}

\begin{document}
\sisetup{
  detect-weight = true,
  detect-inline-weight = math
}
\pagenumbering{gobble}
\maketitle

\begin{abstract}
We investigate how much bias in the initial configuration is required to drive global agreement in synchronous, deterministic majority dynamics on large random $d$-regular graphs. Nodes take values~$\pm 1$ and update their states at each discrete time step to align with the majority of their neighbors. Using the \bdcm, we estimate the minimal fraction of initial $+1$ nodes required to achieve a $+1$ consensus in $p$ time steps. Our analysis predicts that for $d\geq4$ an initial global minority of $+1$ nodes is sufficient to quickly steer the entire system toward consensus on $+1$. 

We then investigate whether such initial conditions can be determined explicitly for a given large random regular graph. To this end, we introduce a new algorithm, which we name \textit{history-passing reinforcement} (HPR), designed to find such initial configurations with a minority of $+1$ nodes. We find, as a main result, that the HPR algorithm finds initial configurations where the minority takes over the majority for $d$-regular random graphs with $d\geq4$. 

The HPR algorithm outperforms standard simulated annealing-based methods, but does not reach the lowest densities predicted by the \bdcm. Rather, the lowest density achievable by the algorithm is near the onset of a \doRSB phase, which we estimate using a \oRSB formulation of the \bdcm. While we focus on the majority dynamics and random $d$-regular graphs, the algorithm can be extended to other dynamical rules and classes of sparse graphs.
\end{abstract}
\glsresetall
\pagenumbering{arabic}

\section{Introduction} \label{sec: intro}
Let $G(V,E)$ be an undirected graph with $n$ nodes $V=\{1,\dots,n\}$ and edges $E=\{(i,j)\,|\,i,j\in V\}$. Each node is associated with a value $x_i\in\{-1,+1\}$, and we define a graph configuration as $\mathbf{x} =\{x_1, \ldots, x_n\}$ with magnetization (or bias)
$m(\mathbf x)=\frac{1}{n}\sum_{i\in V}x_i$.
On this graph we consider the time evolution $\mathbf x^{t} \mapsto \mathbf x^{t+1}$ of a synchronous local \textit{majority dynamics} (with always-stay tie breaking): i.e. each node takes the value of the majority in their neighborhood, and in case of a tie it stays the same.
Formally, for the local magnetization 
$h_i^t = \sum_{j\in\partial i} x_j^t$, we define the update function as
\begin{equation}
    x_i^{t+1} = f(x_i^t,\mathbf{x}_{\partial i}^t) = \begin{cases}
\text{sign}\left(h_i^t\right) &\text{if sign$\left(h_i^t\right)\neq0$}\,,\\
x_i^t &\text{otherwise}\,,
\end{cases} \label{majority dynamics 1}
\end{equation}
where $x_i^t$ denotes the state of node $i$ at time step $t$, $\partial_i$ denotes the set of neighbors of node $i$, and $\mathbf{x}_{\partial i}^t$ contains the states of these neighbors.

This setting is motivated by a variety of spreading processes, such as the diffusion of opinions or rumors in social networks or the transmission of diseases in epidemiology.
In applications, one often asks how to influence such dynamics to steer the collective outcome, i.e. which individuals to target for persuasion or the initial seeding of information or immunization, as in~\cite{maximizing_spread,min_cont_set, altarelli_optimizing_2013, power_of_few}. From a control-and-optimization viewpoint, this is a strategic bribery problem that motivates the central question of our work on majority dynamics:
\begin{center}
    \textit{What is the smallest initial bias that leads to a positive consensus after a fixed number of time steps?}
\end{center}
We consider this question for typical instances of randomly sampled graphs $G$, and in the large system limit $n \to \infty$.
We define $m_G^*(p)$ as the minimal initial bias such that for a graph $G$ there exists an initial configuration $\mathbf x^1$ with $m(\mathbf x^1)=m_G^*(p)$, and this configuration evolves to a $+1$ consensus in $p$ steps ($m(\mathbf x^p)=1$). 

To this end, we study the number of initial configurations with a given bias that lead to a consensus after a fixed number of steps $p$, for typical instances of random $d$-regular graphs. We also investigate whether these initial configurations can be found on concrete graph instances using efficient algorithms.

In the following, we both establish precise estimations of the  minimal biases for random $d$-regular graphs, and provide an algorithm that empirically finds corresponding initial configurations. While our method extends to other families of random graphs, we focus on \rrgs, where every node has exactly degree $d$:
\begin{itemize}
    \item We analyze the problem using the \bdcm~\cite{BDCM}, which allows for the study of short trajectories on sparse random graphs. Under the replica symmetric assumption, we find that for \rrgs, for $d\geq 4$ and $p\geq 3$ (or $d\geq8$, $p\geq2$) there exist initializations with $+1$ in the minority which under majority dynamics evolve into $+1$ consensus.
    \item To find these initial conditions with low initial magnetization for a particular RRG, we introduce an algorithm that we name \hpr. This algorithm is based on the \bpr algorithm~\cite{chavas_survey-propagation_2005, braunstein_encoding_2007} which we extend to the dynamical setting. The \hpr algorithm successfully finds solutions with negative $m_{\rm init}$ that go into positive consensus for $d\ge 4$. 
    We also numerically study the performance of a \sa algorithm on this problem. We find that \sa does not find initializations with negative $m_{\rm init}$, and that \hpr outperforms it in all the considered cases. However, \hpr does not reach the lowest possible bias given by the replica symmetric BDCM. 
    \item Using the \oRSB formulation of the \bdcm, we find that replica symmetry breaks before reaching the minimal bias predicted by the \rs BDCM. Nevertheless, the onset of the \oRSB phase remains in the negative-magnetization regime. Thus, BDCM still predicts the existence of strategic minorities leading to consensus. Moreover, the onset of the \doRSB phase roughly corresponds to the lowest algorithmically reachable $m_{\rm init}$.
\end{itemize}
The code to reproduce our results, including the \hpr algorithm is publicly available\footnote{\url{https://github.com/SPOC-group/history-passing}}.
The work is structured as follows: We begin in Section~\ref{sec:related work} with an overview of related work. 
Section~\ref{sec:BDCM} introduces the \bdcm formalism, both under the \rs assumption and its \oRSB extension. 
In Section~\ref{sec:BDCM for enhancing consensus}, we apply this framework to majority dynamics on $d$-regular random graphs, deriving results for the initial bias required for consensus. 
Section~\ref{sec:HPR} then presents the \hpr algorithm, designed to construct strategic initializations that drive the system to consensus. 
Finally, Section~\ref{sec:hpr-results} evaluates its performance, comparing it to simulated annealing as well as to the \rs and \oRSB predictions.

\section{Related work}\label{sec:related work}

\paragraph{Opinion dynamics}
The study of opinion dynamics spans multiple disciplines, including but not limited to sociology, mathematics, physics, and computer science~\cite{flache_models_2017, n_zehmakan_spread_2021, castellano_statistical_2009, becchetti_consensus_2020}. Various update rules (e.g. the threshold, Galam, dissemination of culture, Sznajd, and majority models) and network topologies (e.g. fully connected, Erdős–Rényi, regular, and scale-free, sometimes dynamically changing) have been analyzed~\cite{granovetter_threshold_1978, galam_majority_1986, galam_sociophysics_2008, axelrod_dissemination_1997, sznajd-weron_opinion_2000, kanoria_majority_2011}. We also refer the reader to~\cite{castellano_statistical_2009, xia_opinion_2011, grabisch_survey_2020} for reviews. 
The problem of finding an initial configuration that turns a small initial bias into a majority on one side is practically relevant in the context of bribing problems~\cite{maximizing_spread,min_cont_set, altarelli_optimizing_2013, power_of_few, becchetti_consensus_2020}, and related to gerrymandering~\cite{gerrymander}.
From a technical point of view, our work aligns most closely with statistical physics methods applied to opinion dynamics, a field known as sociophysics~\cite{galam_sociophysics_2012}.

\textcite{kanoria_majority_2011} study the majority dynamics with stochastic tie breaking on infinite regular trees and provide a lower bound on the initial number of $+1$ opinions such that a consensus of $+1$ is reached for $T\rightarrow \infty$. In~\cite{Majority_model_RRG}, the authors consider majority dynamics on $d$-\rrgs, and they study how a typical small initial majority is guaranteed to reach full consensus for high $d$'s. However, these works only concern the average case, as the initialization is drawn uniformly at random from the possible initializations with a given density of opinions. In contrast, we aim to minimize the number of $+1$ opinions, so an adversarial perspective is more relevant. The work~\cite{zehmakan_majority_2021} investigates such an adversarial approach: an attacker can choose the $+1$ nodes, and guarantees are given when no such choice leads to a $+1$ consensus. However, the dynamic rule of~\cite{zehmakan_majority_2021} is different, with the addition of ``neutral'' nodes taking the opinion of the majority. A broader discussion of opinion dynamics in related settings can be found in~\cite{PhDthesis_zehmakan2019}. Reference~\cite{Whom_to_befriend} investigates the ``Minimum Links Problem'', which seeks to achieve global consensus by adding the minimum number of new nodes and links. The work~\cite{min_cont_set} studies the minimal contagious (or percolating) set problem on \rrgs, which is the same optimization setting as our case, but with threshold dynamics, where nodes that reach state $+1$ remain in it indefinitely. We also refer to this work for references to many other studies on extremal properties of graphs.

\paragraph{The Mpemba effect}
The problem we analyze can be interpreted as an instance of the Mpemba effect. This effect was first observed for water, where hot water can freeze faster than cold water under certain conditions~\cite{mpemba1969cool, jeng2006mpemba}. This phenomenon has been observed and studied in diverse settings, including spin glasses~\cite{lu2017nonequilibrium, baity2019mpemba} and colloidal systems~\cite{kumar2020exponentially}. The standard effect describes how a system starting ``hotter'' (further from equilibrium) may relax faster than a ``colder'' system. The strong version of the effect occurs when this anomalous relaxation is exponentially faster~\cite{klich2019mpemba}. In our context, the initial magnetization $m(\mathbf{x}^1)$ is analogous to the temperature, where initializations with lower magnetization are ``hotter'' with respect to the $+1$ consensus state. Our objective of finding initial configurations with low magnetization that nevertheless reach a $+1$ consensus is equivalent to identifying initial states that exhibit a Mpemba-like effect. These ``hot'' initial states follow specific efficient relaxation pathways that are inaccessible to ``cold'' initial states chosen at random. Therefore, the algorithm that we introduce is a tool for discovering initializations that leverage anomalous relaxation pathways.

\paragraph{The cavity method} The cavity method~\cite{Bethe, geffner_reverend_2022, mezard_sk_1986} is a standard tool in statistical physics. The method yields a system of self-consistent equations whose solutions provide estimates of key observables. Its effectiveness has led to applications in diverse domains, for instance in theoretical computer science, e.g. for constraint satisfaction problems~\cite{mezard_analytic_2002, krzakala_gibbs_2007, montanari_clusters_2008, phys_inf_comp, saad_physics-inspired_2014} or in the modeling of epidemic processes~\cite{karrer_message_2010, altarelli_optimizing_2013, altarelli_patient-zero_2014, braunstein_statistical_2023, ghio_bayes-optimal_2023}. Its dynamical variants, usually called dynamical cavity methods or dynamical message passing, extend the method to out-of-equilibrium systems~\cite{neri_cavity_2009, kanoria_majority_2011, aurell_message-passing_2011, del_ferraro_dynamic_2015, hwang_number_2020, lokhov_dynamic_2015}. Moreover, the previously introduced \bdcm extends the cavity formalism to short trajectories ending in cycles~\cite{BDCM}. The \bdcm is the backbone of our analysis. By adapting the belief propagation reinforcement scheme introduced for the standard cavity method~\cite{chavas_survey-propagation_2005, braunstein_encoding_2007}, we obtain a solver capable of tackling problems in a dynamical setting.

\section{The BDCM and its replica symmetry breaking extension}\label{sec:BDCM}
Our goal is to determine the smallest initial magnetization $m_{\rm init}$ that leads to consensus within a fixed number of time steps. To this end, we use the \bdcm~\cite{BDCM} to compute the entropy of typical initializations satisfying this criterion on large random graphs. The smallest initial bias then comes from finding the smallest $m_{\rm init}$ with a non-zero entropy. 

In this methodological section, we first review the \bdcm for general local deterministic dynamics, as originally stated in~\cite{BDCM}. To refine the replica-symmetric (RS) estimate and capture a potentially clustered structure of satisfying configurations, we formulate it within the \oRSB framework; this has not previously been done for the \bdcm.
While the method is presented in general terms for deterministic dynamics on arbitrary graphs, in Section~\ref{sec:BDCM for enhancing consensus} we apply it specifically to majority dynamics on \rrgs, to answer the question of interest.

Note that the computation for both the RS and 1RSB results are described here as approximations of the corresponding quantities on a given large random graph. This will be crucial when turning towards specific algorithms, such as the history passing reinforcement presented in Section~\ref{sec:HPR}. In Section~\ref {sec:BDCM for enhancing consensus}, we then turn to the thermodynamic limit in which, thanks to self-averaging, the single-graph results can be used to deduce the typical behavior of large random graphs.  

\subsection{Backtracking Dynamical Cavity Method}

\paragraph{Dynamical Processes on Graphs}
For generality in this section we consider a node~$i$ to take value $x_i \in S$ where $S$ is a discrete set of states.
Analogously, a graph configuration is $\mathbf{x} \in S^n$.
In the following, we consider general \textit{local}, \textit{deterministic}, \textit{time-discrete} and \textit{synchronous} dynamics, where all nodes are simultaneously updated with a rule $f_i(x_i,\mathbf{x}_{\partial i})$.
Thus, the local dynamics $f_i:S \times S^{|\partial i|}\rightarrow S$ leads to the global dynamics $F:S^n\rightarrow S^n$ with $\left[F(\mathbf{x})\right]_i = f_i(x_i,\mathbf{x}_{\partial i})$. The subsequent application of $F$, $\mathbf{x}^{t+1}=F(\mathbf{x}^t)$, creates a \textit{trajectory} $\bbar{x}_i=(x_i^1,\dots,x_i^T)$ at the level of each node and $\bbar{\mathbf{x}} = (\mathbf{x}^1,\dots,\mathbf{x}^T)$ at the level of the whole graph, where $t=1,\dots,T-1$. 
Since the configuration space $S^{n}$ is finite and the dynamics is deterministic, all trajectories $\bbar{\mathbf{x}}$ end in a cycle. The pre-periodic part of the trajectory is called the \textit{transient} and the periodic part is the \textit{attractor}. We denote the transient length as $p$ and the attractor length as $c$. The set of all initializations ending in a given attractor is called its \textit{basin of attraction}.

\paragraph{The BDCM idea}
A trajectory $\bbar{\mathbf{x}}$ of length  $T = p+ c$, consisting of a transient of length $p$ followed by a cycle of length $c$, is called a $(p,c)$ \textit{backtracking attractor}. The goal of the \bdcm is to count such backtracking attractors under given constraints, e.g. the number $\mathcal{N}$ of trajectories that end in an attractor of a specific period or start with a specific magnetization. In the thermodynamic limit, $\mathcal{N}$ generally grows exponentially with $n$, so we instead aim to compute the entropy density $s = \frac{1}{n}\log(\mathcal{N})$.
The key idea, following the cavity method, is to interpret $\mathcal{N}$ as the normalization constant of a uniform measure supported only on admissible backtracking attractors. 
Although the underlying object of interest, the backtracking attractor, captures dynamical properties, we approach it through a static formulation. By encoding admissible trajectories into a probability measure, the counting problem reduces to evaluating the normalization constant. Within the \rs ansatz, this normalization can be estimated from a set of fixed-point equations, giving access to an estimate of the typical count in the thermodynamic limit.
The \rs ansatz thus provides the natural starting point for the analysis, but it is limited: the resulting estimate gives only an upper bound on the asymptotical entropy \cite{franz2003replica}, and when the space of solutions has a clustered structure it may not be accurate. We address the latter case via \oRSB, which is presented in Section~\ref{sec: RSB in main}.
We can also refine the constraints on the backtracking attractors by reweighting the probability measure, biasing it toward trajectories with the desired properties. 

\paragraph{A probability measure over backtracking attractors}
More formally, we can define the following measure on backtracking attractors $\bbar{\mathbf{x}} \in S^{p+c}$, where
$\bbar{\mathbf{x}}$ is a valid trajectory of length $p$ that ends in a cycle of length $c$ and we reweight every 
$\bbar{\mathbf{x}}$ according to $K$ observables $\Xi_k$:
\begin{align}
    P(\bbar{\mathbf{x}}) = \frac{1}{Z} 
    e^{-\sum_{k}\lambda_{k}\Xi_{k}(\bsbar{\mathbf{x}})}
    \mathds{1}\left[\mathbf{x}^{p+1}=F(\mathbf{x}^{p+c})\right] 
    \prod_{t=1}^{p+c-1} \mathds{1}\left[\mathbf{x}^{t+1}=F(\mathbf{x}^t)\right]\,.
    \label{prob dist BDCM}
\end{align}
$Z$ is the normalization constant of the probability distribution, and $\mathds{1}\left[\rm s\right]$ is the indicator function\footnote{ $\mathds{1}\left[\rm s\right]$ is $1$ if the boolean statement $s$ is true and $0$ otherwise.}.

The set $\{\Xi_k\}_{k=1}^K$ contains functions that represent extensive observables\footnote{$\Xi(\bbar{\mathbf{x}}) \sim O(n)$ as $n \rightarrow \infty$} on these backtracking attractors, with $\Xi: S^{n(p+c)} \rightarrow \mathbb{R}$. 
They can represent properties such as the initial magnetization or the number of times a node changes value throughout the trajectory.
We denote their intensive counterparts $\{\Xi_k(\bbar{\mathbf{x}})/n\}_{k=1}^K$ as $\{\xi_k\}_{k=1}^K$. 

\paragraph{Computing the entropy density} The \bdcm's objective is to express the number of valid backtracking attractors conditioned on the observable values $\{\xi_k\}_{k=1}^K$, $\mathcal{N}\left(\{\xi_k\}_{k=1}^K\right)$, via the \textit{entropy density}:
\begin{equation}
    s\left(\{\xi_k\}_{k=1}^K\right) = 
    \frac{\log{\mathcal{N}\left(\{\xi_k\}_{k=1}^K\right)}}{n}\,. \label{entropy general def} 
\end{equation}
The entropy $s\left(\{\xi_k\}_{k=1}^K\right)$ is related to the normalization constant $Z$ through the following transformation:
\begin{equation}
\begin{split}
    Z &= \sum_{\{\bsbar{\mathbf{x}}\}} e^{-\sum_{k}\lambda_{k}\Xi_{k}(\bsbar{\mathbf{x}})}
    \mathds{1}\left[\mathbf{x}^{p+1}=F(\mathbf{x}^{p+c})\right] 
    \prod_{t=1}^{p+c-1} \mathds{1}\left[\mathbf{x}^{t+1}=F(\mathbf{x}^t)\right]\\
    &= \sum_{\{\xi_k\}_{k=1}^K} \mathcal{N}\left(\{\xi_k\}_{k=1}^K\right)  e^{-\sum_{k}\lambda_{k}\Xi_{k}(\bsbar{\mathbf{x}})}
    \simeq \int \biggl(\prod_{k=1}^K d\xi_k\biggr) e^{n\left[s\left(\{\xi_k\}_{k=1}^K\right) - \sum_{k=1}^K\lambda_{k}\xi_{k}\right]}\,,
\end{split} \label{Z to saddle}
\end{equation}
where the last expression is valid in the large $n$ limit, in which the intensive quantities take the same values for almost every realization of the graph, the so-called self-averaging of statistical physics of disordered systems \cite{mezard1987spin}. 
Defining the \textit{free entropy density} as $\Phi(\{\lambda_k\}_{k=1}^K) = \log{Z}/n$ and applying the saddle point method, this connects the entropy density to $Z$ in the large $n$ limit as
\begin{equation}
    s(\{\hat{\xi}_k\}_{k=1}^K) = \Phi(\{\lambda_k\}_{k=1}^K) + \sum_k \lambda_k\hat{\xi}_k\,, \label{s via phi general}
\end{equation}
where we denote as $\{\hat{\xi}\}_{k=1}^K$ the maximizer of $s\left(\{\xi_k\}_{k=1}^K\right) - \sum_{k}\lambda_{k}\xi_{k}$. Thus, we have the following conditions for all $k=1,\dots,K$:
\begin{equation}
    \frac{\partial \Phi(\lambda_1,\dots,\lambda_K)}{\partial \lambda_k} = -\hat{\xi}_k = -\frac{1}{n}\langle \Xi_k\rangle_{P(\bsbar{\mathbf{x}})}\,. \label{der phi to xi_k}
\end{equation}
Consequently, if we know $\Phi(\{\lambda_k\}_{k=1}^K)$ for an arbitrary set of Lagrange parameters $\{\lambda_k\}_{k=1}^K$, we can also obtain the entropy density $s\left(\{\xi_k\}_{k=1}^K\right)$ with fixed observables $\{\xi_k\}_{k=1}^K$ from~\eqref{s via phi general}. These observables are enforced by values of $\{\lambda_k\}_{k=1}^K$, as seen in~\eqref{der phi to xi_k}. 

\paragraph{Estimating the normalization constant via Belief Propagation} Computing $\Phi$ or $Z$ on a given graph exactly is intractable in general, as we have an exponential number of trajectories $\bbar{\mathbf{x}}$. However, on large random graphs, we can estimate $\Phi$ using the cavity method~\cite{phys_inf_comp, lecture_notes}. 
For this, we represent the probability distribution~\eqref{prob dist BDCM} as a tree-like graphical model.   
By definition, the dynamics $F(\mathbf{x})$ factorizes on the node neighborhoods. We similarly assume that the global observables $\Xi$ can be factorized over the nodes $V$ via $\Xi(\bbar{\mathbf{x}}) = 1/n\sum_{i \in V} \Tilde{\Xi}(\bbar{x}_i)$ or analogously over the edges $E$\footnote{We denote by $\Tilde{\Xi}(\bbar{x}_i): S^{p+c}\rightarrow \mathbb{R}$ observables depending on single nodes and  by $\Bar{\Xi}(\bbar{x}_i,\bbar{x}_j): S^{2(p+c)}\rightarrow\mathbb{R}$ observables depending on two nodes connected by an edge.}.
Then, we can factorize the whole probability distribution~\eqref{prob dist BDCM} as
\begin{equation}
    P(\bbar{\mathbf{x}}) = \frac{1}{Z} \prod_{i\in V} \underbrace{\left[e^{-\sum_{\Tilde{k}}\lambda_{\Tilde{k}}\Tilde{\Xi}_{\Tilde{k}}(\bsbar{x}_i)} \mathds{1}\left[x_i^{p+1}=f_i(x_i^{p+c},\mathbf{x}^{p+c}_{\partial i})\right] \prod_{t=1}^{p+c-1} \mathds{1}\left[x_i^{t+1}=f_i(x_i^t,\mathbf{x}^t_{\partial i})\right]\right]}_{\mathcal{A}_i(\bsbar{x}_i,\bsbar{\mathbf{x}}_{\partial i})} \prod_{(ij)\in E} \underbrace{\left[e^{-\sum_{\Bar{k}}\lambda_{\Bar{k}}\Bar{\Xi}_{\Bar{k}}(\bsbar{x}_i,\bsbar{x}_j)}\right]}_{a(\bsbar{x}_i,\bsbar{x}_j)}\,, \label{prob dist BDCM factorized}
\end{equation}
where we define the factor nodes $\mathcal{A}_i(\bbar{x}_i,\bbar{\mathbf{x}}_{\partial i})$ and $a(\bbar{x}_i,\bbar{x}_j)$. When we can interpret this distribution as a tree factor graph (described in detail in Appendix~\ref{sec:derivation BDCM}), we can use belief propagation to compute $Z$. We do this by introducing \textit{messages} $\chi^{i \rightarrow j}_{\bsbar{x}_i,\bsbar{x}_j} \in \mathbb{R}$. The messages correspond to the probability of node trajectories $\bbar{x}_i,\bbar{x}_j$ taking on specific values when we restrict ourselves to a particular subpart of the factor graph (on the original graph $G$ this corresponds to focusing on a subgraph containing node $i$ after omitting the edge $(ij)$, i.e. after introducing cavity). For these messages we arrive at the following equations (see Appendix~\ref{sec:derivation BDCM} for details)
\begin{equation}
    \chi^{i \rightarrow j}_{\bsbar{x}_i,\bsbar{x}_j} = \frac{1}{Z^{i \rightarrow j}}\,a(\bbar{x}_i,\bbar{x}_j) \sum_{\bsbar{\mathbf{x}}_{\partial i\backslash j}} \mathcal{A}_i(\bbar{x}_i,\bbar{\mathbf{x}}_{\partial i})\!\!\prod_{k\in\partial i\backslash j}\!\chi^{k \rightarrow i}_{\bsbar{x}_k,\bsbar{x}_i}\,.\label{BDCM eqs}
\end{equation}
Messages that fulfill this recurrence can be used to obtain an estimate of $\Phi$ called the \textit{Bethe free entropy density} $\Phi_B$:
\begin{align}
    n\Phi_B &= \sum_{i\in V} \log(Z^i) - \sum_{(ij)\in E} \log(Z^{ij})\,, \label{phi B}\\
    Z^i &= \sum_{\bsbar{x}_i,\bsbar{\mathbf{x}}_{\partial i}} \mathcal{A}_i(\bbar{x}_i,\bbar{\mathbf{x}}_{\partial i}) \prod_{j\in \partial i} \chi^{j\rightarrow i}_{\bsbar{x}_j,\bsbar{x}_i}\,,\label{Zi def}\\
    Z^{ij} &= \sum_{\bsbar{x}_i,\bsbar{x}_j} \frac{1}{a(\bbar{x}_i,\bbar{x}_j)} \chi^{i\rightarrow j}_{\bsbar{x}_i,\bsbar{x}_j}\chi^{j\rightarrow i}_{\bsbar{x}_j,\bsbar{x}_i}\,.\label{Zij def}
\end{align}
The derivation of these relations can also be found in Appendix~\ref{sec:derivation BDCM}.
The Bethe free entropy density calculated by~\eqref{phi B} from the converged \bdcm messages gives the asymptotically exact value of the free entropy density in the thermodynamic limit (again relying on self-averaging) even for \emph{non-tree} graphs if the \rs assumption holds. This assumption translates into the independence of the incoming messages to any factor node. We describe how to approach the case where the \rs assumption breaks down in the following and in Appendices~\ref{sec:derivation BDCM} and~\ref{sec: 1RSB}.

\subsection{Replica Symmetry Breaking for BDCM} \label{sec: RSB in main}
This section summarizes the one-step replica symmetry breaking version of the \bdcm; detailed derivations are deferred to Appendix~\ref{sec: 1RSB}. The concept of replica symmetry breaking was introduced by Parisi~\cite{Parisi_RSB_first} via the replica method and later formulated within the cavity method~\cite{bethe_lattice_revisted, CM_at_zero}, the approach we adopt here. A comprehensive treatment of the cavity formalism for RSB can be found in~\cite{phys_inf_comp}.

The assumption behind \oRSB is that the space of solutions splits into multiple distinct clusters, and each cluster is associated with a fixed point of the \bdcm equations. Let $P_{\mathrm{1RSB}}(\Tilde{\chi})$ be the following Boltzmann distribution over the fixed points  $\Tilde{\chi}=\{\chi^{i\rightarrow j}, \chi^{j\rightarrow i}\}_{(ij)\in E}$ with $\chi^{i\rightarrow j} = \{\chi^{i\rightarrow j}_{\bsbar{x}_i,\bsbar{x}_j}\}_{(\bsbar{x}_i,\bsbar{x}_j)\in S^{2(p+c)}}$:
\begin{equation}
    P_{\mathrm{1RSB}}(\Tilde{\chi}) = \frac{1}{Z_{\mathrm{1RSB}}(r)}e^{nr\Phi_{\mathrm{int}}(\Tilde{\chi})}\,.\label{prob dist over clusters}
\end{equation}
Here, $r$ is the \textit{Parisi parameter}.  $Z_{\mathrm{1RSB}}$ is the normalization and $\Phi_{\mathrm{int}}(\Tilde{\chi})$ the Bethe free entropy density computed from the fixed point $\Tilde{\chi}$ according to equation~\eqref{phi B}. We call it the \textit{internal free entropy density} as it is the free entropy density of a cluster. We define the \textit{replicated free entropy density} as
\begin{equation}
    \Psi(r)=\frac{\log Z_{\mathrm{1RSB}}(r)}{n}\,. \label{psi def}
\end{equation}
Furthermore, the entropy density of clusters with internal free entropy density~$\Phi_{\mathrm{int}}$ is given by the \textit{complexity} $\Sigma(\Phi_{\mathrm{int}})$
\begin{equation}
    \Sigma(\Phi_{\mathrm{int}}) =\frac{\log{\mathcal{N}(\Phi_{\mathrm{int}})}}{n}\,, \label{complexity def}
\end{equation}
where $\mathcal{N}(\Phi_{\mathrm{int}})$ is the total number of clusters with internal free entropy density $\Phi_{\mathrm{int}}$.

Expressing $Z_{\text{1RSB}}$ as an integral over the internal free entropies and applying the saddle-point method, we obtain in the thermodynamic limit
\begin{equation}
    \Psi(r) = r\hat{\Phi}_{\mathrm{int}} + \Sigma(\hat{\Phi}_{\mathrm{int}})\,, \label{psi =rphi + sigma}
\end{equation}
where $\hat{\Phi}_{\mathrm{int}}$ extremizes $\Psi(r)$. 
The \textit{total free entropy density} is defined as
\begin{equation} 
    \Psi_{\mathrm{tot}} = \Phi_{\mathrm{int}} + \Sigma(\Phi_{\mathrm{int}})\,, \label{psi tot def}
\end{equation}
where $\Phi_{\mathrm{int}}$ maximizes $\Psi_{\mathrm{tot}}$ under the constraint $\Sigma(\Phi_{\mathrm{int}})\geq 0$ (there must exist at least one cluster). Intuitively, \textit{typical solutions} are those that jointly maximize $\Phi_{\mathrm{int}}$ (the size of the clusters) and $\Sigma$ (the number of clusters). Ignoring the constraint $\Sigma(\Phi_{\mathrm{int}})\geq 0$, the total free entropy density is maximized for $r=1$ by $\hat{\Phi}_{\mathrm{int}}$.

If the complexity for $r=1$ is positive, the problem is in the \doRSB phase. The problem admits exponentially many thermodynamically relevant clusters with internal free entropy density $\Phi_{\mathrm{int}} = \hat{\Phi}_{\mathrm{int}}$. 

When the complexity is negative for $r=1$, $r$ must be changed until $\Sigma\geq 0$ to obtain a physically meaningful result. The total entropy is then maximized for $r<1$ such that $\Sigma=0$ (see Appendices~\ref{sec: 1RSB equations} and~\ref{sec: appendix s1RSB results} for additional details) the problem is said to be in the \soRSB phase. The problem admits a sub-exponential number of thermodynamically relevant clusters ($\Sigma=0$).

The \rs and \doRSB \bdcm estimates of the total entropy density~\eqref{phi B} are equivalent. In the presence of \soRSB, the \rs and \soRSB entropy are not equivalent, the later is smaller, and both provide an upper estimate of the exact asymptotic entropy density \cite{franz2003replica}. Note that additional steps of replica symmetry breaking are possible, but we do not check for them.

Computing the observables in the \oRSB setting is again done using the cavity method. This is described in Appendix~\ref{sec: 1RSB equations} together with the population dynamics algorithm (Appendix~\ref{sec: population dynamics}) used to approximate the solution of the \oRSB \bdcm equations, which are now over probability distributions.

\section{Results for Minority Takeover on Random Regular Graphs}\label{sec:BDCM for enhancing consensus}

We return to our initial question: \textit{``For majority dynamics, what is the smallest initial bias that leads to consensus after a fixed number of time steps?''}
To build some intuition for this specific dynamical system on regular graphs, it is useful to recall some of its general properties. In particular, majority dynamics admits attractors of exactly two lengths, $c\in\{1,2\}$~\cite{argument_for_c_leq_2}, and there are four dominant attractor types~\cite{BDCM}: the two $\pm1$ consensus states; a stable period-1 state with a finite fraction of both spins; and a period-2 state where some nodes flip repeatedly. The typical behavior of initial configurations with a fixed initial magnetization on a $4$-regular random graph illustrates the general phenomenology. Initializations with $m_{\rm  init} \lesssim -0.5$ converge\footnote{After possibly $p=2^n$ steps.} to a value of $-1$ in the attractor, and $m_{\rm  init} \gtrsim 0.5$ converge to $+1$. Figure~\ref{fig: typical dynamics +entropy plots} (left) shows the RS estimation of the entropy as function of the initial magnetization. The transition between the mixed state and $+1$ consensus empirically lies at $m^*_{\rm sample}(\infty) \sim 0.46$. This marks the point where upon random sampling of a configuration with $m_{\rm init} > m^*_{\rm sample} $ and a large enough time budget $p$, the nodes will eventually reach $+1$ consensus~\cite{BDCM}.

However, these results do not tell us whether initializations that reach $+1$ consensus and have the magnetization smaller than $m^*_{\rm sample}$  exist, whether they converge quickly, and whether they can be found algorithmically. In this section, we apply the formalism introduced in the previous section to estimate the minimal initial magnetization within the \rs assumption, and we study when this assumption breaks down with the \oRSB formalism.

\subsection{BDCM Equations for Majority Dynamics on Random Regular Graphs}\label{sec: BDCM eqs for majority dynamics on rrg}
To investigate the above questions, we will use the \bdcm to compute the smallest magnetization for which the replica symmetric entropy is non-negative, denoting 
\begin{align}
m_{\rm RS}^*(d,p)\;=\;\argmin_m s_{\rm RS}(m;d,p)\ge 0,\label{eq:lower-bound-rs}
\end{align} 
where  $s_{\rm RS}(m;d,p)$ denotes the replica-symmetric entropy of backtracking attractors that reach $+1$ within fixed $p$ steps.
Since the \rs entropy is an upper bound on the true asymptotic entropy, minimizing over $m_{\rm init}$ gives the lower bound on $m^*(p)$.

The \bdcm probability distribution~\eqref{prob dist BDCM factorized} for the majority dynamics simplifies to
\begin{equation}
    P(\bbar{\mathbf{x}}) = \frac{1}{Z} \prod_{i\in V} \underbrace{\left[e^{-\lambda_{\mathrm{init}}x_i^1} 
    \mathds{1}\left[x_i^{p+1}=+1\right]
    \prod_{t=1}^{p} \mathds{1}\left[x_i^{t+1}=f_i(x_i^t,\mathbf{x}^t_{\partial i})\right]\right]}_{\mathcal{A}(\bsbar{x}_i,\bsbar{\mathbf{x}}_{\partial i})}\,,\label{prob dist BDCM our case}
\end{equation}
where $\lambda_{\mathrm{init}}$ is the Lagrange parameter associated with the initial magnetization $m_\mathrm{init}(\bbar{\mathbf x}) = \frac{1}{n}\sum_{i} x_i^1$.

To simplify the equation further, we use the symmetry of \rrgs. In the thermodynamic limit, the neighborhood around each node becomes identical, and hence the fixed points of the \bdcm equations $\chi^{i \rightarrow j}_{\bsbar{x}_i,\bsbar{x}_j}$~\eqref{BDCM eqs} are the same for all $i,j$~\cite{thesis_Lenka}. Consequently, we can consider one edge-independent set of message values $\chi^{\rightarrow}_{\bsbar{x},\bsbar{y}}$, and find fixed points for all trajectory combinations $(\bbar{x},\bbar{y})\in \{\pm 1\}^{2(p+1)}$ of the following equation
\begin{align}
    \chi^{\rightarrow}_{\bsbar{x},\bsbar{y}} = \frac{1}{Z^{\rightarrow}} \sum_{\bsbar{\mathbf{y}}_{[d-1]}} \mathcal{A}\left(\bbar{x},\bbar{\mathbf{y}}_{[d]}\right) \prod_{\bsbar{y}\in\bsbar{\mathbf{y}}_{[d-1]}} \chi^{\rightarrow}_{\bsbar{y},\bsbar{x}}\,, \label{BDCM  eqs RRG}
\end{align}
where $\bbar{\mathbf{y}}_{[d-1]}$ are the trajectories $(\bbar{y}_1,\dots,\bbar{y}_{d-1})$ of the $d-1$ neighboring messages, and $Z^{\rightarrow}$ is the normalization as before. From the converged message values $\chi^{\rightarrow}_{\bsbar{x},\bsbar{y}}$, we compute the Bethe free entropy density as 
\begin{align}
    \Phi_B &= \log(Z^{\mathrm{fac}}) - \frac{d}{2} \log(Z^{\mathrm{var}})\,, \label{phi B RRG}\\
    Z^{\mathrm{fac}} &= \sum_{\bsbar{x},\bsbar{\mathbf{y}}_{[d]}} \mathcal{A}(\bbar{x},\bbar{\mathbf{y}}_{[d]}) \prod_{y\in \bsbar{\mathbf{y}}_{[d]}} \chi^{\rightarrow}_{\bsbar{y},\bsbar{x}}\,,\label{Zi def RRG}\\
    Z^{\mathrm{var}} &= \sum_{\bsbar{x},\bsbar{y}} \chi^{\rightarrow}_{\bsbar{x},\bsbar{y}}\chi^{\rightarrow}_{\bsbar{y},\bsbar{x}}\,.\label{Zij def RRG}
\end{align}
Note that the obtained quantities are now averages over the ensemble of \rrgs that thanks to self-averaging capture the behavior of typical large random graphs (modulo the RS assumption). Using $\Phi_B$, we can again compute the entropy density $s(m_{\rm{init}})$ using equations~\eqref{s via phi general} and~\eqref{der phi to xi_k}. This quantity provides the exponentially leading number of valid $(p+1)$ backtracking attractors with $m_{\rm{attr}}=1$, conditioned on a fixed value of $m_{\rm{init}}$. Knowing the function $s(m_{\rm{init}})$ allows us to analyze the system's behavior when initialized with a given $m_{\rm{init}}$. In particular, $m_{\rm{init}}$ such that $s(m_{\rm{init}})=0$ is the minimal initial magnetization of trajectories that reach full consensus in $p$ steps, i.e. it is the $m_{\rm RS}^*(d,p)$ from \eqref{eq:lower-bound-rs}.

\subsection{Replica Symmetric Results}\label{sec:rs-lower-bounds}

\begin{figure}
    \centering
    \includegraphics[width=0.49\textwidth]{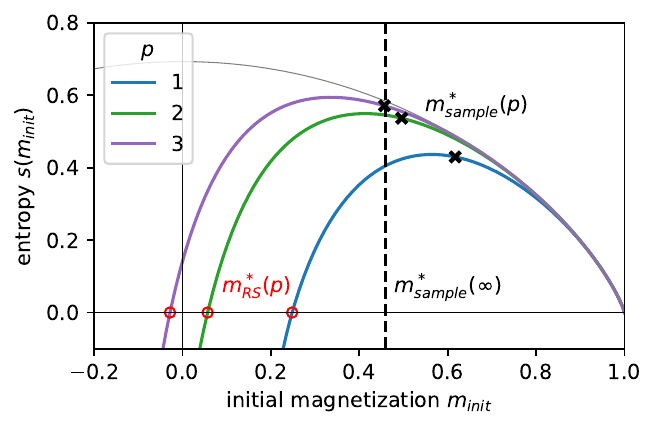}\vspace{-1em}
    \includegraphics[width=0.49\textwidth]{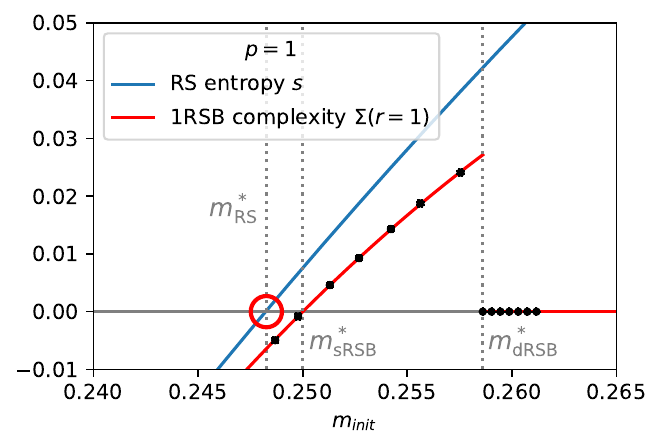}
        \caption{\textbf{Always-stay majority dynamics on $4$-regular random graphs.}
        \textit{(Left)} Replica symmetric entropy on initializations going to a positive consensus in $p$ steps as a function of the initial magnetization obtained from \bdcm for $d=4$ and $p=1,2,3$. We mark the replica symmetric lower bound $m^*_{\rm RS}(p)$, and the magnetization $m_{\rm sample}^*(p)$ as the one at which $+1$ consensus becomes the entropically dominating attractor~\cite{BDCM}. The gray line represents the total entropy of all configurations.
        \textit{(right)} Zoom into the region where the entropy becomes negative for the $p=1$ case. The red line shows the complexity (eq.~\ref{complexity def}) at Parisi parameter $r=1$. The dashed grey lines denote the initial magnetizations at which the \doRSB and \soRSB phases emerge, as well as the point where the \rs entropy crosses $0$.
        }
    \label{fig: typical dynamics +entropy plots}
\end{figure}

The analysis of~\cite{BDCM} shows that running the dynamics until convergence from a random initialization with $m_{\rm init} > m^*_{\rm sample}(d)$ will typically reach $+1$ consensus.
For a fixed $p$, $m^*_{\rm sample}(p,d)$ describes the smallest value for which a randomly sampled backtracking attractor of fixed path length $p$ typically reaches $+1$ consensus\footnote{Under the replica symmetric assumption, but this was checked to be correct for these values of $m_{\rm init}$ in~\cite{BDCM}}.
It can be obtained by comparing the entropy of the backtracking attractors with the four other possible attractors and then determine which one is entropically dominant. This is shown as the dashed line for large $p$ and crosses for $p=1,2,3$ in Figure~\ref{fig: typical dynamics +entropy plots} (left), and also in Table~\ref{tab:1RSB}.

However, while $m^*_{\rm sample}(p,d)$ characterizes typical initializations when $+1$ consensus dominates entropically, we are interested in exploring the more extreme cases: initial configurations that start farthest from the majority yet still reach full $+1$ consensus.
Accordingly, we restrict our analysis to backtracking attractors that lead to consensus and consider the smallest $m_{\rm init}$ under the replica-symmetric assumption, called $m_{\rm RS}^*(d, p)$, for which their entropy is non-negative~\eqref{eq:lower-bound-rs}.
The entropy of these backtracking attractors is shown in Figure~\ref{fig: typical dynamics +entropy plots} (left) as a function of the initial magnetization for $d=4$. We mark the points $m_{\rm RS}^*(p)$ where the entropy becomes negative. This approach predicts the existence of initializations in $4$-\rrgs with negative magnetization (up to $-0.0281(1)$) that evolve into the all-ones attractor within $p=3$ steps under majority dynamics. 

\begin{table}[h]
  \centering
  \small
    \centering
\begin{tabular}{c *{6}{S[table-format=1.3]}}
\toprule
{$d$} & \multicolumn{6}{c}{$p$} \\
\cmidrule(lr){2-7}
 & 1 & 2 & 3 & 4 & 5 & 6 \\
\midrule
3 & 0.369 & 0.188 & 0.106 & 0.065 & 0.046 & 0.033 \\
4 & 0.248 & 0.057 & \bftabnum -0.028 & \bftabnum  -0.076 &  &  \\
5 & 0.261 & 0.057 & \bftabnum -0.040 &  &  &  \\
6 & 0.206 & 0.010 & \bftabnum -0.079 &  &  &  \\
7 & 0.213 & 0.009 & \bftabnum -0.086 &  &  &  \\
8 & 0.179 & \bftabnum -0.015 & \bftabnum -0.100 &  &  &  \\
9 & 0.184 & \bftabnum -0.016 & \bftabnum -0.106&  &  &  \\
10 & 0.161 & \bftabnum -0.029 \\
11 & 0.164 & \bftabnum -0.031 \\
12 & 0.147 & \bftabnum -0.038 \\
\bottomrule
\end{tabular}
\caption{Values for $m_{\mathrm{RS}}^*(d,p)$, the RS minimal initial magnetization leading to consensus in $p$ steps for majority dynamics and always-stay tie breaking on $d$-\rrgs. It is given by the last $m_{\rm{init}}$ for which the \rs \bdcm calculation returns a non-negative entropy density. We highlight the negative values of $m_{\rm{init}}$, which indicate that there might exist initializations with a \emph{minority} of $+1$ opinions that still lead to a full $+1$ consensus.}
\label{tab:bdcm-results}
\end{table}

This is not only much lower than the randomly sampled initializations $m^*_{\rm sample}$, but shows that there exist initializations with a minority of $+1$'s where they become the majority through majority dynamics.
While it is perhaps unsurprising that such initializations should exist for regular but adversarially chosen graphs with carefully designed structure, this result indicates that such initializations exist for typical randomly sampled graphs.

This phenomenon is not restricted to $4$-regular graphs. In Table~\ref{tab:bdcm-results} we show the minimal initial magnetization that achieves zero entropy for different combinations of degrees $d$ and trajectory lengths $p$. 
We highlight the lower bounds $m_{\rm{RS}}^*$ that are negative in bold, as only in those cases minority-initialized configurations can potentially achieve full consensus. This is the case for all $d\geq8$ with trajectory lengths $p=2$ or for $d\geq4$ with $p=3$.
For $p=1$ finding such typical initializations for a smaller than zero bias seems infeasible.

\begin{figure}[h!]
    \centering
    \includegraphics[width=\linewidth]{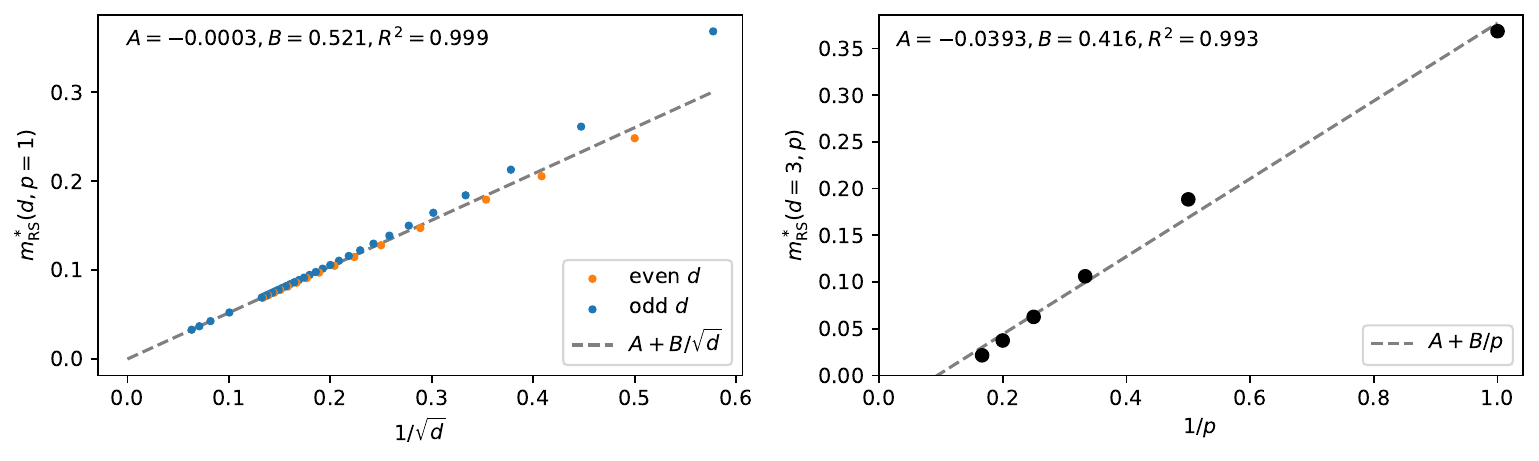}
    \vspace{-1em}
    \caption{Extrapolation of the RS minimal initial magnetization leading to $+1$ consensus under majority dynamics with always-stay tie breaking on $d$-\rrgs. (\textit{Left}) $m^*_{\rm RS}(d)$ for $p=1$ as a function of $1/\sqrt{d}$.  The linear fit for points with $d\geq40$ ($1/\sqrt{d}\leq0.158$) is shown. (\textit{Right}) $m^*_{\rm RS}(p)$ for $d=3$ as a function of $1/p$. All points are used in the fit.}
    \label{fig: extrapolations d3 and p1}
\end{figure}

In Figure~\ref{fig: extrapolations d3 and p1} (left) we extrapolate the finite $d$ results up until $d\leq249$.
Even degree settings for lower $d$s show lower minimal $m_{\mathrm{init}}$. This stems from the tie breaking applied in even degrees.
The initial magnetization follows a scaling as $m_{\mathrm{RS}}^*(d) \sim 1/\sqrt{d}$, eventually reaching zero for $d=\infty$.
This indicates that for $p=1$ $m_{\mathrm{init}}<0$ is not reached. 
This is consistent with the result expected for a fully connected graph with $d=n-1$: More than half of the nodes need to be in state $+1$ to arrive at consensus in a single step, hence we must have $m_{\rm init} > 0$ if we want to arrive there in a single step.

For the case $d=3$ we were not able to directly find a $p$ which admits a negative $m_{RS}^*(p)$. 
An extrapolation scaling as $1/p\to 0$, Figure~\ref{fig: extrapolations d3 and p1} (right), indicates that our results would be compatible with the case $m_{\rm{RS}}^*(p)<0$ for $p\geq 13$. However, given the scarcity of data points, this extrapolation should be treated with particular caution. We leave the problem of finding initializations for which minority takes over majority for $3$-random regular graphs as an open question. 

Importantly, the results in this section are only asymptotically exact under the replica-symmetric assumption. 
In~\cite{BDCM}, stability checks of the \rs equations were performed only for the entropically dominant solutions with $m_{\rm init} > m_{\rm sample}^*(p,d)$.
There, the results indicated that the \rs assumption held true.
However, as the initial magnetization $m_{\rm init}$ decreases, the constraints on the system become increasingly difficult to satisfy, as evidenced by the decreasing entropy shown in Figure~\ref{fig: typical dynamics +entropy plots}. Consequently, it is natural to expect the solution space to become increasingly clustered at low $m_{\rm init}$ values, indicating that the \oRSB assumption may be required.

\subsection{Replica Symmetry Breaking for Small Initial Magnetizations}\label{sec:RSB for majority dynamics}
\begin{table}
 \centering
    \setlength{\tabcolsep}{4pt} 
    \begin{tabular}{ccllll} 
    \toprule
    $p$ &  &\multicolumn{4}{c}{$d$} \\
\cmidrule(lr){3-6}
       &  & $\phantom{-}3$ & $\phantom{-}4$ & $\phantom{-}5$ & $\phantom{-}6$\\ [1ex] 
    \midrule
    $1$ & $m^*_{\mathrm{sample}}$ & $\phantom{-}0.777$ & $\phantom{-}0.617$ & $\phantom{-}0.443$& $\phantom{-}0.398$\\
     &$m^*_{\mathrm{dRSB}}$ & $\phantom{-}0.379$ & $\phantom{-}0.258$ & $\phantom{-}0.274$ & $\phantom{-}0.221$ \\ 
     &$m^*_{\mathrm{sRSB}}$ & $\phantom{-}0.376$ & $\phantom{-}0.250$ & $\phantom{-}0.263$  & $\phantom{-}0.206$  \\ 
     &$m^*_{\mathrm{RS}}$ & $\phantom{-}0.369$ & $\phantom{-}0.247$ & $\phantom{-}0.261$ & $\phantom{-}0.206$\\
     \midrule
     $2$& $m^*_{\mathrm{sample}}$ & $\phantom{-}0.692$ & $\phantom{-}0.496$ & $\phantom{-}0.302$ & $\phantom{-}0.255$ \\
     &$m^*_{\mathrm{dRSB}}$ & $\phantom{-}0.199^*$ & $\phantom{-}0.075^*$ & $\phantom{-}0.081^*$& $\phantom{-}0.039^*$\\
     &$m^*_{\mathrm{sRSB}}$ & $\phantom{-}0.191$ & $\phantom{-}0.064$ & $\phantom{-}0.063$& $\phantom{-}0.015$\\
     &$m^*_{\mathrm{RS}}$ & $\phantom{-}0.188$ & $\phantom{-}0.056$ & $\phantom{-}0.057$ & $\phantom{-}0.010$ \\
     \midrule
     $3$& $m^*_{\mathrm{sample}}$ & $\phantom{-}0.643$ & $\phantom{-}0.457$ & $\phantom{-}0.231$ & $\phantom{-}0.188$ \\
     &$m^*_{\mathrm{dRSB}}$ & $\phantom{-}0.117^*$ & $-0.005^*$ & \phantom{-0.}- & \phantom{-0.}-\\ 
     &$m^*_{\mathrm{RS}}$ & $\phantom{-}0.106$ & $-0.028$ & $-0.039$ & $-0.079$\\
     \bottomrule
    \end{tabular}
    \caption{Thresholds of the initial magnetization where typical initializations cease to go to $+1$ consensus ($m_\mathrm{sample}^*$), where \doRSB appears ($m_\mathrm{dRSB}^*$), where \soRSB appears ($m_\mathrm{sRSB}^*$) and where the \rs entropy crosses zero ($m_\mathrm{RS}^*$). The values $m^*_{\rm{sample}}$ for $d=3,4,5$ are from~\cite{dynamical_PT_GCA}, while the case $d=6$ is discussed in Appendix~\ref{sec: appendix supplementary results m sample}. The values with an asterisk are an upper bound on the \doRSB phase, stemming from the last \rs stable values of $m_{\mathrm{init}}$. For $p=3$ and $d=5,6$, numerical instabilities in the population dynamics procedure rendered the identification of the \doRSB phase inconclusive. Similarly, the \soRSB threshold is not noted for $p=3$ (see Appendix~\ref{sec: appendix s1RSB results}).}
    \label{tab:1RSB}
\end{table}

\begin{figure}[h!]
    \centering
    \includegraphics[width=1.0\linewidth]{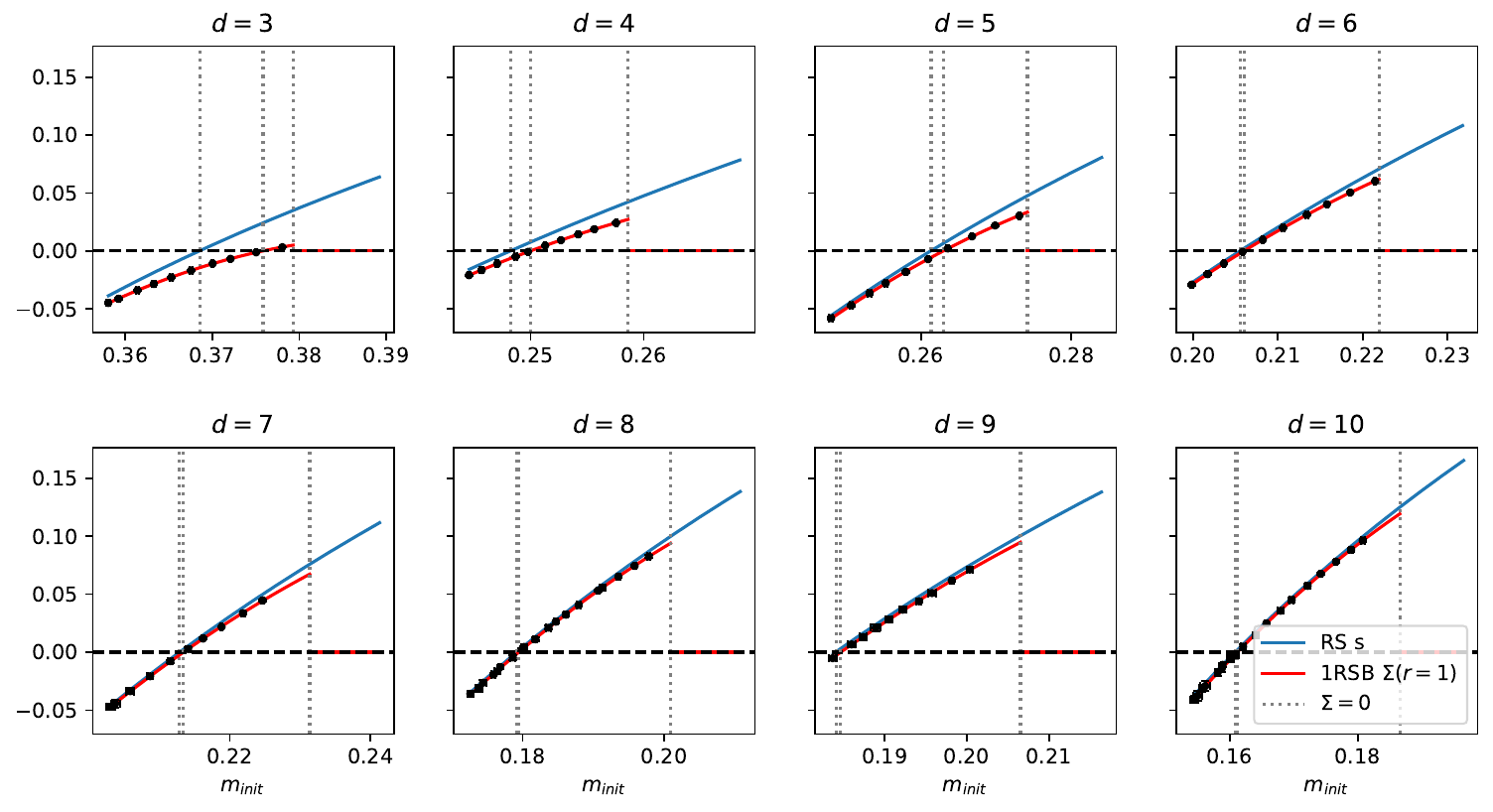}
    \caption{Thresholds of $m^*_{\rm RS}, m^*_{\rm sRSB}$ and $m^*_{\rm dRSB}$ (vertical lines from left to right) for $p=1$ and different values of $d$. Each black point represents the complexity $\Sigma$ obtained from the 1RSB equations for a single $\lambda$, and the error on the sample mean is indicated by error bars in $x$ and $y$ over 10 samples. The red curve is a quadratic fit of this data, which ends at $m^*_{\rm dRSB}$ from Table~\ref{tab:1RSB}.}
    \label{fig:sRSB-vis}
\end{figure}
    
We recall that Section~\ref{sec: RSB in main} introduces the replica symmetry breaking formalism and that Appendix~\ref{sec: 1RSB equations} presents the derivation of the \oRSB BDCM equations. Using the population dynamics procedure described in Appendix~\ref{sec: population dynamics}, we find a numerical solution to these equations and identify the value of $\lambda_{\mathrm{init}}$ at which the \doRSB transition occurs and then compute the corresponding initial magnetization $m_{\mathrm{dRSB}}^*$. Similarly, we find the initial magnetization $m_{\mathrm{sRSB}}^*$. These values are noted in Table~\ref{tab:1RSB}. 

The structural changes of the solution space associated with the 1RSB are captured by the complexity $\Sigma$, defined in~\eqref{complexity def}. The complexity $\Sigma$ becomes strictly positive at $m^*_{\mathrm{dRSB}}$ when the problem enters the \doRSB phase, as exponentially many clusters appear. In this phase, the geometry of the solution space changes, but the observables of the problem (entropy, magnetization, etc.) remain the same as in the \rs phase. The \rs observables of the problem cease to be asymptotically exact in the \soRSB phase, when the complexity at $r=1$ becomes negative. This is illustrated in Figure~\ref{fig: typical dynamics +entropy plots} (right) in the case $p=1, d=4$. The figure shows the \rs entropy and the complexity~\eqref{complexity def} with Parisi parameter $r=1$ as a function of $m_{\mathrm{init}}$. Figure~\ref{fig:sRSB-vis} shows the entropy and complexity curves for other degrees and $p=1$, while higher $p$s are shown in Appendix~\ref{sec: appendix s1RSB results}. 

Another question of interest that can be answered by the \oRSB analysis is whether the solutions are \textit{frozen}. A node trajectory from a given solution is said to be frozen if it takes the same value for all solutions within its cluster. If a cluster contains an extensive number of frozen trajectories, it is said to be \textit{frozen}. The emergence of an extensive number of frozen trajectories in \emph{typical} solutions marks the \textit{rigidity transition}, and this phase is called rigid. The transition at which \emph{all} clusters (including atypical ones) become frozen is known as the \textit{freezing transition}, and the corresponding phase is referred to as frozen~\cite{PT_coloring_hypergraphs}.

We can estimate the fraction of frozen trajectories from the approximate solution obtained from the population dynamics. This is done by counting the number of messages in the population that imply a fixed trajectory value, that is, messages that are ``delta peaked'' at a specific trajectory. Our numerical results show that, as soon as the \doRSB phase is entered, a finite fraction of the messages are almost peaked, in that they assign a probability close to $1$ to a specific trajectory. The obtained probability distributions are shown in Figure~\ref{fig:histograms-1RSB} in Appendix~\ref{sec:numerical evidence freezing}. However, this probability is not exactly $1$, which we attribute to numerical approximations. Alternatively, it is also possible that the messages are in fact only ``almost peaked'' (these are called \textit{quasi-hard fields} in \cite{PT_coloringgraphs_LenkaFlo}). In this case, distinguishing freezing and quasi-freezing is hard to do numerically. Nonetheless, we favor the hypothesis of true freezing, as the threshold closely matches the performance of the \hpr algorithm, as will be discussed below. Moreover, Figure~\ref{fig:sRSB-vis} shows that for larger values of
$d$, the complexity approaches the entropy. This indicates that the solution space becomes highly clustered, with only a small number of solutions contained within each cluster. Although this phenomenon is not necessary for freezing, it is typical as few solutions within each cluster are often related to many frozen variables. Numerical experiments and further discussion on the freezing can be found in Appendix~\ref{sec:numerical evidence freezing}.

\section{History Passing Reinforcement}\label{sec:HPR}

We are inspired by the well-known breakthrough of the survey propagation algorithm \cite{mezard_analytic_2002} that turned a cavity method solution into an algorithm that was able to solve unprecedentedly hard and large random satisfiability instances. Here we use the same idea for the replica symmetric version of the analysis method.  
The \hpr algorithm is a \textit{dynamical} variation of the belief propagation reinforcement procedure~\cite{chavas_survey-propagation_2005, braunstein_encoding_2007} that utilizes the \bdcm described in Section~\ref{sec:BDCM}. For a particular $d$-\rrg, we want to find its initialization that leads to a $+1$ consensus in $p$ steps, but starts with $+1$ nodes in minority. We refer to such an initialization as a \textit{solution}. 

The \bdcm messages $\chi_{\bsbar{x}_i,\bsbar{x}_j}^{i\rightarrow j}$ provide estimates of marginal probabilities for node trajectories. Thus, running the \bdcm equations on a given graph offers guidance on the structure of promising initializations. The \hpr algorithm exploits these marginals by reinforcing messages corresponding to higher-probability trajectory values, while suppressing those with lower probability. For example, if a node is likely to start in state $-1$, this information is propagated to its neighbors through reinforced updates. Iterative reinforcement in this way sharpens the marginals to better satisfy the constraints, gradually biasing them toward the desired initial node values.

Concretely, this leads to the addition of biases $b^i_{x^1_i}$ to every message $\chi^{i \rightarrow j}_{\bsbar{x}_i,\bsbar{x}_j}$  in the \bdcm fixed-point equation:
\begin{equation}
    \chi^{i \rightarrow j}_{\bsbar{x}_i,\bsbar{x}_j} = \frac{1}{Z^{i \rightarrow j}} \sum_{\bsbar{\mathbf{x}}_{\partial i\backslash j}} \mathcal{A}_i(\bbar{x}_i,\bbar{\mathbf{x}}_{\partial i})\!\!\prod_{k\in\partial i\backslash j}\!b^k_{x^1_k}\color{black}\chi^{k \rightarrow i}_{\bsbar{x}_k,\bsbar{x}_i}\,. \label{biased BDCM}
\end{equation}
We consider binary states, so we have two scalar bias values $b^i_1$ and $b^i_{-1}$ for each message. The following heuristic is used to update the bias values:
\begin{equation}
    \begin{split}
b^i_1&=\pi,\phantom{1-}\hspace{0.48cm} b^i_{-1} = 1-\pi,\hspace{0.3cm} \text{if}\,\, \mu^i_{x_i^1=1} < \mu^i_{x_i^1=-1}\,, \\
     b^i_1&=1-\pi,\hspace{0.4cm} b^i_{-1} =\pi,\phantom{1-}\hspace{0.38cm} \text{if}\,\, \mu^i_{x_i^1=1} \geq \mu^i_{x_i^1=-1}\,, \label{biases update general case}
    \end{split}
\end{equation}
where $\pi\in [0,1/2]$ is a hyperparameter. The marginal $\mu^i_{x_i^1=s}$, with $s\in\{\pm1\}$ is the \bdcm estimate on the marginal probability that $x_i^1=s$ for the variable node $(\bbar{x}_i, \bbar{x}_j)$.  Note that setting $\pi=1/2$ results in the unmodified \bdcm equations due to normalization, and lowering $\pi$ leads to stronger reinforcement. The marginals are computed as
\begin{equation}
    \mu^i_{x_i^1=s} = \frac{1}{Z_{\mu}}\prod_{j\in\partial i}Z^{ij}_{x_i^1=s}\,, \label{marginals def}
\end{equation}
where $Z_{\mu}$ is the normalization of the marginals: $Z_{\mu}=\sum_{s\in\{\pm 1\}}\prod_{j\in\partial i}Z^{ij}_{x_i^1=s}$, and $Z^{ij}_{x_i^1=s}$ is defined as
\begin{equation}
    Z^{ij}_{x_i^1=s} = \frac{1}{Z_c} \sum_{\bsbar{x}_i,\bsbar{x}_j} \mathds{1}\left[x_i^1=s\right] \chi^{i \rightarrow j}_{\bsbar{x}_i,\bsbar{x}_j}\chi^{j \rightarrow i}_{\bsbar{x}_j,\bsbar{x}_i}\,, \label{Zij conditioned definition}
\end{equation}
where again $Z_c$ is the normalization $Z_c=\sum_{s} Z^{ij}_{x_i^1=s}$. 
Detailed derivations of these relations are given in Appendix~\ref{sec:derivation marginals}.

Furthermore, each bias $b^i_{x_i^1}$ is updated independently according to~\eqref{biases update general case} only with probability 
\begin{equation}
    p(t) = 1 - (1+t)^{-\gamma}\,, \label{bias update probability}
\end{equation}
where $t$ is the current value of the iteration step and $\gamma$ is an additional hyperparameter. Note that other choices of the update scheme~\eqref{biases update general case} and the probability~\eqref{bias update probability} exist. We adopt the scheme from~\cite{zdeborova_constraint_2008} that provided good results in our setting.

The \bdcm probability distribution is set up so that only the graph trajectories respecting the graph dynamics have nonzero measure, and the observables can be tuned via the Lagrange parameters. Hence, by computing the marginals~\eqref{marginals def}, we get estimates of the probability that each node $i$ starts with value $s$ under these conditions. Since biases reflect the relative values of the marginals for each node trajectory~\eqref{biases update general case}, we can use them to obtain our guess of a solution $\mathbf{x}^{\mathrm{sol}}\in S^n$: 
\begin{equation}
    x_i^{\rm{sol}} = \underset{s\in \{\pm 1\}}{\text{argmax}}\hspace{0.2cm} b^i_s\,, \hspace{0.2cm} \forall i=1,\dots,n \label{trial solution}
\end{equation}
so $s\in\{\pm 1\}$ with a larger bias (hence also with higher marginal) is taken as a trial solution value for given $i$.

\begin{algorithm}[t]
\caption{\textbf{H}istory \textbf{P}assing \textbf{R}einforcement\\ (factor graph $(G, S, F, \{\Xi_k\}, \{\lambda_k\},p,c),$ steps threshold $T_{\rm{max}},$ damping $\varepsilon,$ hyperparameters: $\pi, \gamma$)}\label{alg:HPR}
\begin{algorithmic}
\item Initialize uniformly at random biases $b^i_{x_i^1}$ and messages $\chi^{i \rightarrow j}_{\bsbar{x}_i,\bsbar{x}_j}$ on the factor graph and normalize them. 
\State $t \gets 0$
\While{ $t\leq T_{\rm{max}}$}
\State Update all the messages $\chi^{i \rightarrow j}_{\bsbar{x}_i,\bsbar{x}_j}$ according to~\eqref{biased BDCM} with damping $\varepsilon$~\eqref{dampened update}. 
\State Compute the marginals $\mu^i_{x_i^1=s}$ from eq.~\eqref{marginals def}. 
\State Update each bias independently according to~\eqref{biases update general case} with probability $p(t)$~\eqref{bias update probability}.
\State Obtain new trial solution $\mathbf{x}^{\mathrm{sol}}_t$ using~\eqref{trial solution}.
\State $t \gets t+1$
\EndWhile
\State \Return $\mathbf{x}^{\mathrm{sol}}_t$ with lowest initial magnetization and which is a solution.
\end{algorithmic}
\end{algorithm}
After every iteration of the message update equations~\eqref{biased BDCM}, we update the biases~\eqref{biases update general case} with probability~\eqref{bias update probability}. This probability is low at the beginning: we essentially update the messages according to the \bdcm equations and gather information towards the correct initialization of the graph. Progressively, the biases are changed in accordance with the marginals. This reinforces our beliefs about the trial solution, and if this trial solution aligns with the constraints (dynamics and initial magnetization) we get a true solution; if not, the biases continue to change, and hence also the messages until we get this alignment. Figure~\ref{fig: HPR visualization} (left) from appendix~\ref{sec:additional figures HPR} captures this development for a specific example. 

To increase the stability of the algorithm during the message update~\eqref{BDCM eqs}, we introduce a damping parameter $\varepsilon\in(0,1)$ as follows:
\begin{equation}
    \chi^{i \rightarrow j}_{\bsbar{x}_i,\bsbar{x}_j}(\mathrm{final}) = \varepsilon\chi^{i \rightarrow j}_{\bsbar{x}_i,\bsbar{x}_j}(\mathrm{after}) + (1-\varepsilon)\chi^{i \rightarrow j}_{\bsbar{x}_i,\bsbar{x}_j}(\mathrm{before})\,. \label{dampened update}
\end{equation}
The \textit{after} and \textit{before} denote the message value after and before the update~\eqref{BDCM eqs} and \textit{final} is the final updated message value with damping. 
 
\paragraph{Generalization} The \hpr procedure extends naturally beyond the present setting. It can incorporate general hard constraints on trajectory values (not limited to the attractor states considered here). Additional observables and their associated Lagrange multipliers can also be included. Furthermore, the framework applies to systems with more than two node states ($|S|>2$) and to optimization problems involving edge observables $\Bar{\Xi}$. These generalizations require introducing additional hyperparameters ${\pi_k}$ with $\sum_k \pi_k =1$, otherwise the procedure remains unchanged.

\paragraph{Algorithmic complexity} One iteration of \hpr requires updating all messages $\chi^{i \rightarrow j}_{\bsbar{x}_i,\bsbar{x}_j}$ according to Equation~\eqref{biased BDCM}. On a given graph, there are $2|E|\,2^{2(p+c)}$ message values. For sparse graphs where the number of edges grows linearly with the system size $n$, this leads to a linear scaling with $n$ which can easily be parallelized. However, each message update~\eqref{biased BDCM} goes over all trajectory combinations $\bbar{\mathbf{x}}_{\partial i\backslash j}$, amounting to $2^{(d(i)-1)(p+c)}$ terms, where $d(i)$ is the degree of node $i$. The marginals~\eqref{Zij conditioned definition} require a sum over $2^{2(p+c)}$ terms, while the update of the biases~\eqref{biases update general case} and creation of a solution~\eqref{trial solution} again require $\sim n$ operations which can be parallelized. This results in a complexity of $\mathcal{O}(n 2^{(d_{\mathrm{max}}+1)(p+c)})$, with an exponential dependence on $d_{\mathrm{max}}, p, c$.

However, for factor nodes or dynamical rules that do not depend on the ordering of the neighbors, we can devise a dynamic programming scheme~\cite{dynamic_programming_first, dynamic_programming_second} that eliminates the exponential dependence on $d$. 
This is the case for the majority rule~\eqref{majority dynamics 1}, which depends only on the sum of the neighbors at each time step. Thus, the complexity reduces to $\mathcal{O}(n d^{(p+c)})$, where the main limitation comes from the exponential dependence, as the HPR equations can be solved in parallel for each message and node. Appendix~\ref{sec: dynamic programming appendix} describes this scheme and provides an example of dynamic programming applied to \hpr updates in the setting of majority dynamics on \rrgs.  

\section{HPR for Minority Takeover on Random Regular Graphs}\label{sec:hpr-results}

\begin{table}
    \centering
    \begin{tabular}{ccllll} 
    \toprule
    $p$ &  &\multicolumn{4}{c}{$d$} \\
\cmidrule(lr){3-6}
       &  & $\phantom{-}3$ & $\phantom{-}4$ & $\phantom{-}5$ & $\phantom{-}6$\\ [1ex] 
    \midrule
    $1$ & $m^*_{\mathrm{sample}}$ & $\phantom{-}0.777$ & $\phantom{-}0.617$ & $\phantom{-}0.443$& $\phantom{-}0.398$\\
      & $m^*_\mathrm{HPR}$ & $\phantom{-}0.384(1)$ & $\phantom{-}0.263(9)$ & $\phantom{-}0.278(9)$  & $\phantom{-}0.231(5)$\\ 
     & $m^*_\mathrm{dRSB}$ & $\phantom{-}0.379$ & $\phantom{-}0.258$ & $\phantom{-}0.274$ & $\phantom{-}0.221$ \\ 
     \midrule
     $2$& $m^*_{\mathrm{sample}}$ & $\phantom{-}0.692$ & $\phantom{-}0.496$ & $\phantom{-}0.302$ & $\phantom{-}0.255$ \\
     & $m^*_\mathrm{HPR}$ & $\phantom{-}0.210(0)$ & $\phantom{-}0.086(9)$ & $\phantom{-}0.091(4)$ & $\phantom{-}0.049(1)$\\ 
     & $m^*_\mathrm{dRSB}$ & $\phantom{-}0.199^*$ & $\phantom{-}0.075^*$ & $\phantom{-}0.081^*$& $\phantom{-}0.039^*$\\
     \midrule
     $3$& $m^*_{\mathrm{sample}}$ & $\phantom{-}0.643$ & $\phantom{-}0.457$ & $\phantom{-}0.231$ & $\phantom{-}0.188$ \\
     & $m^*_\mathrm{SA} $ & $\phantom{-}0.17(1)$ & $\phantom{-}0.06(1)$ & $\phantom{-}0.04(1)$ & $\phantom{-}0.02(1)$ \\
     &$m^*_\mathrm{HPR}$ & $\phantom{-}0.122(0)$ & $-0.003(9)$ & $-0.004(8)$ & $-0.035(3)$ \\ 
     & $m^*_\mathrm{dRSB}$ & $\phantom{-}0.117^*$ & $-0.005^*$ & \phantom{-0.}- & \phantom{-0.}-\\ 
     \bottomrule
    \end{tabular}
    \caption{Comparison of the best achievable $m_{\rm init}$ with algorithms to the threshold where typical initialization cease to evolve to a $+1$ consensus $m^*_{\rm{sample}}$ and \doRSB threshold for $d$-\rrg. The numbers in brackets denote the standard deviation.
    The values $m^*_{\rm{sample}}$ and $m^*_{\rm{dRSB}}$ are from Table~\ref{tab:1RSB}.
    We ran \hpr for $10,000$ steps and record the best trial solution that reached consensus within $p$ steps. Results are then averages over 10 runs on graphs of size $n=10,000$.
    The values with an asterisk are an upper bound on the \doRSB phase, stemming from the last \rs stable values of $m_{\mathrm{init}}$. Also shown are the minimal values of $m_{\mathrm{init}}$ reached by the \sa procedure. Note that \emph{subzero} values of $m_{\mathrm{init}}$ correspond to states where minority becomes majority. Unlike SA, \hpr is able to explicitly find such states for $d\geq 4, p=3$, 
    as predicted by the \bdcm. } 
    \label{tab: all results}
\end{table}
Our goal is to find an initialization for a specific instance of a large random graph that attains the lowest possible magnetization, while reaching $+1$ consensus within $p$ steps. We follow the procedure described in the \hpr Algorithm~\ref{alg:HPR}, complemented by the dynamic programming implementation in Algorithm~\ref{alg:HPR with dynamic prog}, and discuss the impact of replica symmetry breaking.

Table~\ref{tab: all results} presents the minimal initial magnetization $m^*_{\mathrm{HPR}}$ obtained using \hpr for varying degrees $d$ and path lengths $p$. To obtain these values, we generate  $10$ random instances of $d$-RRG with $n=10^4$ nodes and perform the HPR procedure for each combination of $d$ and $p$. The complete \hpr parameters are listed in Table~\ref{tab:hyperparameters HPR} in Appendix~\ref{sec: appendix parameters HPR and lambda behavior}.
The resulting initial magnetizations leading to consensus are then averaged. In particular, for $d\geq 4$, $p\geq 3$ we observe that \hpr indeed finds initializations with subzero magnetization: initializations where a minority quickly turns into the majority.

The values achieved by the \hpr are significantly lower than the values $m^*_{\rm{sample}}$. We recall that  $m^*_{\rm{sample}}$ denotes the threshold below which the $+1$ consensus stops being the entropically dominating attractor, i.e. the lowest value of $m_{\mathrm{init}}$ for which the dynamics ends in the homogeneous attractor when initialized uniformly at random as $n\rightarrow\infty$. This underscores the impact of strategic initialization.
Moreover, as noted in Section~\ref{sec:related work}, the dynamical difference between strategic and typical initializations can be seen as an Mpemba-like effect. Using \hpr, we can indeed find such initializations that are ``hotter'' (farther from the all-ones consensus) but nevertheless reach equilibrium faster than colder, randomly initialized states.

In Table~\ref{tab: all results} we further compare the \hpr-obtained initial magnetization $m^*_\mathrm{HPR}$ with those obtained via an alternative algorithmic strategy -- simulated annealing (\sa)~\cite{simulated_annealing}. We call the minimal initial magnetization obtained from \sa $m^*_\mathrm{SA}$. We describe the  \sa procedure in the context of majority dynamics on $d$-RRGs in Appendix~\ref{sec: SA approach appendix}. The \sa parameters are summarized in Table~\ref{tab: parameters for SA} and, due to slow convergence, $n=3\,000$ was used. 
Note that, for $p=3$, the  \sa algorithm was unable to reach subzero magnetization initializations, even when the \hpr was. More extensive comparison (including $p\leq3$, different $n$s and a check of a longer path to consensus for obtained initializations) of the SA and HPR performances can be found in Appendices~\ref{app:History Passing}, \ref{sec: SA approach appendix}. 

The analysis above focuses on solutions whose initial configurations reach full consensus within at most $p$ steps. However, both the \hpr and \sa algorithms can generate initial configurations that, while not satisfying this constraint, nonetheless converge to consensus quickly, possibly starting from even a lower $m_{\rm init}$. In this sense, \hpr can serve as a heuristic for identifying configurations that reach consensus after an effective time $T_{\rm eff} > p$, even though the optimization only enforces behavior up to time $p$. We did not explore this direction systematically, since doing so would require controlling the magnetization at time $p$. However, we note that constraining consensus at step $p$ alone already yields configurations that ultimately reach consensus at $T_{\rm eff} > p$ with initial magnetizations lower than $m^*_{\rm HPR}$ at time $p$. These additional results are presented in Appendix~\ref{sec:additional figures HPR}. A similar analysis is conducted for \sa in Appendix~\ref{sec: SA approach appendix}.

The \hpr approaches but does not reach the \rs lower bounds $m^*_{\rm RS}$ (Table~\ref{tab:1RSB}).
While tuning the hyperparameters such as $\lambda_{\mathrm{init}}$ allows the \hpr algorithm to reach lower values of $m^*_{\rm{HPR}}$ at the cost of a larger number of steps (as shown in Figure~\ref{fig: HPR visualization}), these values plateau before reaching $m^*_{\rm RS}$. We discuss this next.

\paragraph{1RSB implications on algorithmic performance of \hpr} 

The Overlap Gap Property (OGP) provides a rigorous explanation for typical-case hardness and has been proven to defeat a broad class of algorithms \cite{Gamarnik_OGP}. Its key feature is a solution space of small, distant clusters, a concept related to the frozen clusters discussed in Section~\ref{sec:RSB for majority dynamics}, which are conjectured to contain algorithmically inaccessible solutions~\cite{zdeborova_constraint_2008, thesis_Lenka, disordered_sys_comp_hard,zdeborova2008locked}. The frozen \doRSB phase often signals the onset of this complex geometry. We provide experimental evidence for this framework in a dynamical setting.

The lowest initial magnetization $m^*_{\rm{HPR}}$ reached by the \hpr algorithm is, in all considered cases, very close to the \doRSB threshold $m^*_{\rm{dRSB}}$, as seen in Table~\ref{tab: all results}. Thus, the \doRSB transition seems to mark the limiting performance of the \hpr algorithm. However, the onset of d1RSB is in general not sufficient to have generic algorithmic hardness \cite{krzakala_gibbs_2007}. In addition, it is postulated that the solutions must be frozen. The study of frozen clusters is discussed in Section~\ref{sec:RSB for majority dynamics} and expanded upon in Appendix~\ref{sec:numerical evidence freezing}. While not conclusive, our numerical results suggest that freezing appears as soon as the \doRSB phase is entered. This is consistent with the performance of \hpr. This also suggests that $m^*_{\rm{d1RSB}}$ marks the initial magnetization under which initial configurations that go to consensus are hard to find algorithmically in a reasonable time.

\section{Conclusion and Future Work}
\glsresetall
In this work, we investigated how the behavior of complex systems depends on initial conditions, focusing on synchronous discrete-time dynamics on sparse graphs. Specifically, we considered the majority dynamics of binary variables on random $d$-regular graphs. In this setting, we used the \bdcm to obtain lower bounds on the initial magnetizations that lead to $+1$ consensus in typical $d$-random regular graphs in the limit where the number of nodes $n$ goes to infinity. To find such low magnetization initializations on a given graph, we introduced the \hpr algorithm. \hpr outperforms the simulated annealing approach and reaches initializations where $+1$ nodes begin as a minority but, within a few steps, become the consensus for all $d\ge 4$ random regular graphs. The existence of such configurations for random $3$-regular graphs is left as an open challenge. 

We complement these results with \oRSB calculations that identify the onset of the \doRSB phase as an accurate marker of the performance limits of the \hpr algorithm. In particular, the lowest initial magnetization values achieved by \hpr consistently remain close to the \doRSB threshold. Additionally, the \oRSB analysis seems to indicate that typical solutions are frozen. Thus, we hypothesize that the \doRSB threshold marks the onset of algorithmic hardness for the considered graph ensemble. Our results advance the understanding of belief propagation based reinforcement algorithms and their limitations in the largely uncharted dynamical regime.

The \hpr algorithm is based on the \bdcm and thus shares the same limitations. In particular, it is computationally constrained to short trajectories (exponential complexity in the length of trajectories), and so it is also restricted to dynamical systems with short attractors. Additionally, in some settings, we observe a sensitive dependence of the required number of iterations on the hyperparameter values. To achieve optimal performance, fine-tuning of these hyperparameters is needed.

\hpr demonstrably provides insights into the difficult study of non-equilibrium properties in dynamical systems. Therefore, we expect it to help address various combinatorial optimization questions in toy models of such complex phenomena as epidemics, ecology, spin glasses, and social dynamics.

Possible extensions of this work include incorporating dynamical rules that vary across individual nodes. This would lead to more realistic models of social phenomena or various bribing scenarios, where not only the initial values of nodes but also their behavior can be influenced. The effectiveness of \hpr on other classes of graphs than random regular remains an open question. Thus, another avenue for future research is applying \hpr to sparse real-world networks or graphs that evolve over time. Additionally, an important extension would be to consider more than two possible node values. Furthermore, we expect that the results produced by the \bdcm and \hpr procedures can offer insights that will aid mathematicians in devising formal proofs of the “power of few”~\cite{power_of_few,CLT_maj_dyn, power_of_one} behavior. This could be facilitated by the fact that the replica symmetric entropy in this problem is equal to the annealed entropy that is typically simpler to treat rigorously. 

\section*{Acknowledgments}
We thank Chris Moore for suggesting the name of the algorithm, Marija Vucelja and Pasquale Calabrese for insightful discussions about the connection to the Mpemba effect.
We are grateful to Aude Maier for her initial explorations on understanding structured initializations in majority dynamics, which partly inspired this work.
MJ would like to thank the national grid infrastructure MetaCentrum for providing computational resources, supported by the Ministry of Education, Youth and Sports of the Czech Republic (project LM2018140).

\printbibliography
\appendix
\glsresetall
\section{Derivation of the BDCM relations} \label{sec:derivation BDCM}

\paragraph{Factor graph and BDCM equations.}
In the main text, we have defined a probability distribution over graph trajectories $\bbar{\mathbf{x}}$ in equation~\eqref{prob dist BDCM}. We have also shown how this probability distribution factorizes into~\eqref{prob dist BDCM factorized} by leveraging the locality of the dynamical rule $f_i:S \times S^{|\partial i|}\rightarrow S$ and the observables $\Tilde{\Xi}(\bbar{x}_i): S^{p+c}\rightarrow \mathbb{R}$,\, $\Bar{\Xi}(\bbar{x}_i,\bbar{x}_j): S^{2(p+c)}\rightarrow\mathbb{R}$. We repeat the factorized form of the \bdcm probability distribution here for convenience:
\begin{equation}
    P(\bbar{\mathbf{x}}) = \frac{1}{Z} \prod_{i\in V} \underbrace{\left[e^{-\sum_{\Tilde{k}}\lambda_{\Tilde{k}}\Tilde{\Xi}_{\Tilde{k}}(\bbar{x}_i)} \mathds{1}\left[x_i^{p+1}=f_i(x_i^{p+c},\mathbf{x}^{p+c}_{\partial i})\right] \prod_{t=1}^{p+c-1} \mathds{1}\left[x_i^{t+1}=f_i(x_i^t,\mathbf{x}^t_{\partial i})\right]\right]}_{\mathcal{A}_i(\bsbar{x}_i,\bsbar{\mathbf{x}}_{\partial i})} \prod_{(ij)\in E} \underbrace{\left[e^{-\sum_{\Bar{k}}\lambda_{\Bar{k}}\Bar{\Xi}_{\Bar{k}}(\bbar{x}_i,\bbar{x}_j)}\right]}_{a(\bsbar{x}_i,\bsbar{x}_j)}\,.
\end{equation}
Now, we describe how we can graphically represent this probability distribution and arrive at the \bdcm equations~\eqref{BDCM eqs}, which are used to obtain an estimate of the free entropy density $\Phi = \log Z / n$. We recall that $Z$ is the normalization/partition function such that $P(\bbar{\mathbf{x}})$ is a proper probability distribution.

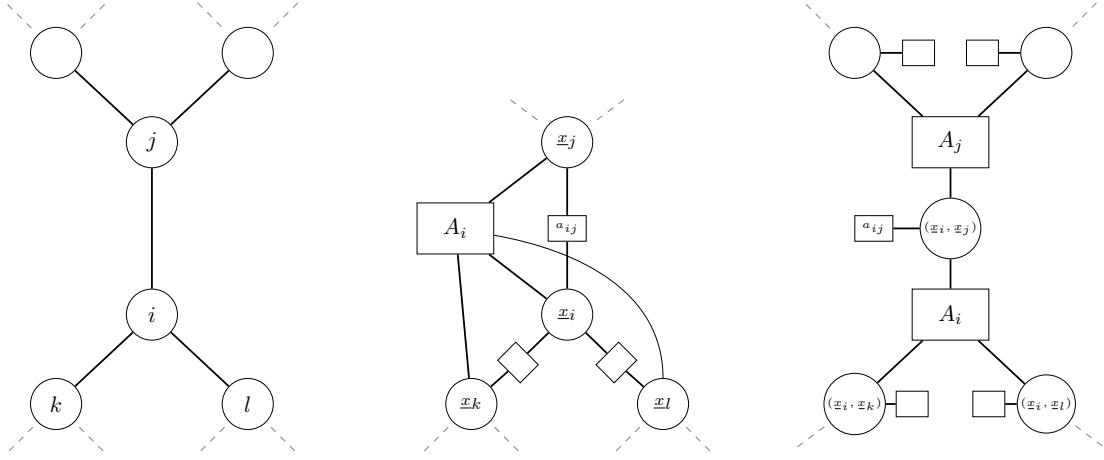
\begin{figure}[h!]
\centering
\resizebox{0.85\textwidth}{!}{
\begin{tikzpicture}[
    rect/.style={draw, rectangle, minimum width=1.2cm, minimum height=0.8cm},
    inv_rect/.style={draw=white, rectangle, minimum width=1.2cm, minimum height=0.8cm},
    rect_a/.style={draw, rectangle, minimum width=0.5cm, minimum height=0.4cm},
    circ/.style={draw, circle, minimum size=0.8cm, inner sep=1pt},
    arrow/.style={-{Latex}, thick},
    edge/.style={thick},
    dashedline/.style={draw=gray, dashed}
]

\node[circ] (jkOG) at (-1.5-9,3.7) {};
\node[circ] (jk1OG) at (1.5-9,3.7) {};
\node[circ] (jOG) at (0-9,2.3) {$j$};

\node[circ] (iOG) at (0-9,-0.4) {$i$};

\node[circ] (leftcOG) at (-1.5-9,-1.8) {$k$};
\node[circ] (rightcOG) at (1.5-9,-1.8) {$l$};


\draw[edge] (jk1OG) -- (jOG);
\draw[edge] (jkOG) -- (jOG);

\draw[edge] (jOG) -- (iOG);

\draw[edge] (iOG) -- (leftcOG);
\draw[edge] (iOG) -- (rightcOG);

\draw[dashedline, shorten <=19pt] (jkOG) -- (-1.85-9,4.05);
\draw[dashedline, shorten <=19pt] (jkOG) -- (-1.15-9,4.05);
\draw[dashedline, shorten <=20pt] (jk1OG) -- (1.15-9,4.05);
\draw[dashedline, shorten <=20pt] (jk1OG) -- (1.85-9,4.05);
\draw[dashedline, shorten <=18pt] (leftcOG)--(-1.8-9,-2.1);
\draw[dashedline, shorten <=18pt] (leftcOG)--(-1.2-9,-2.1);
\draw[dashedline, shorten <=18pt] (rightcOG)--(1.8-9,-2.1);
\draw[dashedline, shorten <=18pt] (rightcOG)--(1.2-9,-2.1);

\pgfmathsetmacro{\xshiftd}{2.5}

\node[circ] (jNF) at (0-\xshiftd,2.3) {$\bsbar{x}_j$};
\node[rect_a] (aijNF) at (0-\xshiftd,0.95) {\tiny$a_{ij}$};

\node[rect] (AiNF) at (-1.75-\xshiftd,0.95) {$A_i$};

\node[circ] (iNF) at (0-\xshiftd,-0.4) {$\bsbar{x}_i$};

\node[circ] (leftcNF) at (-1.5-\xshiftd,-1.8) {$\bsbar{x}_k$};
\node[circ] (rightcNF) at (1.5-\xshiftd,-1.8) {$\bsbar{x}_l$};
\node[rect_a,rotate=-45] (lqNF) at (-0.77-\xshiftd,-1.15) {};
\node[rect_a,rotate=45] (rqNF) at (0.77-\xshiftd,-1.15) {};


\draw[edge] (jNF) -- (aijNF);
\draw[edge] (iNF) -- (aijNF);

\draw[edge] (jNF) -- (AiNF);
\draw[edge] (iNF) -- (AiNF);
\draw[edge] (leftcNF) -- (AiNF);
\draw[-] (rightcNF) to  [out=90, in=-10]  (AiNF);

\draw[edge] (iNF) -- (lqNF);
\draw[edge] (lqNF) -- (leftcNF);
\draw[edge] (iNF) -- (rqNF);
\draw[edge] (rqNF) -- (rightcNF);

\draw[dashedline, shorten <=20pt] (jNF) -- (-0.4-\xshiftd,2.6);
\draw[dashedline, shorten <=20pt] (jNF) -- (0.4-\xshiftd,2.6);
\draw[dashedline, shorten <=18pt] (leftcNF)--(-1.8-\xshiftd,-2.1);
\draw[dashedline, shorten <=18pt] (leftcNF)--(-1.2-\xshiftd,-2.1);
\draw[dashedline, shorten <=18pt] (rightcNF)--(1.8-\xshiftd,-2.1);
\draw[dashedline, shorten <=18pt] (rightcNF)--(1.2-\xshiftd,-2.1);

\pgfmathsetmacro{\xshift}{3.5}

\node[circ] (jk) at (-1.5+\xshift,3.7) {};
\node[rect_a] (ajk) at (-0.5+\xshift,3.7) {};
\node[circ] (jk1) at (1.5+\xshift,3.7) {};
\node[rect_a] (ajk1) at (0.5+\xshift,3.7) {};
\node[rect] (Aj) at (0+\xshift,2.3) {$A_j$};

\node[circ] (ij) at (0+\xshift,0.95) {\tiny$(\bsbar{x}_i,\bsbar{x}_j)$};
\node[rect_a] (aij) at (-1.2+\xshift,0.95) {\tiny$ a_{ij}$};

\node[rect] (Ai) at (0+\xshift,-0.4) {$A_i$};

\node[circ] (leftc) at (-1.5+\xshift,-1.8) {\tiny$(\bsbar{x}_i,\bsbar{x}_k)$};
\node[circ] (rightc) at (1.5+\xshift,-1.8) {\tiny$(\bsbar{x}_i,\bsbar{x}_l)$};
\node[rect_a] (lq) at (-0.6+\xshift,-1.8) {};
\node[rect_a] (rq) at (0.6+\xshift,-1.8) {};


\draw[edge] (ij) -- (aij); 
\draw[edge] (jk) -- (ajk);
\draw[edge] (jk1) -- (ajk1);
\draw[edge] (jk1) -- (Aj);
\draw[edge] (jk) -- (Aj);

\draw[edge] (Aj) -- (ij);
\draw[edge] (ij) -- (Ai);

\draw[edge] (Ai) -- (leftc);
\draw[edge] (Ai) -- (rightc);
\draw[edge] (leftc) -- (lq);
\draw[edge] (rightc) -- (rq);

\draw[dashedline, shorten <=19pt] (jk) -- (-1.85+\xshift,4.05);
\draw[dashedline, shorten <=20pt] (jk1) -- (1.85+\xshift,4.05);
\draw[dashedline, shorten <=18pt] (leftc)--(-1.9+\xshift,-2.1);
\draw[dashedline, shorten <=18pt] (rightc)--(1.9+\xshift,-2.1);

\end{tikzpicture}
}
\caption{Factor graph construction for the \bdcm probability distribution. \textit{(Left.)} Part of the original tree graph $G$. \textit{(Middle.)} Naive construction of the factor graph, where variable nodes (circles) correspond directly to the nodes of $G$. The addition of factor nodes in this manner produces short cycles which is problematic for the \bp procedure. \textit{(Right.)} Factor graph in the edge-dual representation of the graph $G$ with pairs of trajectories $(\bbar{x}_i,\bbar{x}_j)$ as variable nodes. This construction preserves the tree structure.} \label{fig: factor G construction}
\end{figure}

We start by assuming that the graph $G$ is a \emph{tree} and construct a naive \textit{factor graph}, where the node values $\{\bbar{x}_i\}, i\in V$ are present as \textit{variable nodes}, and the functions $\{\mathcal{A}_i(\bbar{x}_i,\bbar{\mathbf{x}}_{\partial i})\},  i\in V$ and $\{a(\bbar{x}_i,\bbar{x}_j)\}, (ij)\in E$ are represented as \textit{factor nodes}. In the factor graph, there is an edge between a variable node and a factor node if and only if the variable $\bbar{x}_i$ is among the arguments of the corresponding factor function. Hence, the factor graph is bipartite. This straightforward construction of the factor graph is depicted in Figure~\ref{fig: factor G construction} (middle). We can see that the factor graph constructed in this way does not preserve the tree structure of the original graph $G$. This construction introduces many short cycles, which is generally problematic for \bp procedures. We remedy this by introducing new variable nodes, the pairs of trajectories $\{(\bbar{x}_i,\bbar{x}_j)\}, (ij)\in E$. Now, in place of \emph{edges} in the original graph $G$, we have variable \emph{nodes} $\{(\bbar{x}_i,\bbar{x}_j)\}$ and their corresponding factor nodes $\{a(\bbar{x}_i,\bbar{x}_j)\}$, which are attached only to these variable nodes. In place of the nodes of $G$, we have the factor nodes $\{\mathcal{A}_i(\bsbar{x}_i,\bsbar{\mathbf{x}}_{\partial i})\}$. Hence, we refer to this construction as the factor graph in the \textit{edge-dual} representation of the graph $G$. We denote by $ij= ji$ the variable nodes $(\bbar{x}_i,\bbar{x}_j)$, and by $\partial ij$ a set of factor nodes that share an edge with the variable node $ij$, \emph{except} the factor node $a(\bbar{x}_i,\bbar{x}_j)$. Hence, $\partial ij$ is composed of the two factor nodes $\mathcal{A}_i(\bsbar{x}_i,\bsbar{\mathbf{x}}_{\partial i})$ and $\mathcal{A}_j(\bsbar{x}_j,\bsbar{\mathbf{x}}_{\partial j})$. The factor nodes $\mathcal{A}_i(\bsbar{x}_i,\bsbar{\mathbf{x}}_{\partial i})$ are denoted as $A_i$, and we write as $\partial A_i$ the set of variable nodes on which the factor node acts. We summarize this in Table~\ref{tab: FG nodes}.

\begin{table}[h!]
    \centering
    \begin{tabular}{c|c|c|c|c} 
    \toprule
     Factor graph node& Type  & Symbol & Neighborhood & Neib.~symbol\\ [1ex] 
    \midrule
     $\mathcal{A}_i(\bbar{x}_i,\bbar{x}_{\partial i})$ & function
     &$A_i$ & $\{(\bbar{x}_i,\bbar{x}_j)\}_{j\in\partial i}$
     &$\partial A_i$\\ 
     $a(\bbar{x}_i,\bbar{x}_j)$ &function &$a_{ij}= a_{ji}$ & $(\bbar{x}_i,\bbar{x}_j)$ & $\partial a_{ij}$ \\ 
     $(\bbar{x}_i,\bbar{x}_j)$ &variable &$ij = ji$ & $\mathcal{A}_i(\bbar{x}_i,\bbar{x}_{\partial i}), \mathcal{A}_j(\bbar{x}_j,\bbar{x}_{\partial j})$ &$\partial ij$\\ 
    \end{tabular}
    \caption{The factor graph is a bipartite graph with variable and factor nodes. This table summarizes the definitions and abbreviations used in the \bdcm context.} \label{tab: FG nodes}
\end{table}

This construction of the factor graph preserves the tree structure of $G$, which plays an important role in the procedure for iterative computation of the normalization $Z$.

\begin{figure}[ht]
\centering
\begin{tikzpicture}[
    rect/.style={draw, rectangle, minimum width=1.2cm, minimum height=0.8cm},
    inv_rect/.style={draw=white, rectangle, minimum width=1.2cm, minimum height=0.8cm},
    rect_a/.style={draw, rectangle, minimum width=0.5cm, minimum height=0.4cm},
    circ/.style={draw, circle, minimum size=0.8cm, inner sep=1pt},
    arrow/.style={-{Latex}, thick},
    edge/.style={thick},
    dashedbox/.style={draw=blue, dashed, rounded corners, inner sep=0.4cm},
    dashedline/.style={draw=gray, dashed},
    dashedbox1/.style={draw=red, dashed, rounded corners, inner sep=0.35cm},
    dashedline/.style={draw=gray, dashed},
    dashedbox2/.style={draw=olive, dashed, rounded corners, inner sep=0.2cm},
    dashedline/.style={draw=gray, dashed}
]

\node[rect] (Ak) at (-2.6,4.8) {$A_k$};
\node[inv_rect] (Ak2) at (2.2,4.8) {};
\node[inv_rect] (inv) at (0,5.2) {};
\node[inv_rect] (inv2) at (-1.2,5.2) {};
\node[circ] (jk) at (-1.5,3.7) {$jk$};
\node[rect_a] (ajk) at (-0.5,3.7) {\tiny$ a_{jk}$};
\node[circ] (jk1) at (1.5,3.7) {};
\node[rect_a] (ajk1) at (0.6,3.7) {};
\node[rect] (Aj) at (0,2.5) {$A_j$};

\node[circ] (ij) at (0,0.9) {$ij$};
\node[rect_a] (aij) at (-0.9,0.9) {\tiny$ a_{ij}$};

\node[rect] (Ai) at (0,-0.7) {$A_i$};

\node[circ] (leftc) at (-1.5,-1.8) {};
\node[circ] (rightc) at (1.5,-1.8) {};
\node[rect_a] (lq) at (-0.6,-1.8) {};
\node[rect_a] (rq) at (0.6,-1.8) {};

\draw[arrow] (jk) -- node[left, pos=0.66] {\scalebox{0.72}{$V^{jk\rightarrow A_j}_{\bsbar{x}_j,\bsbar{x}_k}$}}  (Aj);
\draw[arrow] (Aj) -- node[right, near start ] {\scalebox{0.72} {$R^{A_j\rightarrow ij}_{\bsbar{x}_j,\bsbar{x}_i}$}} (ij);
\draw[arrow] (ij) -- node[right, near start] {\scalebox{0.72}{$V^{ij\rightarrow A_i}_{\bsbar{x}_i,\bsbar{x}_j}$}}  (Ai);

\draw[edge] (Ai) -- (leftc);
\draw[edge] (Ai) -- (rightc);
\draw[edge] (leftc) -- (lq);
\draw[edge] (rightc) -- (rq);
\draw[edge] (ij) -- (aij); 
\draw[edge] (Ak) -- (jk);
\draw[edge] (jk) -- (ajk);
\draw[edge] (jk1) -- (ajk1);
\draw[edge] (jk1) -- (Aj);

\draw[dashedline, shorten <=19pt] (Ak) -- (-3,5.3);
\draw[dashedline, shorten <=19pt] (Ak) -- (-2.3,5.3);
\draw[dashedline, shorten <=20pt] (jk1) -- (1.85,4.05);
\draw[dashedline, shorten <=18pt] (leftc)--(-1.9,-2.1);
\draw[dashedline, shorten <=18pt] (rightc)--(1.9,-2.1);

\node[dashedbox2, fit=(jk)(ajk)(Ak)(inv2), label=right:{$\mathcolor{olive}{G_{jk}}$}] (Gjk) {};
\node[dashedbox1, fit=(Aj)(jk)(ajk)(Ak)(jk1)(ajk1)(Ak2)(inv), inner ysep=0.43cm, label=right:{$\mathcolor{red}{G_{A_j}}$}] (GAj) {};
\node[dashedbox, fit=(Aj)(Ak)(Ak2)(ij)(aij)(jk)(ajk)(jk1)(ajk1)(inv),inner xsep=1.2cm,inner ysep=0.52cm, label=right:{$\mathcolor{blue}{G_{ij}}$}] (Gij) {};

\end{tikzpicture}
\caption{Derivation of the \bp scheme in the \bdcm setting. The tree structure of the factor graph is utilized to rewrite partition function $Z$ via the partial partition functions $V$ and $R$, for which we obtain the local iterative schemes~\eqref{V via R} and~\eqref{R via V}.} \label{fig: BP derivation}
\end{figure}
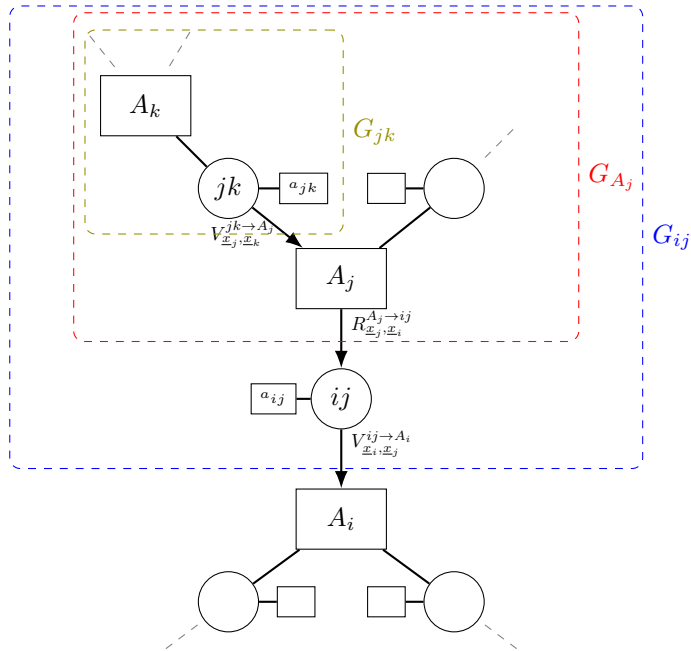

Choosing a factor node $\mathcal{A}_i(\bbar{x}_i,\bbar{\mathbf{x}}_{\partial i})$ for arbitrary $i\in G$ and rooting the factor graph in it, we can write the normalization $Z$ as
\begin{equation}
    \begin{split}
        Z &= \sum_{\{\bsbar{\mathbf{x}}\}} P(\bbar{\mathbf{x}}) = \sum_{\{\bsbar{\mathbf{x}}\}_{G}}\, \prod_{i\in G} \mathcal{A}_i(\bbar{x}_i,\bbar{\mathbf{x}}_{\partial i}) \prod_{(ij)\in G} a(\bbar{x}_i,\bbar{x}_j) =\\
        &= \sum_{\{\bsbar{x}_i,\bsbar{\mathbf{x}}_{\partial i}\}} \mathcal{A}_i(\bbar{x}_i,\bbar{\mathbf{x}}_{\partial i}) \prod_{ij\in\partial A_i} \left( \sum_{\{\bsbar{\mathbf{x}}\}_{G_{ij}}\backslash\{\bsbar{x}_i,\bsbar{x}_{j}\}}\,\prod_{k\in G_{ij}} \mathcal{A}_k(\bbar{x}_k,\bbar{\mathbf{x}}_{\partial k}) \prod_{(kl)\in G_{ij}} a(\bbar{x}_k,\bbar{x}_l) \right)\,. \label{Z rooting at A_i}
    \end{split}
\end{equation}
 In the last equality, we use the tree structure of the factor graph to write the rest of the partition function as a product over the tree branches that stem from $\mathcal{A}_i(\bbar{x}_i,\bbar{\mathbf{x}}_{\partial i})$. We denote these neighboring branches as $\{G_{ij}\}, ij\in\partial A_i$, when the branch \textit{starts} with the variable node $(\bbar{x}_i,\bbar{x}_j)$ (it does not include $A_i$). $\{\bsbar{\mathbf{x}}\}_{G_{ij}}$ designates the ensemble of all graph trajectories restricted to the subgraph $G_{ij}$. For a graphical depiction of this, see Figure~\ref{fig: BP derivation}. Note that what truly makes these branches independent is the fact that the values of the variable nodes $(\bbar{x}_i,\bbar{x}_j)$, for $ij\in\partial A_i$ are fixed -- the sum over these values is in front of the entire product. The rest of the values of the variable nodes in $G_{ij}$, i.e. $\{\bbar{\mathbf{x}}\}_{G_{ij}}\backslash\{\bbar{x}_i,\bbar{x}_{j}\}$, can be summed independently in all branches $ij\in\partial A_i$. We refer to the independent sums through these branches -- taken with the fixed variable node value $(\bbar{x}_i,\bbar{x}_j)$ in~\eqref{Z rooting at A_i} -- as \textit{partial partition functions}, and denote them by $V^{ij\rightarrow A_i}_{\bsbar{x}_i,\bsbar{x}_j}$. Thus, we write~\eqref{Z rooting at A_i} as
\begin{equation}
    Z = \sum_{\{\bsbar{x}_i,\bsbar{\mathbf{x}}_{\partial i}\}} \mathcal{A}_i(\bbar{x}_i,\bbar{\mathbf{x}}_{\partial i}) \prod_{ij\in\partial A_i} V^{ij\rightarrow A_i}_{\bsbar{x}_i,\bsbar{x}_j} \,. \label{Z as prod V}
\end{equation}
Note that $V^{ij\rightarrow A_i}_{\bsbar{x}_i,\bsbar{x}_j}$ “goes” from a variable node to a factor node. The subscript indicates the fixed value of the variable node $(\bbar{x}_i,\bbar{x}_j)$. 
Let us examine the structure of the partial partition functions $V^{ij\rightarrow A_i}_{\bsbar{x}_i,\bsbar{x}_j}$. We can take the factor node $a(\bbar{x}_i,\bbar{x}_j)$ outside the sum because the values of the variable node $ij$ are fixed:
\begin{equation}
    V^{ij\rightarrow A_i}_{\bsbar{x}_i,\bsbar{x}_j} = a(\bbar{x}_i,\bbar{x}_j)\hspace{-0.5cm} \sum_{\{\bsbar{\mathbf{x}}\}_{G_{ij}}\backslash\{\bsbar{x}_i,\bsbar{x}_{j}\}}\,\prod_{k\in G_{ij}} \mathcal{A}_k(\bbar{x}_k,\bbar{\mathbf{x}}_{\partial k}) \prod_{(kl)\in G_{ij}\backslash(ij)} a(\bbar{x}_k,\bbar{x}_l)\,. \label{V}
\end{equation}
By doing so, we recognize that we are summing only over variable node values of a graph rooted at $A_j$ (see Figure~\ref{fig: BP derivation}). We denote this graph rooted at $A_j$ by $G_{A_j}$, and introduce another partial partition function going from factor node to variable node with fixed variable node value $(\bbar{x}_i,\bbar{x}_j)$:
\begin{equation}
    R^{A_j\rightarrow ij}_{\bsbar{x}_j,\bsbar{x}_i} = \sum_{\{\bsbar{\mathbf{x}}\}_{G_{A_j}}}\,\prod_{k\in G_{A_j}} \mathcal{A}_k(\bbar{x}_k,\bbar{\mathbf{x}}_{\partial k}) \prod_{(kl)\in G_{A_j}} a(\bbar{x}_k,\bbar{x}_l)\,. \label{R def}
\end{equation}
We immediately have 
\begin{equation}
    V^{ij\rightarrow A_i}_{\bsbar{x}_i,\bsbar{x}_j} = a(\bbar{x}_i,\bbar{x}_j) R^{A_j\rightarrow ij}_{\bsbar{x}_j,\bsbar{x}_i}\,. \label{V via R}
\end{equation}
Note that the values of $(\bbar{x}_i,\bbar{x}_j)$ in $R^{A_j\rightarrow ij}_{\bsbar{x}_j,\bsbar{x}_i}$ are present in the factor node $A_j$, in which we rooted the graph $G_{A_j}$. Additionally, the fixed value of $\bbar{x}_j$ also partially fixes the variable nodes $(\bbar{x}_k,\bbar{x}_j)= kj\in A_j\backslash ij$. Therefore, similarly as above for $Z$, we write $R^{A_j\rightarrow ij}_{\bsbar{x}_j,\bsbar{x}_i}$ as
\begin{equation}
    R^{A_j\rightarrow ij}_{\bsbar{x}_j,\bsbar{x}_i} =\sum_{\{\bsbar{\mathbf{x}}_{\partial j\backslash i}\}} \mathcal{A}_j(\bbar{x}_j,\bbar{\mathbf{x}}_{\partial j})
    \prod_{jk\in\partial A_j\backslash ij}\,\left( a(\bbar{x}_j,\bbar{x}_k)
    \sum_{\{\bsbar{\mathbf{x}}\}_{G_{jk}}\backslash\{\bsbar{x}_j,\bsbar{x}_k\}}\,\prod_{\hspace{0.1cm}l\in G_{jk}} \mathcal{A}_l(\bbar{x}_l,\bbar{\mathbf{x}}_{\partial l}) \prod_{(lm)\in G_{jk}\backslash (jk)} a(\bbar{x}_l,\bbar{x}_m) \right)\,, \label{rewriting R}
\end{equation}
where $G_{jk}$ is defined as before with $G_{ij}$ and is also depicted in Figure~\ref{fig: BP derivation}. We recognize~\eqref{V} in~\eqref{rewriting R} and get
\begin{equation}
    R^{A_j\rightarrow ij}_{\bsbar{x}_j,\bsbar{x}_i} = \sum_{\{\bsbar{\mathbf{x}}_{\partial j\backslash i}\}} \mathcal{A}_j(\bbar{x}_j,\bbar{\mathbf{x}}_{\partial j})
    \prod_{jk\in\partial A_j\backslash ij} V^{jk\rightarrow A_j}_{\bsbar{x}_j,\bsbar{x}_k}\,. \label{R via V}
\end{equation}
Substitution of~\eqref{R via V} into~\eqref{V via R} results in an iterative scheme for obtaining values of the partial partition functions $V$:
\begin{equation}
    V^{ij\rightarrow A_i}_{\bsbar{x}_i,\bsbar{x}_j} = a(\bbar{x}_i,\bbar{x}_j) \sum_{\{\bsbar{\mathbf{x}}_{\partial j\backslash i}\}} \mathcal{A}_j(\bbar{x}_j,\bbar{\mathbf{x}}_{\partial j})
    \prod_{jk\in\partial A_j\backslash ij} V^{jk\rightarrow A_j}_{\bsbar{x}_j,\bsbar{x}_k}\,. \label{BDCM eqs for V}
\end{equation}
Partition functions typically grow exponentially with $n$, as they are a sum over an exponential number of terms. Then the iterative scheme~\eqref{BDCM eqs for V} tends to be numerically unstable and it is better to transition to normalized \textit{messages}
\begin{equation}
    \chi^{i\rightarrow j}_{\bsbar{x}_i,\bsbar{x}_j} = \frac{V^{ij\rightarrow A_j}_{\bsbar{x}_i,\bsbar{x}_j}}{\sum_{\{\bsbar{x}'_i,\bsbar{x}'_j\}} V^{ij\rightarrow A_j}_{\bsbar{x}'_i,\bsbar{x}'_j}} \,. \label{mes def}
\end{equation}
We use the convention that $i\rightarrow j$ designates a message that begins in variable node $ij$ and goes in the direction of the factor node $A_j$. Thus, within this convention, we refer only to the nodes of the original graph $G$. The message $\chi^{i\rightarrow j}_{\bsbar{x}_i,\bsbar{x}_j}$ is the probability that the variable node $ij$ takes the value $\bbar{x}_i,\bbar{x}_j$ when restricted to subgraph $G_{ji}$ (the subgraph does not include $A_j$). This is the probability in the presence of a cavity -- hence the name cavity method~\cite{lecture_notes}.

We use the iterative scheme for the $V$s to write
\begin{equation}
    \begin{split}
        \chi^{i\rightarrow j}_{\bsbar{x}_i,\bsbar{x}_j} &\stackrel{\eqref{mes def}\&\eqref{BDCM eqs for V}}{=} \frac{a(\bbar{x}_i,\bbar{x}_j) \sum_{\{\bsbar{\mathbf{x}}_{\partial i\backslash j}\}} \mathcal{A}_i(\bbar{x}_i,\bbar{\mathbf{x}}_{\partial i})
        \prod_{ik\in\partial A_i\backslash ij} V^{ik\rightarrow A_i}_{\bsbar{x}_k,\bsbar{x}_i}}{\sum_{\{\bsbar{x}_i,\bsbar{x}_j\}} a(\bbar{x}_i,\bbar{x}_j) \sum_{\{\bsbar{\mathbf{x}}_{\partial i\backslash j}\}} \mathcal{A}_i(\bbar{x}_i,\bbar{\mathbf{x}}_{\partial i})
        \prod_{ik\in\partial A_i\backslash ij} V^{ik\rightarrow A_i}_{\bsbar{x}_k,\bsbar{x}_i}} 
        \cdot \frac{\prod_{ik\in\partial A_i\backslash ij} \sum_{\{\bsbar{x}'_k,\bsbar{x}'_i\}} V^{ik\rightarrow A_i}_{\bsbar{x}'_k,\bsbar{x}'_i}}{\prod_{ik\in\partial A_i\backslash ij} \sum_{\{\bsbar{x}'_k,\bsbar{x}'_i\}} V^{ik\rightarrow A_i}_{\bsbar{x}'_k,\bsbar{x}'_i}}\\
        &\hspace{0.5cm}= \frac{a(\bbar{x}_i,\bbar{x}_j) \sum_{\{\bsbar{\mathbf{x}}_{\partial i\backslash j}\}} \mathcal{A}_i(\bbar{x}_i,\bbar{\mathbf{x}}_{\partial i})
        \prod_{ik\in\partial A_i\backslash ij} \frac{V^{ik\rightarrow A_i}_{\bsbar{x}_k,\bsbar{x}_i}}{\sum_{\{\bsbar{x}'_k,\bsbar{x}'_i\}} V^{ik\rightarrow A_i}_{\bsbar{x}'_k,\bsbar{x}'_i}}}{\sum_{\{\bsbar{x}_i,\bsbar{x}_j\}} a(\bbar{x}_i,\bbar{x}_j) \sum_{\{\bsbar{\mathbf{x}}_{\partial i\backslash j}\}} \mathcal{A}_i(\bbar{x}_i,\bbar{\mathbf{x}}_{\partial i})
        \prod_{ik\in\partial A_i\backslash ij} \frac{V^{ik\rightarrow A_i}_{\bsbar{x}_k,\bsbar{x}_i}}{\sum_{\{\bsbar{x}'_k,\bsbar{x}'_i\}} V^{ik\rightarrow A_i}_{\bsbar{x}'_k,\bsbar{x}'_i}}}\,.
    \end{split}
\end{equation}
Writing $ik\in A_i\backslash ij$ as $k\in\partial i\backslash j$, we obtain the \bdcm equations~\eqref{BDCM eqs}
\begin{equation}
    \chi^{i\rightarrow j}_{\bsbar{x}_i,\bsbar{x}_j} = \frac{1}{Z^{i\rightarrow j}} a(\bbar{x}_i,\bbar{x}_j) \sum_{\{\bsbar{\mathbf{x}}_{\partial i\backslash j}\}} \mathcal{A}_i(\bbar{x}_i,\bbar{\mathbf{x}}_{\partial i})
    \prod_{k\in\partial i\backslash j} \chi^{k\rightarrow i}_{\bsbar{x}_k,\bsbar{x}_i}\,,\label{BDCM eqs again}
\end{equation}
where
\begin{equation}
    Z^{i\rightarrow j}= \sum_{\{\bsbar{x}_i,\bsbar{x}_j\}} a(\bbar{x}_i,\bbar{x}_j) \sum_{\{\bsbar{\mathbf{x}}_{\partial i\backslash j}\}} \mathcal{A}_i(\bbar{x}_i,\bbar{\mathbf{x}}_{\partial i})
    \prod_{k\in\partial i\backslash j} \chi^{k\rightarrow i}_{\bsbar{x}_k,\bsbar{x}_i}\,.\label{Zi->j mes def}
\end{equation}

\paragraph{Bethe free entropy density.} Knowing the values of the messages $\chi^{i\rightarrow j}_{\bsbar{x}_i,\bsbar{x}_j}$ at a fixed point, we can compute various local quantities of interest, such as the marginal probabilities of node trajectories. We can also obtain global quantities, such as the free entropy density $\Phi$. In tree graphs, estimates based on the messages are exact. We continue our discussion keeping the assumption that $G$ is a tree. We begin by writing the partition function as in~\eqref{Z as prod V}:
\begin{equation}
    Z = \sum_{\{\bsbar{x}_i,\bsbar{\mathbf{x}}_{\partial i}\}} \mathcal{A}_i(\bbar{x}_i,\bbar{\mathbf{x}}_{\partial i}) \prod_{ij\in\partial A_i} V^{ij\rightarrow A_i}_{\bsbar{x}_i,\bsbar{x}_j} \stackrel{\eqref{mes def}}{=} 
    \biggl(\sum_{\{\bsbar{x}_i,\bsbar{\mathbf{x}}_{\partial i}\}} \mathcal{A}_i(\bbar{x}_i,\bbar{\mathbf{x}}_{\partial i}) \prod_{ij\in\partial A_i} \chi^{j\rightarrow i}_{\bsbar{x}_j,\bsbar{x}_i}\biggr) 
    \biggl(\prod_{ij\in\partial A_i} \sum_{\{\bsbar{x}'_i,\bsbar{x}'_j\}} V^{ij\rightarrow A_i}_{\bsbar{x}'_i,\bsbar{x}'_j}\biggr)\,. \label{Z start of the chain}
\end{equation}
The expression inside the first parentheses serves as a definition for the quantity $Z^i$:
\begin{equation}
    Z^i = \sum_{\{\bsbar{x}_i,\bsbar{\mathbf{x}}_{\partial i}\}} \mathcal{A}_i(\bbar{x}_i,\bbar{\mathbf{x}}_{\partial i}) \prod_{j\in\partial i} \chi^{j\rightarrow i}_{\bsbar{x}_j,\bsbar{x}_i}\,, \label{Z^i def}
\end{equation}
where we use the same notation as above for $ij\in\partial A_i$. This is the definition~\eqref{Zi def} in the main text.

Now, we would like to express the terms in the second parentheses in~\eqref{Z start of the chain} through expression involving only the messages $\{\chi\}$, and not the partial partition functions $\{V\}$. However, looking at the definition~\eqref{mes def} of the messages $\chi$, we see that this effort is quite futile, as replacing the sum of $V$s introduces another $V$. Instead, we utilize the following relation for $Z^{i \rightarrow j}$:
\begin{equation}
    Z^{i\rightarrow j}= \sum_{\{\bsbar{x}_i,\bsbar{x}_j\}} a(\bbar{x}_i,\bbar{x}_j) \sum_{\{\bsbar{\mathbf{x}}_{\partial i\backslash j}\}} \mathcal{A}_i(\bbar{x}_i,\bbar{\mathbf{x}}_{\partial i})
    \prod_{ik\in\partial A_i\backslash ij} \chi^{k\rightarrow i}_{\bsbar{x}_k,\bsbar{x}_i} \stackrel{\eqref{mes def}\&\eqref{BDCM eqs for V}}{=} 
    \frac{\sum_{\{\bsbar{x}_i,\bsbar{x}_j\}} V^{ij \rightarrow A_j}_{\bsbar{x}_i,\bsbar{x}_j}}{\prod_{ik\in\partial A_i\backslash ij} \sum_{\{\bsbar{x}'_k,\bsbar{x}'_i\}} V^{ik\rightarrow A_i}_{\bsbar{x}'_k,\bsbar{x}'_i}}\,. \label{Z^i to j and V's}
\end{equation}
This gives a recurrent procedure for replacing $\sum_{\{\bsbar{x}'_i,\bsbar{x}'_j\}} V^{ij\rightarrow A_i}_{\bsbar{x}'_i,\bsbar{x}'_j}$ with $Z^{j\rightarrow i} \prod_{jk\in\partial A_j\backslash ij} \sum_{\{\bsbar{x}'_k,\bsbar{x}'_j\}} V^{jk\rightarrow A_j}_{\bsbar{x}'_k,\bsbar{x}'_j}$. Hence, we write~\eqref{Z start of the chain} as
\begin{equation}
    Z \stackrel{\eqref{Z^i to j and V's}}{=} Z^i \prod_{ij\in\partial A_i} \biggl(Z^{j\rightarrow i} \prod_{jk\in\partial A_j\backslash ij} \sum_{\{\bsbar{x}'_k,\bsbar{x}'_j\}} V^{jk\rightarrow A_j}_{\bsbar{x}'_k,\bsbar{x}'_j}\biggr) \stackrel{\eqref{Z^i to j and V's}}{=}
    Z^i \prod_{ij\in\partial A_i} \biggl[Z^{j\rightarrow i} \prod_{jk\in\partial A_j\backslash ij}\biggl(Z^{k\rightarrow j} \prod_{kl\in\partial A_k\backslash jk} \dots \biggr) \biggr]\,. \label{Z chain 1}
\end{equation}
In equation~\eqref{Z chain 1} (in fact, already in~\eqref{Z start of the chain}), we rooted the factor graph at an \textit{arbitrary} factor node $A_i$. We see that~\eqref{Z chain 1} follows the factor graph structure. We ``go'' from $A_i$ via its neighboring edges to factor nodes $A_j$, and then follow their neighboring edges (except the ones from which we arrived) and so on, going through the entire factor graph. This nested evaluation pattern is perhaps seen even better if we lighten the notation using the convention introduced above. In this form, the partition function reads $Z =  Z^i \prod_{j\in\partial i} \biggl[Z^{j\rightarrow i} \prod_{k\in\partial j\backslash i}\biggl(Z^{k\rightarrow j} \prod_{l\in\partial k\backslash j} \dots \biggr) \biggr]$. Since the factor graph is a tree, this recurrent process ends in leaves, where partial partition functions or messages are readily evaluated (for example, $R^{A_{\mathrm{leaf}}\rightarrow ab} = \mathcal{A}_{\mathrm{leaf}}(\bbar{x}_a,\bbar{x}_b)$ and $V^{ab \rightarrow A_{\mathrm{leaf}}}_{\bsbar{x}_a,\bsbar{x}_b} = a(\bbar{x}_a,\bbar{x}_b)\mathcal{A}_{\mathrm{leaf}}(\bbar{x}_a,\bbar{x}_b)$). \\
Finally, we replace the normalizations $Z^{j\rightarrow i}$ with another quantity that is also expressed via the messages $\chi$, but it is undirected. For that we introduce $Z^{ij}= Z^{ji}$ as in~\eqref{Zij def}:
\begin{equation}
    Z^{ij} = \sum_{\{\bsbar{x}_i,\bsbar{x}_j\}} \frac{1}{a(\bbar{x}_i,\bbar{x}_j)} \chi^{i\rightarrow j}_{\bsbar{x}_i,\bsbar{x}_j}\chi^{j\rightarrow i}_{\bsbar{x}_j,\bsbar{x}_i} = \frac{Z^i}{Z^{i \rightarrow j}} = \frac{Z^j}{Z^{j \rightarrow i}}\,, \label{Z^ij via Z^i and Z^i to j} 
\end{equation}
where we used the \bdcm update equation~\eqref{BDCM eqs again} to rewrite one of the messages and the definition of $Z^i$, equation~\eqref{Z^i def}. Hence, we arrive at 
\begin{equation}
\begin{split}
        Z &=  Z^i \prod_{j\in\partial i} \biggl[Z^{j\rightarrow i} \prod_{k\in\partial j\backslash i}\biggl(Z^{k\rightarrow j} \prod_{l\in\partial k\backslash j} \dots \biggr) \biggr] \stackrel{\eqref{Z^ij via Z^i and Z^i to j}}{=} 
    Z^i \prod_{j\in\partial i} \biggl[\frac{Z^j}{Z^{ij}} \prod_{k\in\partial j\backslash i}\biggl(\frac{Z^k}{Z^{jk}} \prod_{l\in\partial k\backslash j} \dots \biggr) \biggr]\\
    &= \frac{\prod_{i\in V} Z^i}{\prod_{(ij)\in E} Z^{ij}}\,.
\end{split} \label{Z via prod of Zi and zij}
\end{equation}
From this we obtain the relation for the Bethe free entropy density~\eqref{phi B} as
\begin{equation}
    \Phi_B = \frac{\log Z}{n} = \frac{1}{n}\left( \sum_{i\in V}\log Z^i - \sum_{(ij)\in E}\log Z^{ij} \right)\,. \label{derived phi_B}
\end{equation}
The term in the first sum expresses the change in $Z$ when factor node $A_i$ (and the corresponding edges) is added to the system. The term in the second sum corresponds to the change in $Z$ that results from the addition of the edge $(ij)$. Together, this describes the Bethe free entropy~\eqref{derived phi_B} as a sum of `site' free entropies $\Phi_i = \log Z^i$ for all $i$ minus a correction due to a double summation over the edges. Hence, we need to subtract the ``edge'' free entropies $\Phi_{ij} = \log Z^{ij}$ for all edges $(ij)$ ~\cite{lecture_notes}.

\paragraph{\bp on general graphs.} We based the entire formulation of the \bdcm equations~\eqref{BDCM eqs again} and the Bethe free entropy~\eqref{derived phi_B} on the assumption that the original graph $G$ is a tree graph. In this setting, the derived formulas are exact. However, it was realized that applying the \bp procedures on locally tree-like graphs can lead to correct results~\cite{phys_inf_comp, lecture_notes}. We note here that \rrgs are examples of a setting where using the \bp procedure can lead to exact results, as these graphs typically do not contain short loops. In this general setting, we initialize the messages randomly and iterate the \bdcm equations until convergence rather than evaluating the messages from leaves to the root and back.

More generally, \bp procedures on general graphs are exact when the following holds: removing any factor node from a factor graph leaves the adjacent variable nodes independent with respect to the resulting distribution~\cite{phys_inf_comp}. This translates into the independence of the incoming messages to any factor node. We refer to this condition as the \textit{replica symmetric assumption}. The replica symmetric assumption may not hold when the factor graph contains short loops or when the variables are correlated over large distances (the notion of distance between two variable nodes can be given by the number of factor nodes along a path connecting them). For graphs that are locally tree-like, long-distance correlations are responsible for the breakdown of BP. These long-range correlations often signal a phase transition, where solutions cluster in a well-separated region of the space of solutions. This phenomenon is referred to as \oRSB. We describe this in more detail in Section~\ref{sec: 1RSB}.

\subsection{Supplementary results for $m^*_{\rm{sample}}$ with $d=6$}
\label{sec: appendix supplementary results m sample}

\begin{figure}
    \centering
    \includegraphics[width=0.5\linewidth]{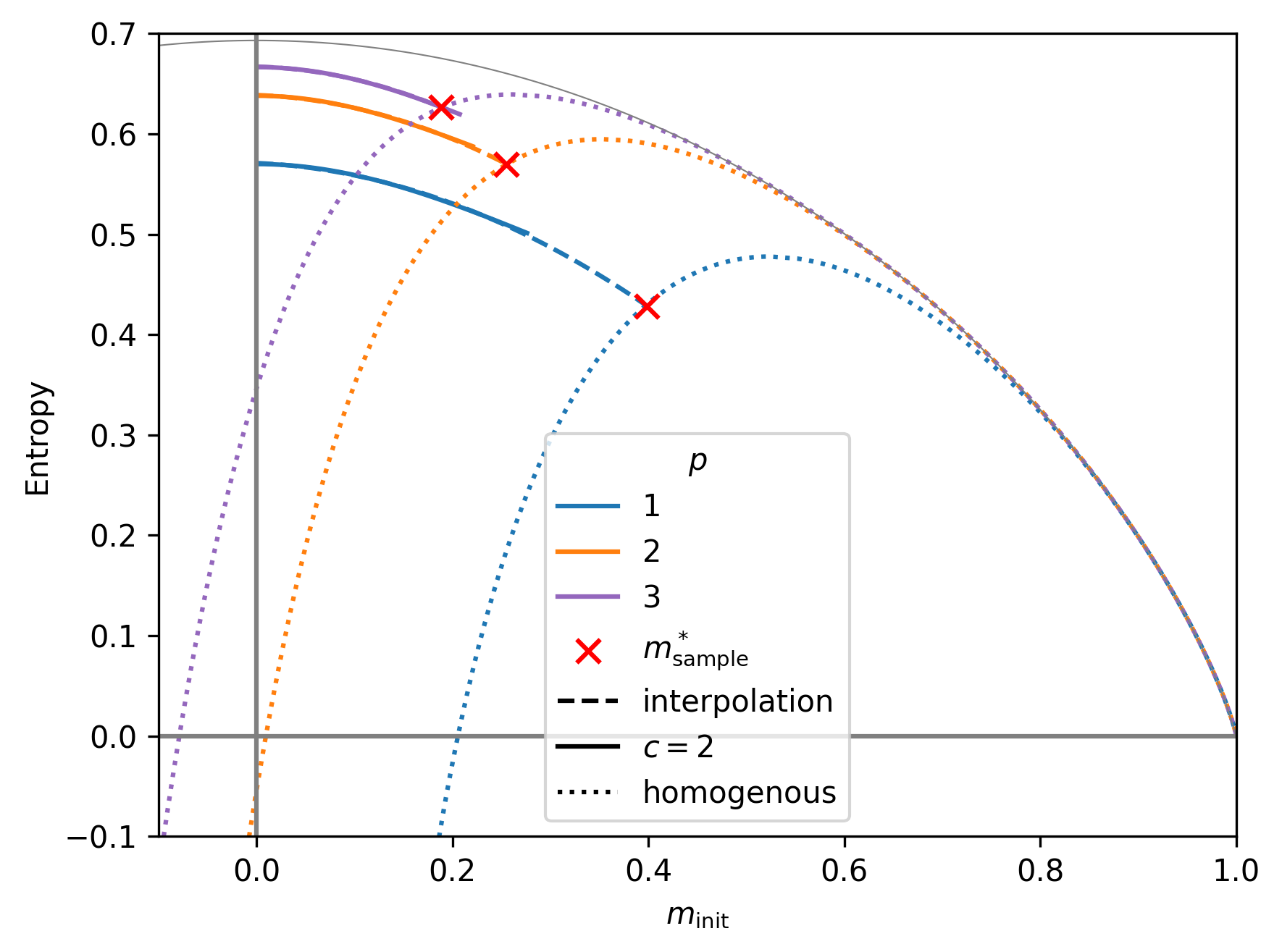}
    \caption{RS entropy for the attractor that is a $c=2$ cycle and the homogeneous $+1$ point attractor, for $d=6$ with the always-stay tie-breaking rule and $p=1,2,3$. The transition where the dominating entropy changes is marked with a red point, $m^*_{\rm sample}$. Where the line is dashed we extrapolated the entropy for $c=2$ using a quadratic function.}
    \label{fig:d=6_m_sample}
\end{figure}

While the results for $m^*_{\rm sample}$ for $d=3,4,5$ are available in~\cite{BDCM,dynamical_PT_GCA}, the results for $d=6$ were not available. 
Hence, we computed them as the intersection between the entropy for the $c=2$ cycles and the trajectories reaching the homogeneous +1 attractor for $p=1,2,3$. In Figure~\ref{fig:d=6_m_sample} we show that in order to find the intersection, we had to extrapolate the $c=2$ entropy using a quadratic fit. This is because the equations becomes unstable as one reaches the fixed point of the homogeneous +1 attractor. This leads us to the values for $m^*_{\rm sample}$ reported in Table~\ref{tab:1RSB}.

\section{One-step Replica Symmetry Breaking} \label{sec: 1RSB}
In this section, we provide an overview-level treatment of the \oRSB mechanism, with an emphasis on the \bdcm formalism. We interpret the setting of Section~\ref{sec:BDCM for enhancing consensus} as an example of a dynamical constraint satisfaction problem. Accordingly, we refer to initializations with a given $m_{\mathrm{init}}$ that evolve to full consensus under majority dynamics on sparse random graphs as \emph{solutions} of this problem. The constraints are: adherence to the dynamical rule, the trajectory being a valid $(p,c)$ backtracking attractor, a fixed trajectory value in the attractor, and the prescribed initial magnetization.

\subsection{Definition and intuition of \oRSB}
We introduce the \oRSB through the lens of the clustering phenomenon appearing in the space of possible trajectories.
To define clusters, we introduce a metric on the solution space. For any two solutions $\mathbf{s}_1,\mathbf{s}_2 \in S^n$, the  \textit{Hamming distance} $d_H(\mathbf{s}_1,\mathbf{s}_2)$ is defined as the number of node values in $\mathbf{s}_2$ that differ from $\mathbf{s}_1$. We say that two solutions $\mathbf{s}_1, \mathbf{s}_2$ are \textit{finitely close}, when, for large $n$, we have $d_H(\mathbf{s}_1,\mathbf{s}_2)=O(1)$. In other words, the two solutions differ at a finite (i.e. non-growing) number of variables. We define a \textit{cluster} $C\subset S^n$ as a maximal set of solutions, where for any two solutions $\mathbf{s}_A,\mathbf{s}_B\in C$, we have a path of \emph{solutions} $\mathcal{S} = \{\mathbf{s}_1,\mathbf{s}_2,\dots\}\subset C$ connecting them, in such a way that any two consecutive solutions on the path $\mathbf{s}_A,\mathbf{s}_1,\mathbf{s}_2,\dots,\mathbf{s}_N,\mathbf{s}_B$ are finitely close in the Hamming distance.

In the \rs phase, the \bdcm is exact and the solutions form one thermodynamically relevant cluster (we describe thermodynamical relevance below). More generally, \bdcm remains exact when removing any factor node from a factor graph leaves the adjacent variable nodes independent with respect to the resulting distribution~\cite{phys_inf_comp}, or equivalently, when the incoming messages to any factor node are independent. This is straightforwardly the case for a tree. However, since there are only long cycles ($\sim \log(n)$ or longer) through a typical node in large random graphs \emph{and} no long-range correlations, the \bdcm is still asymptotically exact~\cite{phys_inf_comp, lecture_notes}. In other words, settings with graphs that contain long cycles on which the message correlations decay quickly are \rs. An example of this are \textit{weakly interacting} models such as CSPs with a low number of constraints compared to the number of variables (the corresponding factor graph is sparse: $|\partial A_i|\ll n$) and/or large temperatures (small values of the Lagrange parameters)~\cite{cavitymethodexactsolutions}. When there is a global symmetry (e.g., $\mathbb{Z}_2$ symmetry in the Ising model), there is a finite set of thermodynamically relevant clusters, but we still describe this situation as \rs.

When long-range correlations are present, solutions might be gathered in different clusters. Intuitively, when we have a large density of constraints, transforming one solution into another cannot \emph{always} be done in finitely many variable changes. Few local changes can immediately propagate to an extensive distance in order to still satisfy all the constraints. When there is no path of incremental changes through solutions to other solutions (as defined above), we have clusters. The replica symmetry is broken. When the solution space is accurately described by the decomposition of the \rs cluster into smaller clusters (with no further internal structure), we say that there is a \textit{one-step replica symmetry breaking}~\cite{Cedric_pap}. The \oRSB phase is associated with two scenarios, dynamical and static \oRSB.

In the \doRSB scenario, we have exponentially many thermodynamically relevant clusters, each associated with a \bp fixed point. The \rs Bethe free entropy density is still exact. If one adds constraints or changes the Lagrange parameters, the number of relevant clusters can become sub-exponential: we have \soRSB. Long-range correlations are stronger than in the \doRSB phase, and \bp procedures lead to an incorrect estimate of the free entropy density~\cite{krzakala_gibbs_2007}. However, within the clusters \bp is still correct -- this is the assumption that holds both in dynamical and static \oRSB. Furthermore, we still assume the one-to-one correspondence between clusters and \bp fixed points.

As the number of constraints increases further, the clusters vanish and the problem becomes unsatisfiable (UNSAT). 

\subsection{\oRSB equations}  \label{sec: 1RSB equations} First, we repeat for convenience the definitions of the main \oRSB quantities described in Section~\ref{sec:BDCM}. We introduce a Boltzmann distribution over the clusters associated with fixed points $\Tilde{\chi}=\{\chi^{i\rightarrow j}, \chi^{j\rightarrow i}\}_{(ij)\in E}$ with $\chi^{i\rightarrow j} = \{\chi^{i\rightarrow j}_{\bsbar{x}_i,\bsbar{x}_j}\}_{(\bsbar{x}_i,\bsbar{x}_j)\in S^{2(p+c)}}$:
\begin{equation}
    P_{\mathrm{1RSB}}(\Tilde{\chi}) = \frac{1}{Z_{\mathrm{1RSB}}}e^{nr\Phi_{\mathrm{int}}(\Tilde{\chi})}\,.
\end{equation}
$r$ is the \textit{Parisi parameter},  $Z_{\mathrm{1RSB}}$ is the normalization and $\Phi_{\mathrm{int}}(\Tilde{\chi})$ is the Bethe free entropy density computed from the fixed point $\Tilde{\chi}$ according to equation~\eqref{phi B}. We call it the \textit{internal} free entropy density as it is the free entropy density of a cluster. We also define the \textit{replicated} free entropy density:
\begin{equation}
    \Psi(r)=\frac{\log(Z_{\mathrm{1RSB}}(r))}{n}\,.
\end{equation}
Furthermore, the exponentially leading number of clusters with $\Phi_{\mathrm{int}}$ is given by the \textit{complexity} $\Sigma(\Phi_{\mathrm{int}})$:
\begin{equation}
    \Sigma(\Phi_{\mathrm{int}}) =\frac{\log{\mathcal{N}(\Phi_{\mathrm{int}})}}{n}\,, 
\end{equation}
where $\mathcal{N}(\Phi_{\mathrm{int}})$ is the total number of clusters with internal free entropy density $\Phi_{\mathrm{int}}$.
Using the complexity, we express the partition function $Z_{\mathrm{1RSB}}$ through an integral over all possible internal free entropy densities as
\begin{equation}
    Z_{\mathrm{1RSB}} = e^{n\Psi(r)} = \sum_{\Tilde{\chi}} e^{nr\Phi_{\mathrm{int}}(\Tilde{\chi})} = \sum_{\Phi_{\mathrm{int}}}  \mathcal{N}(\Phi_{\mathrm{int}}) e^{nr\Phi_{\mathrm{int}}(\Tilde{\chi})} \simeq \int\dd\Phi_{\mathrm{int}} e^{n(r\Phi_{\mathrm{int}} + \Sigma(\Phi_{\mathrm{int}}))}\,. \label{Z 1RSB via sum phi_int}
\end{equation}
Next, we proceed in a similar way as in Section~\ref{sec:BDCM}. Utilizing the saddle-point method with $n\rightarrow\infty$, equation~\eqref{Z 1RSB via sum phi_int} becomes
\begin{equation}
    \Psi(r) = r\hat{\Phi}_{\mathrm{int}} + \Sigma(\hat{\Phi}_{\mathrm{int}})\,, \label{eq:Legendre Psi}
\end{equation}
where $\hat{\Phi}_{\mathrm{int}}$ extremizes $\Psi(r)$. Moreover, we have 
\begin{equation}
    \left.\frac{\partial \Sigma(\Phi_{\mathrm{int}})}{\partial \Phi_{\mathrm{int}}}\right\vert_{\Phi_{\mathrm{int}} = \hat{\Phi}_{\mathrm{int}}} = -r\,. 
\end{equation}
If we could sample solutions uniformly at random, we would see with probability approaching $1$ as $n\rightarrow\infty$ only the solutions from clusters whose $\Phi_{\mathrm{int}}$ maximizes the measure~\eqref{prob dist over clusters}. In other words, this $\Phi_{\mathrm{int}}$ maximizes the \textit{total free entropy density} 
\begin{equation} 
    \Psi_{\mathrm{tot}} = \Phi_{\mathrm{int}} + \Sigma(\Phi_{\mathrm{int}})\,, 
\end{equation}
which gives the total (unbiased by $r$) number of solutions. Typical solutions are those that jointly maximize $\Phi_{\mathrm{int}}$ (size) and $\Sigma$ (number of clusters). The total free entropy density is maximized (with respect to $\Phi_{\mathrm{int}}$) by $\hat{\Phi}_{\mathrm{int}}$ for $r=1$, since $\Psi_{\mathrm{tot}} = \Psi(1)$ and $\hat{\Phi}_{\mathrm{int}}$ is the maximizer of $\Psi(r)$ for all $r$.

If the complexity for $r=1$ is positive, we are in the \doRSB. We have \emph{exponentially} many \textit{thermodynamically relevant} clusters, i.e. their $\Phi_{\mathrm{int}} = \hat{\Phi}_{\mathrm{int}}$. A negative complexity is not physical (we assume that solutions exist; hence there needs to be at least one cluster). Thus, when the complexity is negative for $r=1$, we need to maximize the total free entropy with the constraint $\Sigma\geq 0$. We refer to $\Phi_{\mathrm{int}}$ satisfying this as $\Phi_s$, and we are in the \soRSB phase. That said, the constraint $\Sigma\geq 0$ cannot be enforced locally and is not part of the \bdcm equations. Hence, $\Phi_s$ is obtained by computing the complexity as a function of $\Phi_{\mathrm{int}}$, and $\Phi_s$ is equal to $\Phi_{\mathrm{int}}$ for which $\Sigma(\Phi_{\mathrm{int}})=0$. We can get $\Phi_{\mathrm{int}}$ for a given $r$ as
\begin{equation}
    \Phi_{\mathrm{int}} = \frac{\partial \Psi(r)}{\partial r}\,, \label{phi_int from psi appendix}
\end{equation}
as seen from the \oRSB probability distribution~\eqref{prob dist over clusters} and the definition of $\Psi(r)$~\eqref{psi def} (we assume self-averaging in the thermodynamic limit). Thus, we can get the complexity from~\eqref{psi =rphi + sigma} and by varying $r\in[0,1]$, we can find $\Phi_s$.

There is, however, a familiar obstacle: obtaining $\Psi$ requires computing $Z_{\mathrm{1RSB}}$, which is generally intractable. To address this, we devise one more BP procedure to estimate it. Again, we suppose that the original graph $G$ is a tree and we can construct the factor graph as in section~\ref{sec:derivation BDCM}. We then rewrite the \bdcm update rule~\eqref{BDCM eqs} as
\begin{equation}
    \chi^{i\rightarrow j}_{\bsbar{x}_i,\bsbar{x}_j} = \frac{1}{Z^{i\rightarrow j}} a(\bbar{x}_i,\bbar{x}_j) \sum_{\{\bsbar{\mathbf{x}}_{\partial i\backslash j}\}} \mathcal{A}_i(\bbar{x}_i,\bbar{\mathbf{x}}_{\partial i})
    \prod_{k\in\partial i\backslash j} \chi^{k\rightarrow i}_{\bsbar{x}_k,\bsbar{x}_i} = \mathcal{F}(\{\chi^{k\rightarrow i}_{\bsbar{x}_k,\bsbar{x}_i}\}_{k\in\partial i\backslash j})\,. \label{BDCM updt with F}
\end{equation}
To lighten the notation, we write the message $\chi^{i\rightarrow j}_{\bsbar{x}_i,\bsbar{x}_j}$ without the subscript, which refers to all components of the message. Hence, if~\eqref{BDCM updt with F} is true for all combinations of variable node trajectories $(\bbar{x}_i,\bbar{x}_j)\in S^{2(p+c)}$, we have
\begin{equation}
    \chi^{i\rightarrow j} = \mathcal{F}(\{\chi^{k\rightarrow i}\}_{k\in\partial i\backslash j})\,. \label{BDCM updt with F shortened}
\end{equation}
We refer to all the messages on the graph $\{\chi^{i\rightarrow j}, \chi^{j\rightarrow i}\}_{(ij)\in E}$ simply as $\chi$. When all messages satisfy~\eqref{BDCM updt with F shortened}, i.e. when we have a \bp fixed point, we denote $\chi$ as $\Tilde{\chi}$ as before.

\begin{figure}[ht]
\centering
\resizebox{0.85\textwidth}{!}{
\begin{tikzpicture}[
    rect/.style={draw, rectangle, minimum width=1.2cm, minimum height=0.8cm},
    inv_rect/.style={draw=white, rectangle, minimum width=1.2cm, minimum height=0.8cm},
    rect_a/.style={draw, rectangle, minimum width=0.5cm, minimum height=0.4cm},
    circ/.style={draw, circle, minimum size=0.8cm, inner sep=1pt},
    arrow/.style={-{Latex}, thick},
    edge/.style={thick},
    dashedline/.style={draw=gray, dashed}
]

\node[circ] (jkOG) at (-1.5-9,3.7) {};
\node[circ] (jk1OG) at (1.5-9,3.7) {};
\node[circ] (jOG) at (0-9,2.3) {$j$};

\node[circ] (iOG) at (0-9,-0.4) {$i$};

\node[circ] (leftcOG) at (-1.5-9,-1.8) {$k$};
\node[circ] (rightcOG) at (1.5-9,-1.8) {$l$};


\draw[edge] (jk1OG) -- (jOG);
\draw[edge] (jkOG) -- (jOG);

\draw[edge] (jOG) -- (iOG);

\draw[edge] (iOG) -- (leftcOG);
\draw[edge] (iOG) -- (rightcOG);

\draw[dashedline, shorten <=19pt] (jkOG) -- (-1.85-9,4.05);
\draw[dashedline, shorten <=19pt] (jkOG) -- (-1.15-9,4.05);
\draw[dashedline, shorten <=20pt] (jk1OG) -- (1.15-9,4.05);
\draw[dashedline, shorten <=20pt] (jk1OG) -- (1.85-9,4.05);
\draw[dashedline, shorten <=18pt] (leftcOG)--(-1.8-9,-2.1);
\draw[dashedline, shorten <=18pt] (leftcOG)--(-1.2-9,-2.1);
\draw[dashedline, shorten <=18pt] (rightcOG)--(1.8-9,-2.1);
\draw[dashedline, shorten <=18pt] (rightcOG)--(1.2-9,-2.1);

\pgfmathsetmacro{\xshift}{0.01}

\node[circ] (jk) at (-1.5+\xshift,3.7) {};
\node[rect_a] (ajk) at (-0.5+\xshift,3.7) {};
\node[circ] (jk1) at (1.5+\xshift,3.7) {};
\node[rect_a] (ajk1) at (0.5+\xshift,3.7) {};
\node[rect] (Aj) at (0+\xshift,2.3) {$\delta_{j,i}$};

\node[circ] (ij) at (0+\xshift,0.95) {\scalebox{0.5}{$\chi^{i\rightarrow j}, \chi^{j\rightarrow i}$}};
\node[rect_a] (aij) at (-1.2+\xshift,0.95) {\scalebox{0.5}{$(Z^{ij})^{-r}$}};

\node[rect] (Ai) at (0+\xshift,-0.4) {$\delta_{i,j}$};

\node[circ] (leftc) at (-1.5+\xshift,-1.8) {\scalebox{0.5}{$\chi^{i\rightarrow k}, \chi^{k\rightarrow i}$}};
\node[circ] (rightc) at (1.5+\xshift,-1.8) {\scalebox{0.5}{$\chi^{i\rightarrow l}, \chi^{l\rightarrow i}$}};
\node[rect_a] (lq) at (-0.6+\xshift,-1.8) {};
\node[rect_a] (rq) at (0.6+\xshift,-1.8) {};


\draw[edge] (ij) -- (aij); 
\draw[edge] (jk) -- (ajk);
\draw[edge] (jk1) -- (ajk1);
\draw[edge] (jk1) -- (Aj);
\draw[edge] (jk) -- (Aj);

\draw[edge] (Aj) -- (ij);
\draw[edge] (ij) -- (Ai);

\draw[edge] (Ai) -- (leftc);
\draw[edge] (Ai) -- (rightc);
\draw[edge] (leftc) -- (lq);
\draw[edge] (rightc) -- (rq);

\draw[dashedline, shorten <=19pt] (jk) -- (-1.85+\xshift,4.05);
\draw[dashedline, shorten <=20pt] (jk1) -- (1.85+\xshift,4.05);
\draw[dashedline, shorten <=18pt] (leftc)--(-1.9+\xshift,-2.1);
\draw[dashedline, shorten <=18pt] (rightc)--(1.9+\xshift,-2.1);

\draw[arrow] (leftc) -- node[left, near end] {\scalebox{0.72}{$P^{ik\rightarrow \delta_{i,j}}$}}  (Ai);
\draw[arrow] (rightc) -- node[right, near end] {\scalebox{0.72}{\,$P^{il\rightarrow \delta_{i,j}}$}}  (Ai);
\draw[arrow] (Ai) -- node[left, near end] {\scalebox{0.72}{$P^{\delta_{i,j}\rightarrow ij}$}}  (ij);
\draw[arrow] (ij) -- node[left, pos=0.5] {\scalebox{0.72}{$P^{ij\rightarrow \delta_{j,i}}$}}  (Aj);

\end{tikzpicture}
}
\caption{Factor graph construction for the \oRSB probability distribution~\eqref{prob dist over mes prod_(ij)}. \textit{(Left.)} Part of the original tree graph $G$. \textit{(Right.)} Factor graph with message pairs as variable nodes, and with $(Z^{ij})^{-r}$s~\eqref{Zij def} and $\delta_{i,j}$s~\eqref{delta_ij def} as factor nodes. \bp update for the beliefs $P^{ij\rightarrow \delta_{j,i}}$ is given by $(Z^{ij})^{-r}P^{\delta_{i,j}\rightarrow ij}$, hence we can write self-consistent \bp equations only for the beliefs going into the factor nodes as~\eqref{BP eqs for 1RSB}. In the main text we shorten the notation from $P^{ij\rightarrow \delta_{j,i}}$ to $P^{i\rightarrow j}$.}  \label{fig: factor G 1RSB}
\end{figure}
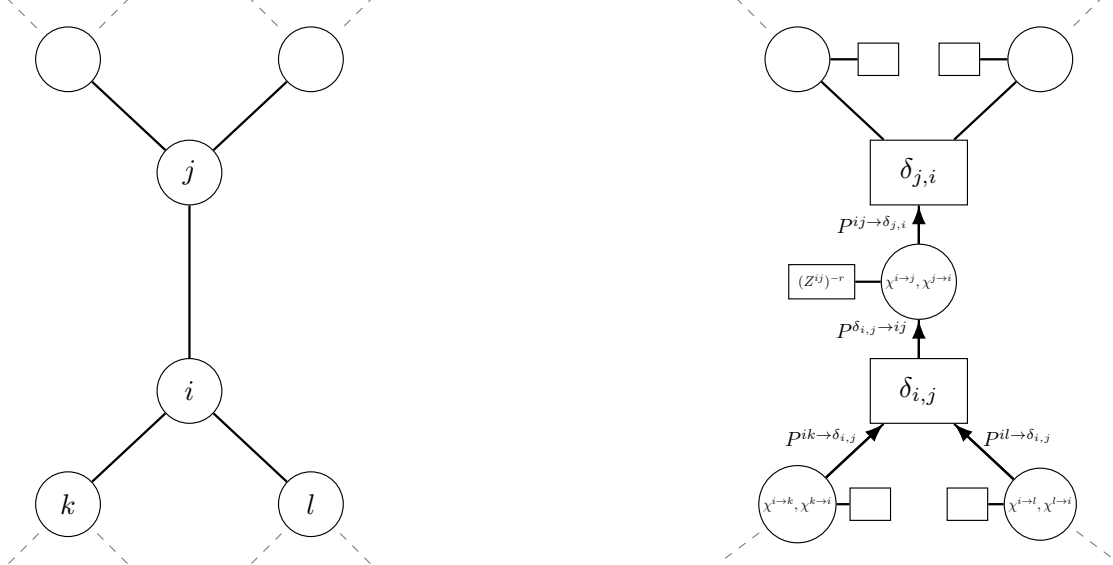

In \oRSB we suppose that the measure $P$~\eqref{prob dist BDCM} can be approximated by the messages inside the clusters. Thus, we obtain the internal free entropy density $\Phi_{\mathrm{int}}$ as the Bethe free entropy density~\eqref{phi B}. Consequently, we express the \oRSB partition function as
\begin{equation}
\begin{split}
    Z_{\mathrm{1RSB}} &= \sum_{\Tilde{\chi}} e^{nr\Phi_{\mathrm{int}}(\Tilde{\chi})} = \int\dd\chi e^{r\left(\sum_i\log Z^i - \sum_{(ij)}\log Z^{ij}\right)} \prod_i\prod_{j\in\partial i}\delta\left(\chi^{i\rightarrow j} - \mathcal{F}(\{\chi^{k\rightarrow i}\}_{k\in\partial i\backslash j})\right)\\
    &= \int\dd\chi \prod_i\left(Z^i\right)^r \prod_{(ij)}\left(Z^{ij}\right)^{-r} \prod_i\prod_{j\in\partial i}\delta\left(\chi^{i\rightarrow j} - \mathcal{F}(\{\chi^{k\rightarrow i}\}_{k\in\partial i\backslash j})\right)\,, \label{Z1RSB via the integral first}
\end{split}
\end{equation}
where the integral goes over all possible message values with the condition that the messages are normalized: $\sum_{\{\bbar{x}_i,\bbar{x}_j\}} \chi^{i\rightarrow j}_{\bsbar{x}_i,\bsbar{x}_j} = 1$. $\delta()$ is the Dirac delta. We can interpret equation~\eqref{Z1RSB via the integral first} as the partition function of the probability distribution
\begin{equation}
    \Tilde{P}(\chi) = \frac{1}{\Tilde{Z}} \prod_i\left(Z^i\right)^r \prod_{(ij)}\left(Z^{ij}\right)^{-r} \prod_i\prod_{j\in\partial i}\delta\left(\chi^{i\rightarrow j} - \mathcal{F}(\{\chi^{k\rightarrow i}\}_{k\in\partial i\backslash j})\right)\,. \label{prob dist over mes}
\end{equation}
When the graph $G$ is a tree, we have: $|V|=|E|+1$. Hence, we can associate every edge with one unique node, except for one edge, which is associated with two nodes. We then formulate~\eqref{prob dist over mes} as
\begin{equation}
    \Tilde{P}(\chi) = \frac{1}{\Tilde{Z}} \prod_{(ij)} \left(\frac{Z^i}{Z^{ij}}\right)^r \delta\left(\chi^{i\rightarrow j} - \mathcal{F}(\{\chi^{k\rightarrow i}\}_{k\in\partial i\backslash j})\right)
    \delta\left(\chi^{j\rightarrow i} - \mathcal{F}(\{\chi^{k\rightarrow j}\}_{k\in\partial j\backslash i})\right)\,. \label{prob dist over mes prod_(ij)}
\end{equation}
We note that $Z^i/Z^{ij} = Z^{i\rightarrow j}$ as shown in~\eqref{Z^ij via Z^i and Z^i to j}. We construct a graphical representation of~\eqref{prob dist over mes prod_(ij)} in the form of a factor graph, similarly as in section~\ref{sec:derivation BDCM}. This time, for every edge of the original graph, we have pairs of messages $(\chi^{i \rightarrow j}, \chi^{j \rightarrow i})$ as variable nodes. In our notation, we have neglected the dependence of the $Z^i$s and $Z^{ij}$s on the messages. Note that $Z^{ij}$ is a function of only $\chi^{i \rightarrow j}$ and $\chi^{j \rightarrow i}$, as seen in~\eqref{Zij def}. Thus, the corresponding factor nodes in the factor graph are attached only to the variable nodes $(\chi^{i \rightarrow j}, \chi^{j \rightarrow i})$. $Z^i$s~\eqref{Zi def} are functions of $\{\chi^{k\rightarrow i}\}_{k\in\partial i}$. Hence, they are parts of the factor nodes $\delta_{i,j}$ that represent the delta functions in~\eqref{prob dist over mes prod_(ij)}: 
\begin{equation}
    \delta_{i,j}= (Z^i)^r \delta\left(\chi^{i\rightarrow j} - \mathcal{F}(\{\chi^{k\rightarrow i}\}_{k\in\partial i\backslash j})\right)\,. \label{delta_ij def}
\end{equation}
In place of a node $i$ in the original graph, we now have a factor node $\delta_{i,j}$ (and $\delta_{j,i}$ for $j$). The structure of the factor graph is the same as before (see Figure~\ref{fig: factor G 1RSB}).

We again introduce partial partition functions on the factor graph, beliefs $P^{i\rightarrow j}$ conditioned on the message values $\chi^{i\rightarrow j}$ and obtain the \bp equations for them:
\begin{equation}
\begin{aligned}
    P^{i\rightarrow j}(\chi^{i\rightarrow j}) = \frac{1}{\mathcal{Z}^{i\rightarrow j}} \int\{\dd\chi^{k\rightarrow i}\}_{k\in\partial i\backslash j} \left(\frac{Z^i}{Z^{ij}}\right)^r 
    &\delta\left(\chi^{i\rightarrow j} - \mathcal{F}(\{\chi^{k\rightarrow i}\}_{k\in\partial i\backslash j})\right) \times {} \\
    {} &\times \prod_{k\in\partial i\backslash j} P^{k\rightarrow i}(\chi^{k\rightarrow i})\,, \label{BP eqs for 1RSB}
\end{aligned}
\end{equation}
where $\mathcal{Z}^{i\rightarrow j}$ is the normalization. We lighten the notation by not explicitly denoting the dependence on the messages and by merging the differentials with the beliefs $P^{k\rightarrow i}$: 
\begin{equation}
    \{\dd\chi^{k\rightarrow i}\}_{k\in\partial i\backslash j} \prod_{k\in\partial i\backslash j} P^{k\rightarrow i}(\chi^{k\rightarrow i}) = \prod_{k\in\partial i\backslash j} \dd P^{k\rightarrow i}\,. \label{better notation, dP^ki}
\end{equation}
To compute the replicated free entropy density, we follow the same procedure used to obtain~\eqref{phi B} (since the factor graph has the same structure in both cases):
\begin{equation}
    \Psi(r) = \frac{1}{n}\left(\sum_{i\in V}\log\mathcal{Z}^i - \sum_{(ij)\in E}\log\mathcal{Z}^{ij}\right)\,
    \label{psi_bethe}
\end{equation}
where 
\begin{equation}
    \mathcal{Z}^i = \int\prod_{k\in\partial i}\dd P^{k\rightarrow i} (Z^i)^r\,, \label{Zi def 1RSB}
\end{equation}
and 
\begin{equation}
    \mathcal{Z}^{ij} = \int\dd P^{i\rightarrow j}\dd P^{j\rightarrow i} (Z^{ij})^{r}\,. \label{Zij def 1RSB}
\end{equation}

On \rrgs, we can again use the fact that every node has the same number of neighbors (see Section~\ref{sec: BDCM eqs for majority dynamics on rrg}) to simplify the equations and obtain the average values over the ensemble of \rrgs. The \oRSB equations then read
\begin{equation}
\begin{aligned}
    P^{\rightarrow}(\chi^{\rightarrow}) = \frac{1}{\mathcal{Z}^{\rightarrow}} \int \prod_{k=1, \hdots, d-1}\dd P^{\rightarrow}\left(\chi_k^{\rightarrow }\right) \left(\frac{Z^{\rm{fac}}}{Z^{\rm{var}}}\right)^r 
    &\delta\left(\chi^{\rightarrow } - \mathcal{F}\left(\{\chi^{\rightarrow }_k\}_{k=1, \hdots, d-1}\right)\right) 
    \,, \label{BP eqs for 1RSB locally equivalent}
\end{aligned}
\end{equation}
where $Z^{\rm{fac}}$ and $Z^{\rm{var}}$ are defined in eqs.~\eqref{Zi def RRG} and \eqref{Zij def RRG}. $\Psi(r)$ is given by
\begin{equation}
    \Psi(r) = \log\mathcal{Z}^{\rm{fac}} - \frac{d}{2}\log\mathcal{Z}^{\rm{var}}\,, \label{eq:replicated free entropy RRG}
\end{equation}
where 
\begin{equation}
    \mathcal{Z}^{\rm{fac}} = \int\prod_{k=1, \hdots, d-1}\dd P^{\rightarrow }(\chi^\rightarrow_k) (Z^{\rm{fac}})^r\,, \label{Zi def 1RSB RRG}
\end{equation}
and 
\begin{equation}
    \mathcal{Z}^{\rm{var}} = \int\dd P^{\rightarrow}(\chi^\rightarrow_1)\dd P^{\rightarrow}(\chi^\rightarrow_2)(Z^{\rm{var}})^{r}\,. \label{Zij def 1RSB RRG}
\end{equation}
Again we did not note the dependence on the messages. The internal free entropy and complexity can again be obtained from eqs~\eqref{phi_int from psi appendix} and the complexity from \eqref{eq:Legendre Psi}. 

\subsection{Population Dynamics} \label{sec: population dynamics}
Here we present a widely used procedure for solving the \oRSB equations~\eqref{BP eqs for 1RSB}. Note that the population dynamics numerical scheme can be used for any distributional equations of this type, including \rs \bp. In the context of the cavity method, it was developed in~\cite{bethe_lattice_revisted}. The main idea is to represent the probability distribution by $N$ (ideally) i.i.d. samples from it. These $N$ samples are called a \textit{population}. Then, we update the population so that it converges to the fixed point of the \oRSB equations. In the \oRSB eqs.~\eqref{BP eqs for 1RSB locally equivalent}, we have a continuous probability distribution (represented by a population) over messages, which are themselves a discrete probability distribution over variable node trajectories. We describe the variant of population dynamics that we use below, and other treatments are available in e.g.~\cite{thesis_Lenka,Cedric_pap,phys_inf_comp}.  
\begin{enumerate}
    \item Initialize a set of $N$ messages (the population). The messages are initialized uniformly at random on \textit{hard-fields}: messages are set to $1$ for some trajectories combination $\bbar{x}_i,\bbar{x}_j$ and $0$ for all others.  The hard-fields initialization is justified by the reconstruction on trees setting~\cite{reconstruction_trees} and further commented on in~\cite{Cedric_pap}. If the population is instead initialized from a random uniform distribution, it typically converges to the RS fixed point, resulting in a delta-function distribution centered at the RS fixed-point value.
    \item  Choose uniformly at random $M=\varepsilon N$ messages from the population. We call this $M$ messages a \emph{subpopulation}. The $\varepsilon$ represents damping and makes the procedure numerically more stable.
    \item Update each message in the subpopulation using the \bdcm update rule~\eqref{BDCM updt with F shortened}. The \bdcm update is computed by sampling $d-1$ (in the case of a \rrg) messages uniformly at random from the \emph{entire} population. 
    \item Compute the normalization $Z^{i\rightarrow j}=Z^i/Z^{ij}$~\eqref{Z^ij via Z^i and Z^i to j} for each message of the updated subpopulation. Number the messages of the subpopulation and their normalizations with $k=1,2,\dots,M$. We call $Z_k$ the $Z^{i\rightarrow j}$ normalization of the $k$th message.
    \item Create a new subpopulation of $M$ messages sampled from the updated subpopulation from steps 3 and 4. Sample the messages into the new subpopulation according to the distribution $P(k)=Z_k^r/\sum_{i=1}^M Z_i^r$. Replace the $M$ originally sampled messages with this new subpopulation.
    \item Repeat from step 2 or stop if the population converged.  
\end{enumerate}
The probability distribution over messages is then obtained from the normalized histogram of the population: the number of samples with a given message value is proportional to the probability of that message value estimated by the \oRSB fixed point. Once the histogram ceases to change significantly, we say that the
population converged. Observables can then be computed from the converged population, as we discuss below.

In this process, we approximate the distributions by the populations and we also approximate the \oRSB update~\eqref{BP eqs for 1RSB}. Step 3 ensures that all messages satisfy the \bdcm update, as mandated by the delta function in~\eqref{BP eqs for 1RSB}. By doing steps 4 and 5 we weight the messages with $(Z^i/Z^{ij})^r$. Thus, the integral is effectively replaced by the representation of the distributions by populations. Indeed, the $\prod_{k\in\partial i\backslash j} \dd P^{k\rightarrow i}$ is approximated by the random choice of update messages in 3., with the additional fact that each message value can appear multiple times. Consequently, the messages sampled in step 3 are selected with probabilities proportional to their \oRSB beliefs.

\paragraph{Observables.} The replicated free entropy density~\eqref{eq:replicated free entropy RRG} is estimated from the converged population. We take $\sim N$ messages from the population and approximate the integrals~\eqref{Zi def 1RSB RRG} and~\eqref{Zij def 1RSB RRG} with them. For better precision, this is repeated several times (possibly for multiple iterations of the population dynamics, where there may be an interval between the iterations used for computation), and then the average is taken. In general, any observables that depend on messages $\{\chi^{\rightarrow }\}$ or on \oRSB beliefs $\{P^{\rightarrow }\}$ can be estimated this way. 

In population dynamics, the \doRSB phase is recognized when the population for $r=1$ converges to a different than \rs fixed point, i.e. the solution to \oRSB equations~\eqref{BP eqs for 1RSB locally equivalent} is not trivial. Moreover, both the internal free entropy density (which is the same as the \rs estimation) and complexity are positive.

In the \soRSB phase, the free entropy density is different from that estimated by the \rs \bdcm~\eqref{phi B}. The complexity is null and $\Phi_s$ is determined by the procedure described in Section~\ref{sec: 1RSB equations} and illustrated in Figure~\ref{fig:sRSB-vis}.

\subsection{On the \soRSB results} \label{sec: appendix s1RSB results}
Table~\ref{tab:bdcm-results} from the main text presents the \soRSB results for $p=1$. To correctly determine the \soRSB phase, we need to explore a range of Parisi parameter values as described in Section~\ref{sec: 1RSB equations}. Figure~\ref{fig:sRSB-vis} shows the complexity as a function of the initial magnetization. The \soRSB initial magnetization is obtained when the complexity becomes positive. Obtaining the $m_{\mathrm{init}}$ values that mark the \soRSB phase in the cases of some $p=2$ and $p=3$ requires large arrays of messages to model the distributions. We were not able to get sufficiently stable populations to estimate the \soRSB initial magnetization in these cases.

\begin{figure}[H]
    \centering
    \includegraphics[width=1.0\linewidth]{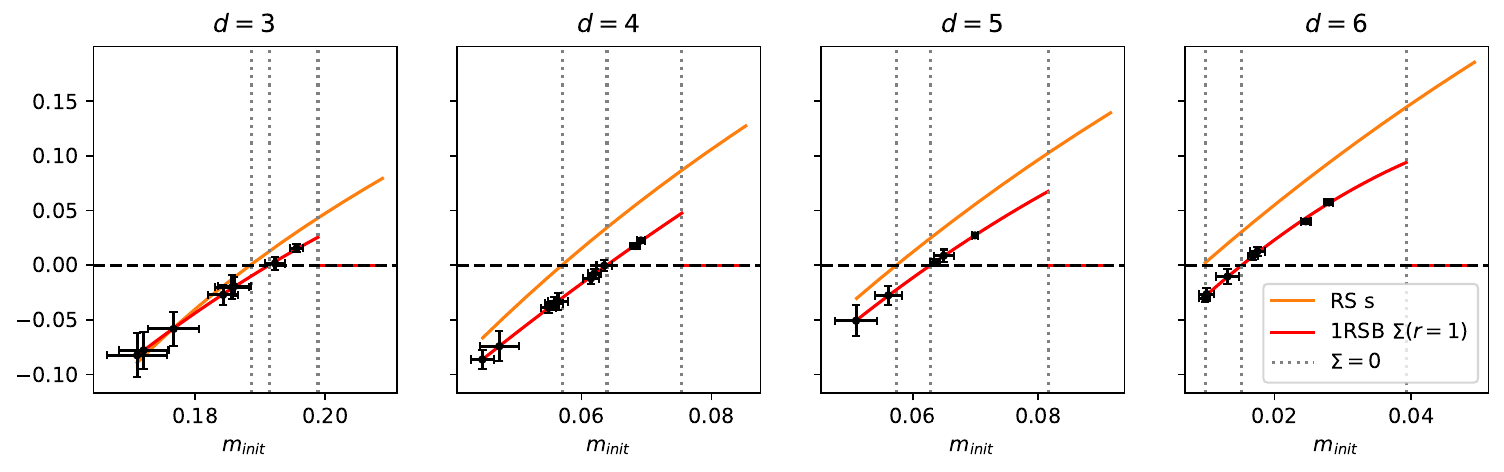}
    \caption{Thresholds of $m^*_{RS}, m^*_{sRSB}$ and $m^*_{dRSB}$ (from left to right) for $p=2$ and different values of $d$. Each black point represents the complexity $\Sigma$ obtained from the 1RSB equations for a single $\lambda$, and the error on the sample mean is indicated by error bars in $x$ and $y$ over 10 differently initialized runs of the population dynamics. For each run, the complexity is estimated as an average of complexity obtained from the last 100 iterations in the process. The red curve is a quadratic fit of this data, which ends at $m^*_{dRSB}$ from Table~\ref{tab:1RSB}.}
    \label{fig:sRSB-vis-p=2}
\end{figure}

\begin{figure}[H]
    \centering
    \includegraphics[width=0.5\linewidth]{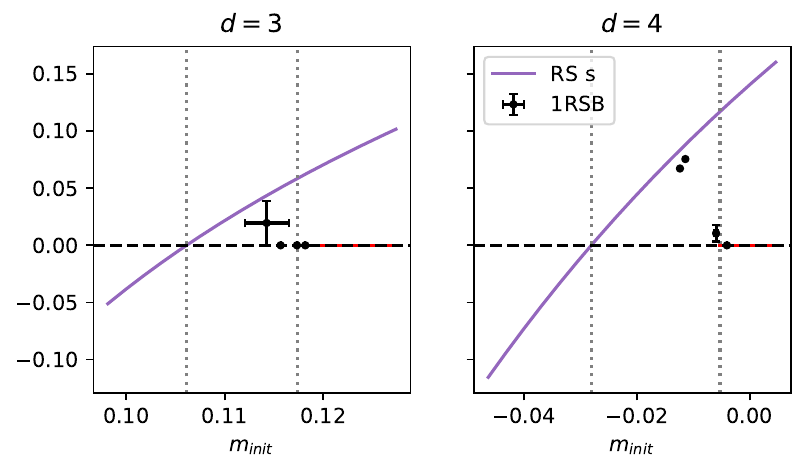}
    \caption{Thresholds of $m^*_{RS}$ and $m^*_{dRSB}$ (from left to right) for $p=3$ and $d=3,4$. Each black point represents the complexity $\Sigma$ obtained from the 1RSB equations for a single $\lambda$, and the error on the sample mean is indicated by error bars in $x$ and $y$ over 2 samples, if no error bars are present, only one sample was generated.}
    \label{fig:sRSB-vis-p=3}
\end{figure}

In Figure~\ref{fig:sRSB-vis-p=2} and~\ref{fig:sRSB-vis-p=3} we show the s1RSB results for $p=2,3$ that we were able to obtain. Due to the increasing instability in the equations, it proved very difficult to compute them numerically for increasing values of $d$ and $p$, which resulted in only very unreliable results on the s1RSB transition for $p=3$ in Figure~\ref{fig:sRSB-vis-p=3}.

\begin{figure}[h!]
    \centering
    \includegraphics[width=0.30\textwidth]{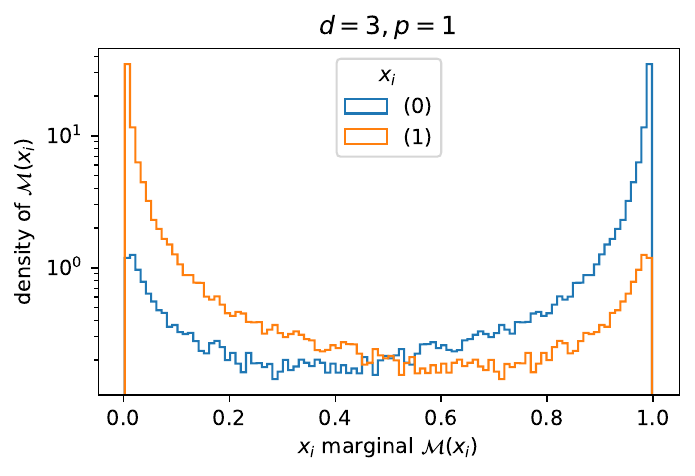}
    \includegraphics[width=0.30\textwidth]{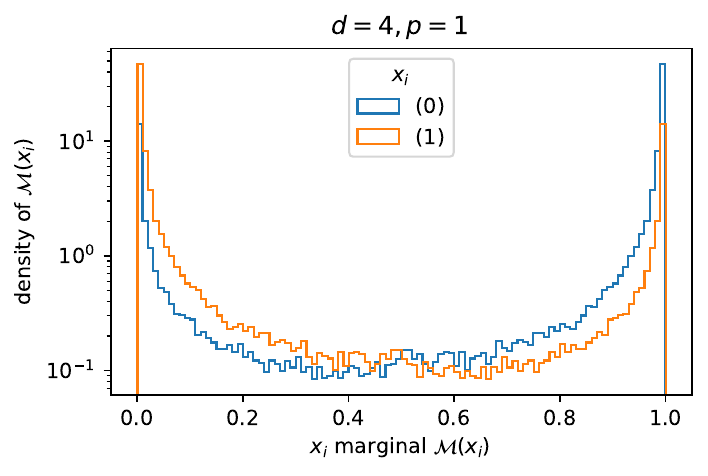}
    \includegraphics[width=0.30\textwidth]{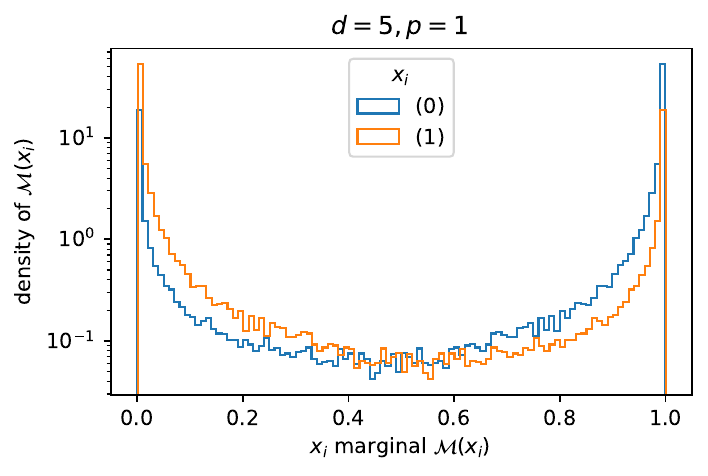}\\
    \includegraphics[width=0.30\textwidth]{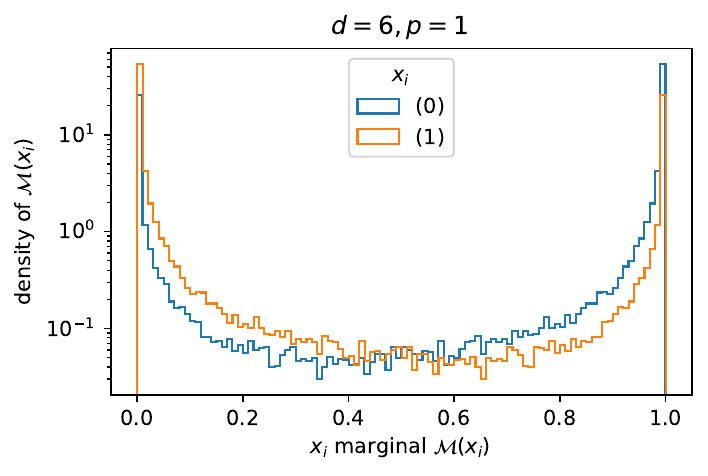}
    \includegraphics[width=0.30\textwidth]
    {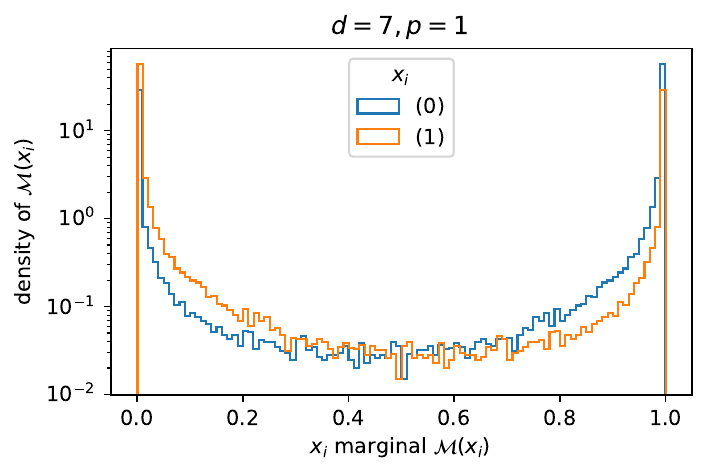} \hspace{0.3\textwidth}
    \includegraphics[width=0.3\textwidth]{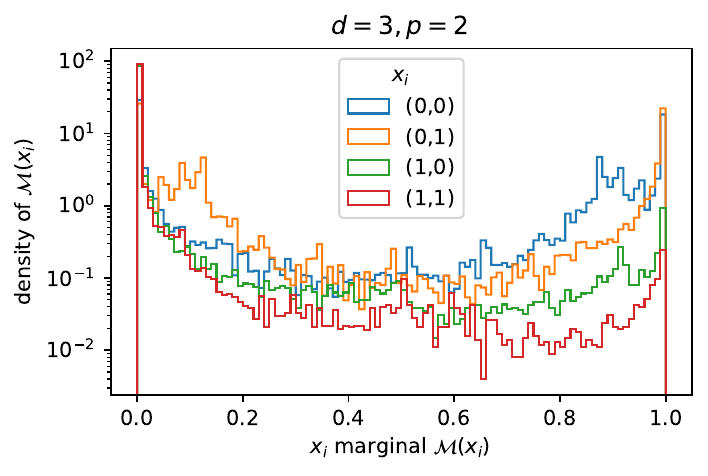}
    \includegraphics[width=0.3\textwidth]{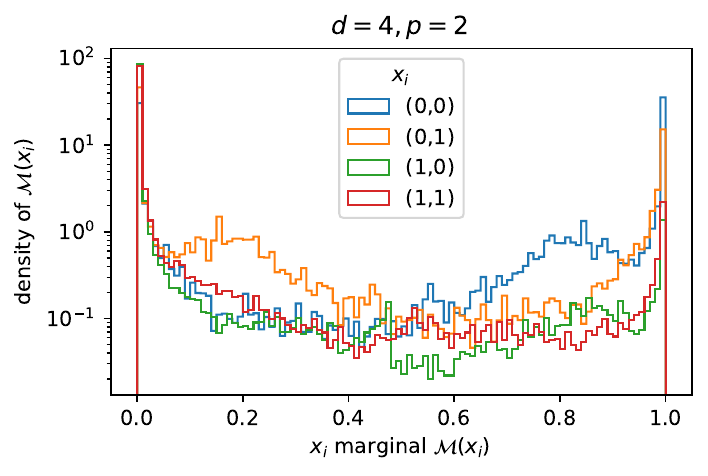}
    \includegraphics[width=0.3\textwidth]{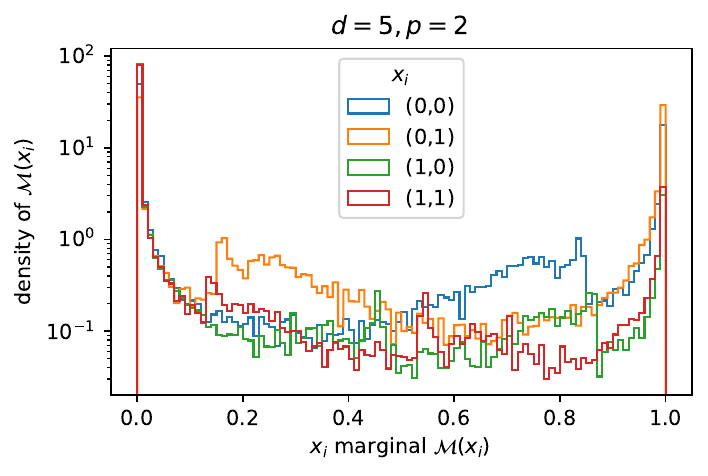}\\
    \includegraphics[width=0.3\textwidth]{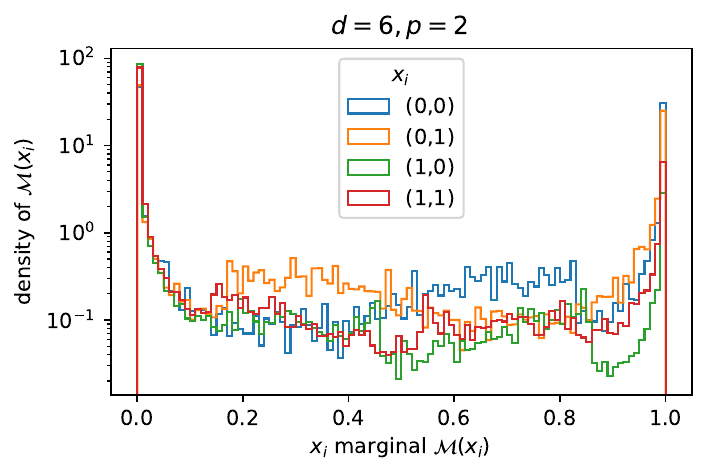}
    \hspace{0.6\textwidth}\hspace{0.3\textwidth}
    \caption{For each combination of $d,p$ we select the 1RSB fixed point which has the largest $m_\mathrm{init}$ with a complexity $\Sigma > 0$, and thereby has \oRSB behavior in Table~\ref{tab: all results}. For the associated equilibrated population of $100\,000$ messages, we show the density over the values of the marginals of the outgoing node, i.e. the density over $\mathcal{M}(\underline{x}_i) = \sum_{\{\underline x_j\}}\chi^\to_{\underline x_i,\underline x_j}$.}
    \label{fig:histograms-1RSB}
\end{figure}

\subsection{Numerical evidence of freezing for $p=1, d=4$}\label{sec:numerical evidence freezing}
To provide numerical evidence for the predicted frozen phase, we analyze the converged message population to identify variables with deterministically fixed values. We focus on the $p=1, d=4$ case, which offers greater numerical precision compared to systems with larger $p$. For these parameters, the problem is equivalent to a CSP where each $-1$ node must have strictly more than two $+1$ neighbors, and each $+1$ node must have at least two $+1$ neighbors. This equivalence permits more extensive simulations. We use the code provided in~\cite{Cedric_pap}, which implements the population dynamics for this CSP. We performed simulations with a population of $10^7$ messages, iterated $10^4$ times with a damping factor $\varepsilon=0.2$.

Our results show the emergence of a frozen phase at an initial magnetization $m_{\rm{init}}= 0.258(2)$. The number in parentheses is the last common digit for simulations just before and after the onset of the rigid phase. This result is consistent with the one obtained from our implementation of the population dynamics. Within this phase, we find that approximately $60 \%$ of the variables are frozen. This value is obtained by the fraction of messages in the converged population that assign a probability $>0.99$ to a given node. Although freezing requires this probability to be exactly 1, the use of a threshold is necessitated by the finite precision of our numerical method, which is limited by population size, iteration count, and floating-point arithmetic. Consequently, our analysis cannot definitively distinguish between variables that are genuinely frozen and those that are ``almost frozen''. A similar situation arises in certain settings for the coloring problem of random graphs, and is discussed more in depth in~\cite{PT_coloringgraphs_LenkaFlo}. Resolving this distinction remains an open question for future work. 

For larger values of $p$, the use of smaller populations makes the numerical detection of this frozen phase even more unclear. Figure \ref{fig:histograms-1RSB} shows the marginal probability histograms for $p=1,2$ and various $d$ in the \doRSB phase. The histograms peak at $0$ and $1$, as expected for frozen variables. However, as discussed above, this is not enough to ensure the presence of frozen variables.

\section{History Passing}\label{app:History Passing}
\subsection{Derivation of the marginals equation in HPR} \label{sec:derivation marginals}
Here we derive the relation for the marginals~\eqref{marginals def}. The marginal probability distribution is defined as 
\begin{equation}
\mu^i_{x_i^1=s} = \sum_{\{\bsbar{\mathbf{x}}\}}\mathds{1}\left[x_i^1=s\right]P(\bbar{\mathbf{x}})\,.
\end{equation}
We rewrite it similarly as the partition function $Z$ in~\eqref{Z rooting at A_i} and obtain 
\begin{equation}
    \mu^i_{x_i^1=s} = \frac{1}{Z} \sum_{\{\bsbar{x}_i,\bsbar{\mathbf{x}}_{\partial i}\}} \mathds{1}\left[x_i^1=s\right]  \mathcal{A}_i(\bbar{x}_i,\bbar{\mathbf{x}}_{\partial i}) \prod_{ij\in\partial A_i} V^{ij\rightarrow A_i}_{\bsbar{x}_i,\bsbar{x}_j}\,. \label{almost marginal/Zij_C}
\end{equation}
Furthermore, multiplying~\eqref{almost marginal/Zij_C} by unity, we have
\begin{equation}
    \begin{split}
        \mu^i_{x_i^1=s} &= \frac{1}{Z} \sum_{\{\bsbar{x}_i,\bsbar{\mathbf{x}}_{\partial i}\}} \mathds{1}\left[x_i^1=s\right]  \mathcal{A}_i(\bbar{x}_i,\bbar{\mathbf{x}}_{\partial i}) \prod_{ij\in\partial A_i} V^{ij\rightarrow A_i}_{\bsbar{x}_i,\bsbar{x}_j}
        \cdot \frac{\prod_{ij\in\partial A_i} \sum_{\{\bsbar{x}'_i,\bsbar{x}'_j\}} V^{ij\rightarrow A_i}_{\bsbar{x}'_j,\bsbar{x}'_i}}{\prod_{ij\in\partial A_i} \sum_{\{\bsbar{x}'_i,\bsbar{x}'_j\}} V^{ij\rightarrow A_i}_{\bsbar{x}'_j,\bsbar{x}'_i}}\\
        &\stackrel{\eqref{mes def}}{=} \frac{1}{\Tilde{Z}_c}  \sum_{\{\bsbar{x}_i,\bsbar{\mathbf{x}}_{\partial i}\}} \mathds{1}\left[x_i^1=s\right]  \mathcal{A}_i(\bbar{x}_i,\bbar{\mathbf{x}}_{\partial i}) \prod_{ij\in\partial A_i} \chi^{j\rightarrow i}_{\bsbar{x}_j,\bsbar{x}_i}\,,
    \end{split}
        \label{almost Zij_C}
\end{equation}
where we have defined  $1/\Tilde{Z}_c =  \prod_{ij\in\partial A_i} \sum_{\{\bsbar{x'}_i,\bsbar{x'}_j\}} V^{ij\rightarrow A_i}_{\bsbar{x'}_j,\bsbar{x'}_i}/Z$. In~\eqref{almost Zij_C}, with $ij\in A_i \Leftrightarrow j\in\partial i$, we recognize the \bdcm fixed point equation~\eqref{BDCM eqs}
\begin{equation}
    \begin{split}
    \mu^i_{x_i^1=s} &= \frac{1}{\Tilde{Z}_c} \sum_{\{\bsbar{x}_i,\bsbar{x}_j\}} \mathds{1}\left[x_i^1=s\right] \frac{Z^{i\rightarrow j}}{a(\bbar{x}_i,\bbar{x}_j)}\left(\frac{a(\bbar{x}_i,\bbar{x}_j)}{Z^{i\rightarrow j}} \sum_{\{\bsbar{\mathbf{x}}_{\partial i\backslash j}\}} \mathcal{A}_i(\bbar{x}_i,\bbar{\mathbf{x}}_{\partial i})\prod_{k\in\partial i\backslash j}\chi^{k\rightarrow i}_{\bsbar{x}_k,\bsbar{x}_i}\right)\chi^{j\rightarrow i}_{\bsbar{x}_j,\bsbar{x}_i} = \\
    &= \frac{Z^{i\rightarrow j}}{\Tilde{Z}_c} \sum_{\{\bsbar{x}_i,\bsbar{x}_j\}} \frac{\mathds{1}\left[x_i^1=s\right]}{a(\bbar{x}_i,\bbar{x}_j)} \chi^{i\rightarrow j}_{\bsbar{x}_i,\bsbar{x}_j}\chi^{j\rightarrow i}_{\bsbar{x}_j,\bsbar{x}_i} = Z^{ij}_{x_i^1=s}\,,
     \end{split} \label{Zij_C derivation}
\end{equation}
and we get~\eqref{Zij conditioned definition} with $Z_c = Z^{i\rightarrow j}/\Tilde{Z}_c$. Note that we can pick \textit{any} neighboring node $j\in\partial i$ and use the \bdcm equation in~\eqref{Zij_C derivation}. This is the case for the \textit{converged} \bdcm messages on the tree factor graphs (or more generally, in settings where the \rs assumption holds).

However, in the \hpr, we have biases that break the symmetry. By choosing $j\in\partial i$ we get the message pair $\chi^{i\rightarrow j}_{\bsbar{x}_i,\bsbar{x}_j}\chi^{j\rightarrow i}_{\bsbar{x}_j,\bsbar{x}_i}$, where a particular, edge dependent, set of biases is present through the biased message updates~\eqref{biased BDCM}. Furthermore, in the biased \bdcm update equation~\eqref{biased BDCM}, we have a nonlinear relation between the messages and marginals constructed from them, where the message update depends on the marginals via the (changing) biases. Hence, during the \hpr run, the messages do not have to converge to the fixed point values because these values change with the values of changing biases (these are in fact the settings in which the \hpr is effective), as can be seen in Figure~\ref{fig: HPR visualization} (left). Thus, $Z^{ij}_{x_i^1=s}$ for different neighbors $j$ can differ. To get the marginal estimates from the messages, we collect all the available information and combine it into one edge-independent marginal estimate. We arrive at the following product over neighboring edges   
\begin{equation}
    \mu^i_{x_i^1=s} = \frac{1}{Z_{\mu}}\prod_{j\in\partial i}Z^{ij}_{x_i^1=s}\,,
\end{equation}
which is~\eqref{marginals def}. This is in fact sufficient, since for the bias update~\eqref{biases update general case} we ultimately only need to compare the current marginals for node $i$ starting with $\pm1$.

\subsection{Time Dynamics}\label{sec:additional figures HPR}

\begin{figure}[H]
    \centering
    \begin{minipage}{.48\textwidth}
        \includegraphics[width=1\linewidth]{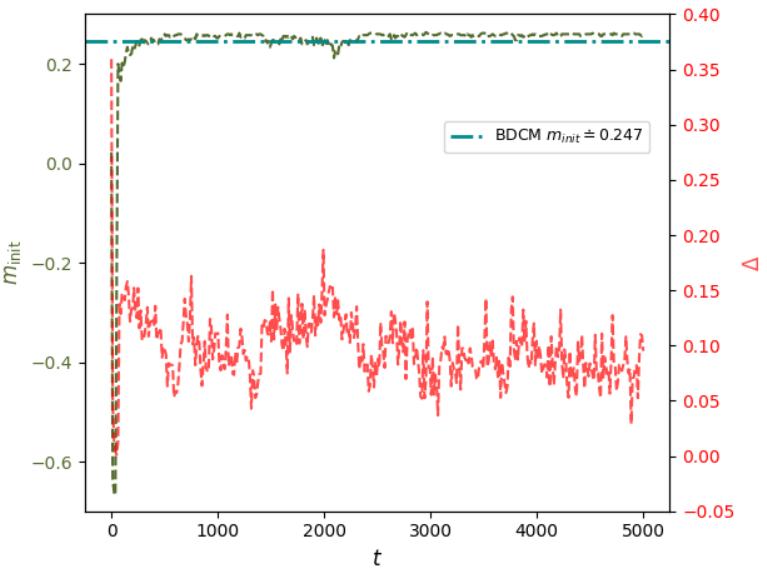}
    \end{minipage}
    \begin{minipage}{.48\textwidth}
        \includegraphics[width=1\linewidth]{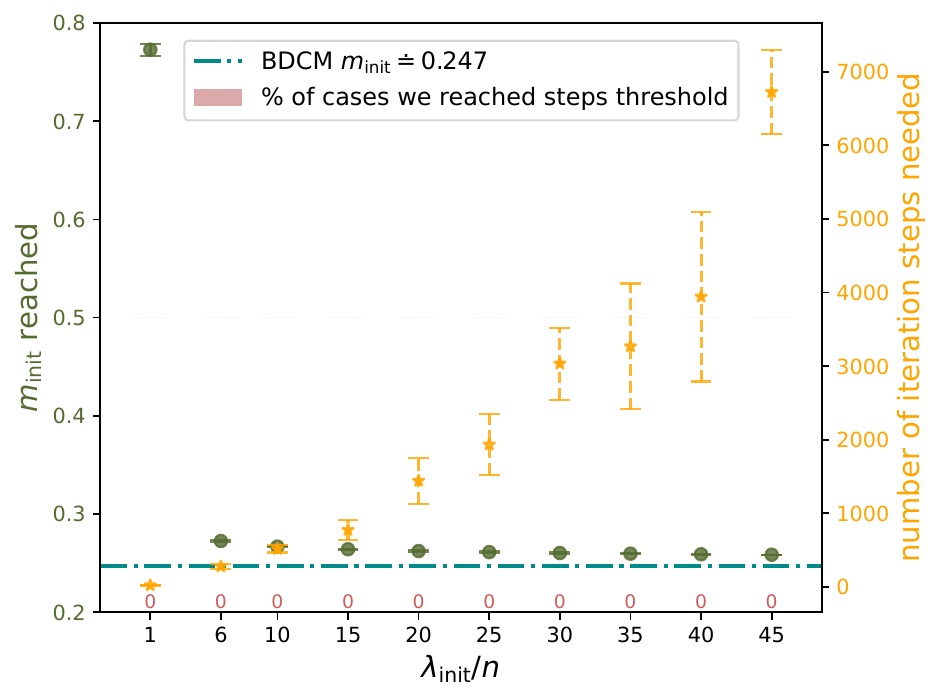}
    \end{minipage}
    \caption{\textit{(Left)} Evolution of biases as reflected in the initial magnetization $m(\mathbf{x}^{\mathrm{sol}})$ and convergence of messages measured by maximum message change $\Delta$ in the \hpr algorithm. We see that the $m(\mathbf{x}^{\mathrm{sol}})$ can quickly approach the final value returned by the algorithm, but the messages (hence, also the biases) continue to change for some time, before finding the true solution.  Parameters used: $n=1000$, $d=4$, $p=1=c$, $\pi=0.3$, $\gamma =0.1$, $\varepsilon=0.4$, $\lambda_{\rm{init}}=25$. \textit{(Right)} Obtained initial magnetization $m_{\rm{init}}$ and number of iterations needed (we stop the HPR algorithm when we reach the solution for the first time) as function of the Lagrange parameter $\lambda_{\rm{init}}$ for $n=10 000$, $d=4$, $p=1=c$, $\pi=0.3$, $\gamma =0.1$ and $\varepsilon=0.4$. Error bars depict the standard deviation, the number of repetitions for each $\lambda_{\rm{init}}$ was 10, except for $\lambda_{\rm{init}} = 40,45$ where it was 5 and 2 respectively. The reached minimal magnetizations plateau before reaching the $m^*_{\mathrm{RS}}$ value.} 
    \label{fig: HPR visualization}
\end{figure}

\paragraph{Convergence.}
Figure~\ref{fig: HPR visualization} shows supplementary results about the \hpr algorithm, completing the discussion in section~\ref{sec:HPR}. The maximum message change $\Delta$ is defined as
\begin{equation}
    \Delta=\max_{i,j,\bsbar{x}_i,\bsbar{x}_j} \left|\chi^{i \rightarrow j}_{\bsbar{x}_i,\bsbar{x}_j}(\mathrm{after})-\chi^{i \rightarrow j}_{\bsbar{x}_i,\bsbar{x}_j}(\mathrm{before})\right|.
\end{equation}

Empirically, we observe that the HPR performs the best when the messages cease to converge quickly to the BP-like fixed point (as in Figure~\ref{fig: HPR visualization} (Left)), which is consistent with the discussion of BP reinforcement in \cite{thesis_Lenka}. This regime can be found by fine-tuning the $\pi$ and $\lambda$.  

\begin{figure}[h!]
    \centering
    \includegraphics[width=\linewidth]{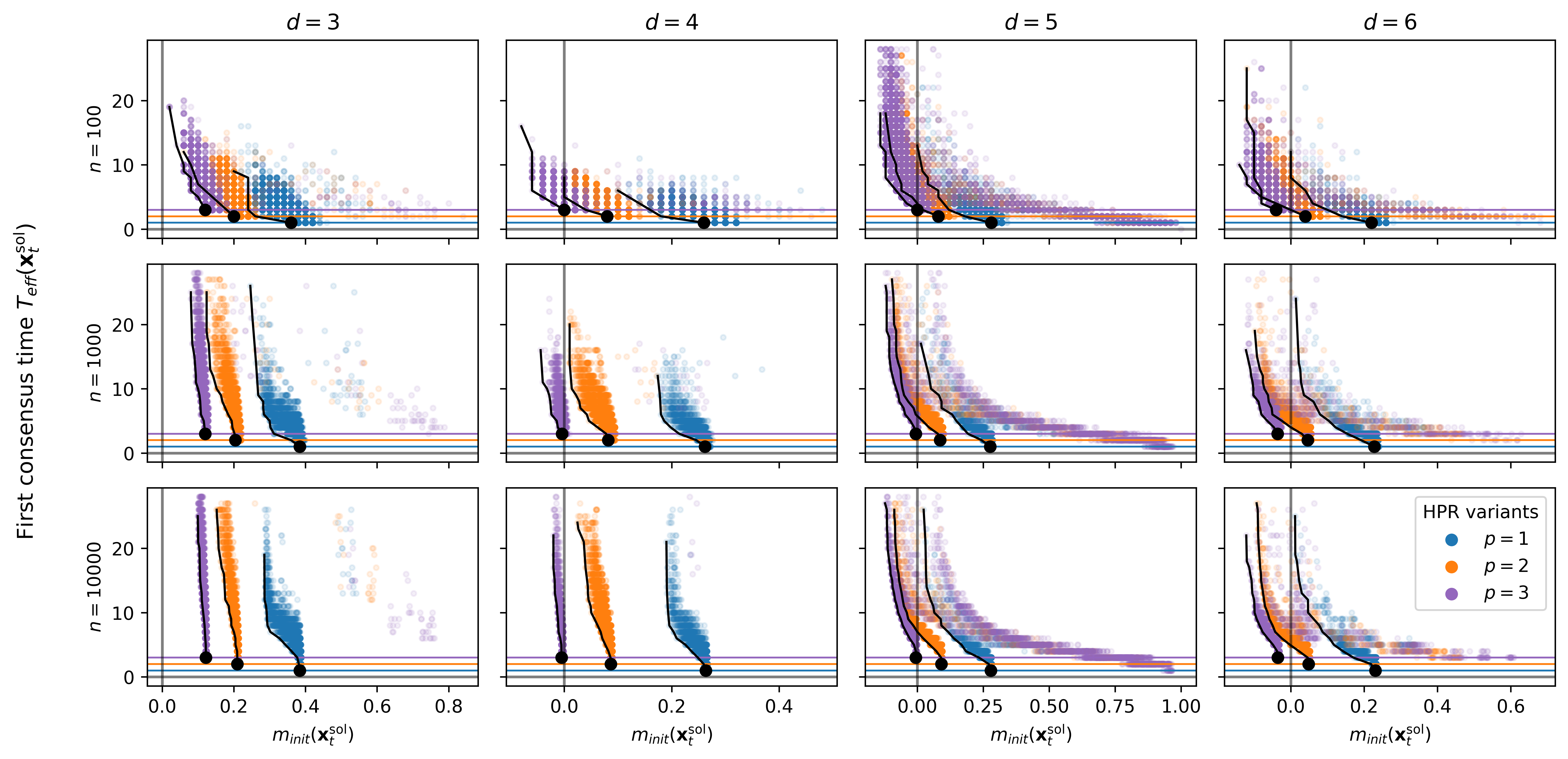}
    \caption{Values of $m_{\rm init}(\mathbf x^{\rm sol}_t)$ and the time to consensus $T_{\rm eff}(\mathbf x^{\rm sol}_t)$ under the majority rule, for all $\mathbf x^{\rm sol}_t$ from the trajectories of HPR runs with $d=3,4,5,6$. We exclude every $\mathbf x^{\rm sol}_t$ that failed to reach consensus within $28$ timesteps. We ran the HPR for 10 random graphs and plot all results. The black lines are the lowest $m_{\rm init}$ for each $T_{\rm eff}>p$ with $p=1,2,3$ over the 10 runs and black dots indicate the best $m_{\rm init}$  for a given $T_{\rm eff}=p$. Note that this reports different values than Table~\ref{tab: all results}, where we reported the average best initialization.}
    \label{fig:time-ev-1}
\end{figure}

\begin{figure}[H]
    \centering
    \includegraphics[width=\linewidth]{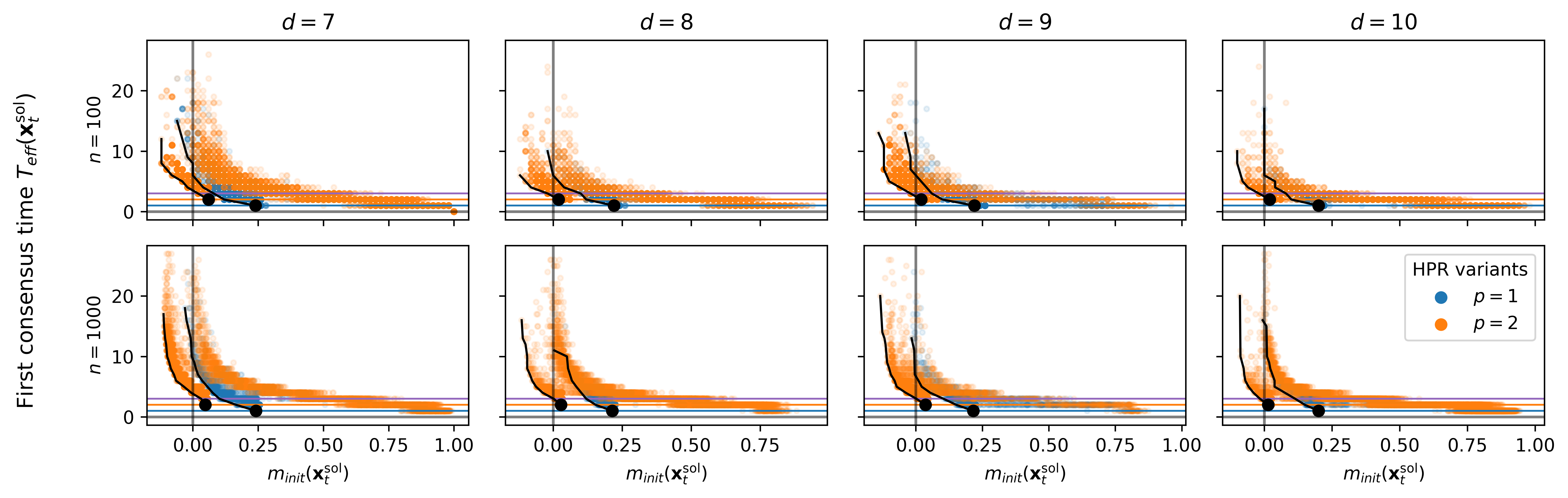}
    \caption{Similar to Figure~\ref{fig:time-ev-1} for $d=7,8,9,10$. }
    \label{fig:time-ev-2}
\end{figure}

\begin{figure}[H]
    \centering
    \includegraphics[width=0.5\linewidth]{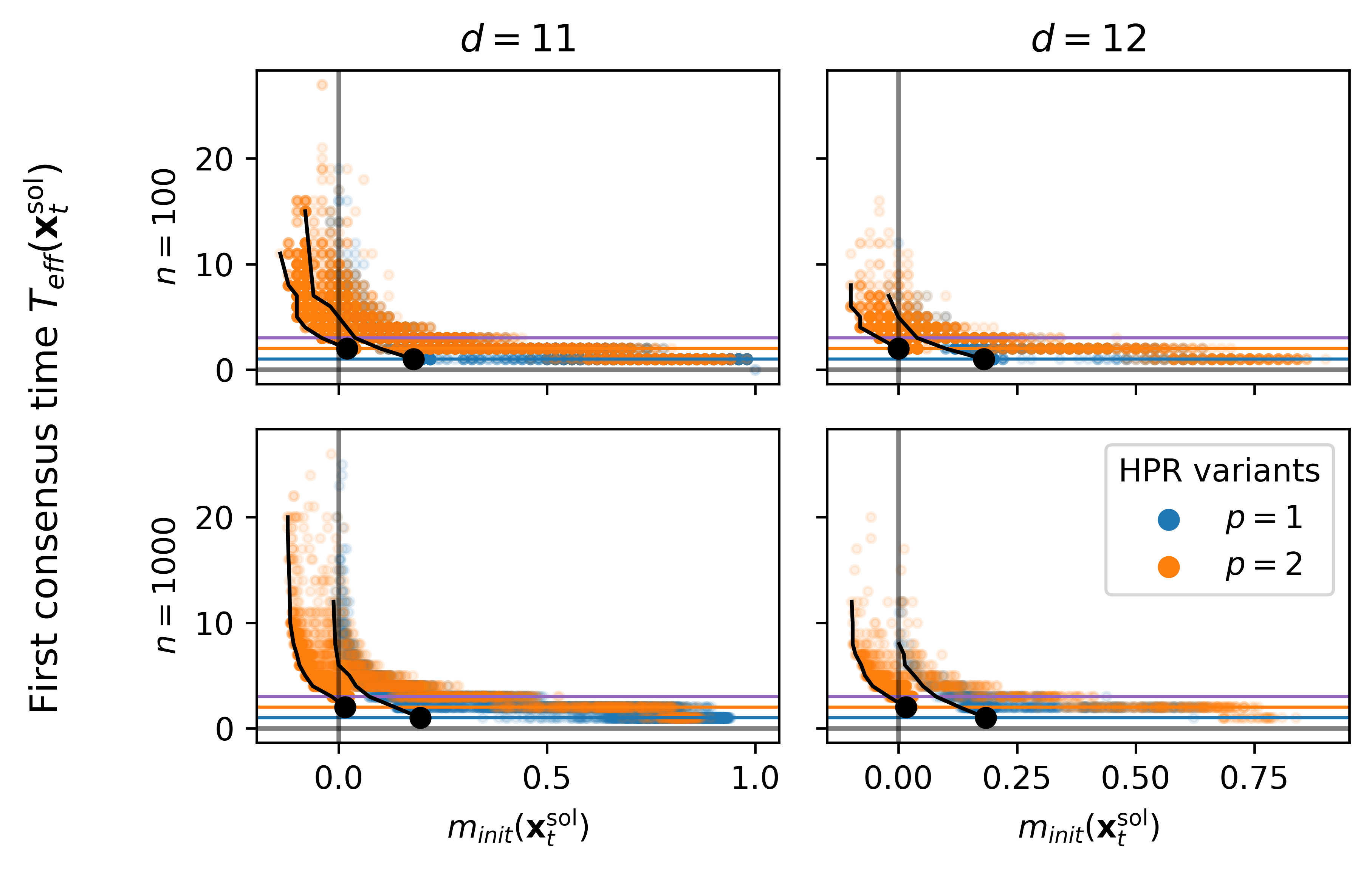}
    \caption{Similar to Figure~\ref{fig:time-ev-1} for $d=11,12$.}
    \label{fig:time-ev-3}
\end{figure}

\paragraph{Time evolution of $\mathbf x^{\rm sol}$.} In Figures~\ref{fig:time-ev-1},~\ref{fig:time-ev-2} and~\ref{fig:time-ev-3}, we show all of the values of  $m_{\rm init}(\mathbf x^{\rm sol}_t)$ and the time to consensus $T_{\rm eff}(\mathbf x^{\rm sol}_t)$ under the majority rule, for all $\mathbf x^{\rm sol}_t$ from the trajectories of HPR runs.
This shows that the algorithm not only finds configurations that converge of the desired kind, but also that it first finds those that converge a little bit less fast, sometimes with a lower initial magnetization. 
This trade-off is clearly visible on the plots for different values of $p$ and $d$.

\subsection{Dynamic programming scheme for HPR} \label{sec: dynamic programming appendix}
We show how dynamic programming significantly reduces the algorithmic complexity of Algorithm~\ref{alg:HPR} for the \hpr procedure. We repeat the general \hpr message update equation here for convenience:
\begin{equation}
    \chi^{i \rightarrow j}_{\bsbar{x}_i,\bsbar{x}_j} = \frac{1}{Z^{i \rightarrow j}}\,a(\bbar{x}_i,\bbar{x}_j) 
    \sum_{\bsbar{\mathbf{x}}_{\partial i\backslash j}} \mathcal{A}_i(\bbar{x}_i,\bbar{\mathbf{x}}_{\partial i})\!\!\prod_{k\in\partial i\backslash j}\!b^k_{x^1_k}\color{black}\chi^{k \rightarrow i}_{\bsbar{x}_k,\bsbar{x}_i}\,.\label{biased update HPR appendix}
\end{equation}
The factor node $\mathcal{A}_i(\bbar{x}_i,\bbar{\mathbf{x}}_{\partial i})$ depends on the neighboring trajectories of node $i$ only via the dynamical rule $f_i(x_i,\mathbf{x}_{\partial i})$. In the following, we consider a dynamical rule that depends only on the sum of the neighboring values: $f_i(x_i,\mathbf{x}_{\partial i}) = f_i(x_i,x_j, \varrho_{\partial i\backslash j})$, where $\varrho_{\partial i\backslash j} = \sum_{k\in\partial i\backslash j} x_k$. This is the case for the majority dynamics rule~\eqref{majority dynamics 1} from Section~\ref{sec:BDCM for enhancing consensus}. Thus, we can write the majority dynamics rule as:
\begin{equation}
    f_i(x_i,x_j, \varrho_{\partial i\backslash j}) = \mathrm{sign}\biggl( 2(x_j + \varrho_{\partial i\backslash j})) +x_i \biggr)\,. \label{majority dynamics via rho}
\end{equation}
Hence, for the factor node $\mathcal{A}_i$ we can write 
\begin{equation}
    \mathcal{A}_i(\bbar{x}_i,\bbar{\mathbf{x}}_{\partial i}) = \mathcal{A}_i(\bbar{x}_i,\bbar{x}_j,\bbar{\varrho}_{\partial i\backslash j})\,,
\end{equation}
where $\bbar{\varrho}_{\partial i\backslash j}$ is the \textit{sum trajectory} $(\varrho^1_{\partial i\backslash j},\dots,\varrho^T_{\partial i\backslash j})$ with $\varrho^t_{\partial i\backslash j} = \sum_{k\in\partial i\backslash j} x^t_k$. This is useful because a large number of trajectory combinations correspond to the same value of $\bbar{\varrho}_{\partial i\backslash j}$ and we can rewrite the message update~\eqref{biased update HPR appendix} as
\begin{equation}
    \chi^{i \rightarrow j}_{\bsbar{x}_i,\bsbar{x}_j} = \frac{1}{Z^{i \rightarrow j}}\, a(\bbar{x}_i,\bbar{x}_j) \sum_{\bsbar{\varrho}_{\partial i\backslash j}}  \mathcal{A}_i(\bbar{x}_i,\bbar{x}_j,\bbar{\varrho}_{\partial i\backslash j}) 
    \underbrace{\sum_{\bsbar{\mathbf{x}}_{\partial i\backslash j}}
    \mathds{1}\left[\sum_{k\in\partial i\backslash j} \bbar{x}_k=\bbar{\varrho}_{\partial i\backslash j}\right] \prod_{k\in\partial i\backslash j}\!b^k_{x^1_k}\chi^{k \rightarrow i}_{\bsbar{x}_k,\bsbar{x}_i} }_{L_{\Tilde{D}=d(i)-1}(\bbar{x}_i,\,\bbar{\varrho}_{\partial i\backslash j})}\,,
\end{equation}
where $L_{\Tilde{D}}(\bbar{x}_i,\,\bbar{\varrho}_{\partial i\backslash j})$ can be computed recursively. We start with $\Tilde{D}=1$, when
\begin{equation}
    L_{1}(\bbar{x}_i,\bbar{\varrho}_{1}) = b^{k_1}_{x^1_{k_1}}\chi^{k_1 \rightarrow i}_{\bsbar{x}_{k_1}=\bsbar{\varrho}_{1},\bsbar{x}_i}\,, \label{L_1}
\end{equation}
for some $k_1\in\partial i\backslash j$. Indeed, we aim to obtain the sum of the products of the biased neighboring messages, $L_{\Tilde{D}=d(i)-1}(\bbar{x}_i,\,\bbar{\varrho}_{\partial i\backslash j})$, but we start with just one message, and hence for the sum trajectory of neighboring ``nodes'' we have directly
$\bbar{\varrho}_{1}= \bbar{x}_{k_1}$. Then, we add neighboring messages according to
\begin{equation}
    L_{\Tilde{D}}(\bbar{x}_i,\bbar{\varrho}_{\Tilde{D}}) = \sum_{\bsbar{x}_{k_{\Tilde{D}}}} \sum_{\bsbar{\varrho}_{\Tilde{D}-1}} \mathds{1}\left[\bbar{x}_{k_{\Tilde{D}}} + \bbar{\varrho}_{\Tilde{D}-1} = \bbar{\varrho}_{\Tilde{D}}\right] L_{\Tilde{D}-1}(\bbar{x}_i,\bbar{\varrho}_{\Tilde{D}-1})\,\, b^{k_{\Tilde{D}}}_{x^1_{k_{\Tilde{D}}}}\chi^{k_{\Tilde{D}} \rightarrow i}_{\bsbar{x}_{k_{\Tilde{D}}},\bsbar{x}_i}\,. \label{L_D from L_D-1}
\end{equation}
In this way, we use all neighboring nodes in $\partial i\backslash j$ uniquely for the \textit{final} $\Tilde{D}=d(i)-1$. Note that we use~\eqref{L_D from L_D-1} to obtain $L_{\Tilde{D}}(\bbar{x}_i,\bbar{\varrho}_{\Tilde{D}})$ for each combination of $\bbar{\varrho}_{\Tilde{D}}$.

In the case of $d$-\rrgs the final value of $D=d(i)-1$ is the same for every node, $d(i)=d\, \forall i \in V$. Hence, we need to compute and keep only the final values $L_{\Tilde{D}=d-1}(\bbar{x}_i,\,\bbar{\varrho}_{\partial i\backslash j})$ for each message (see Algorithm~\ref{alg:HPR with dynamic prog} for the algorithmic procedure in this case). If the node degree varies, we can use the same dynamic procedure as described here, but we need to store $L_{\Tilde{D}=d(i)-1}(\bbar{x}_i,\,\bbar{\varrho}_{\partial i\backslash j})$ for the given range of node degrees.

The new update scheme
\begin{equation}
    \chi^{i \rightarrow j}_{\bsbar{x}_i,\bsbar{x}_j} = \frac{1}{Z^{i \rightarrow j}}\, a(\bbar{x}_i,\bbar{x}_j) \sum_{\bsbar{\varrho}_{\partial i\backslash j}}  \mathcal{A}_i(\bbar{x}_i,\bbar{x}_j,\bbar{\varrho}_{\partial i\backslash j}) 
   L_{d(i)-1}(\bbar{x}_i,\,\bbar{\varrho}_{\partial i\backslash j}) \label{biased BDCM sum over sums traj}
\end{equation}
is summing over $d(i)^{p+c}$ terms (the sum of $d(i)-1$ binary values can result in $d(i)$ outcomes) instead of $2^{(d-1)(p+c)}$ terms. Hence, the exponential dependence on $d(i)$ is removed. Naturally, a similar dynamic programming scheme can also be devised for other permutation-symmetric dynamical rules.
\begin{algorithm}[h!]
\caption{\textbf{H}istory \textbf{P}assing \textbf{R}einforcement -- dynamic programming for majority dynamics on $d$-\rrgs\\ (factor graph $(G, S, F, \{\Xi_k\}, \{\lambda_k\},p,c),$ steps threshold $T_{\rm{max}},$ damping $\varepsilon,$ hyperparameters: $\pi, \gamma$)}\label{alg:HPR with dynamic prog}
\begin{algorithmic}
\item Initialize uniformly at random biases $b^i_{x_i^1}$ and messages $\chi^{i \rightarrow j}_{\bsbar{x}_i,\bsbar{x}_j}$ on the factor graph and normalize them. 
\item  Initialize the sum trajectory values $\{\bbar{\varrho}_{1}\}$ with single node trajectory values $\bbar{x}\in S^{p+c}$, as $\bbar{\varrho}_{1}=\bbar{x}$.
\State $t \gets 0$
\While{ $t\leq T_{\rm{max}}$}
\State Compute $L_{1}(\bbar{x}_i,\bbar{\varrho}_{1})$ for all $\{\bbar{x}_i\}$ and $\{\bbar{\varrho}_{1}\}$ according to~\eqref{L_1}.
\For{$\Tilde{D}\in\{2,\dots,d-1\}$}
\State For all $\{\bbar{\varrho}_{\Tilde{D}}\}$ and $\{\bbar{x}_i\}$ compute $L_{\Tilde{D}}(\bbar{x}_i,\bbar{\varrho}_{\Tilde{D}})$ according to~\eqref{L_D from L_D-1}.
\EndFor
\State Update all the messages $\chi^{i \rightarrow j}_{\bsbar{x}_i,\bsbar{x}_j}$ according to~\eqref{biased BDCM sum over sums traj} with damping $\varepsilon$~\eqref{dampened update}. 
\State Compute the marginals $\mu^i_{x_i^1=s}$ from eq.~\eqref{marginals def}. 
\State Update each bias independently according to~\eqref{biases update general case} with probability $p(t)$~\eqref{bias update probability}.
\State Obtain new trial solution $\mathbf{x}^{\mathrm{sol}}_t$ using~\eqref{trial solution}.
\State $t \gets t+1$
\EndWhile
\State\Return $\mathbf{x}^{\mathrm{sol}}_t$ with lowest initial magnetization and which is a solution.
\end{algorithmic}
\end{algorithm}

\newpage
\subsection{HPR results for small $n$}
In addition to the experiments in the main text, we add HPR results for small $n=100$ and $n=1000$ in Tables~\ref{tab:hpr-n=100} and~\ref{tab:hpr-n=1000}.
In Table~\ref{tab:hpr-large-d} we provide additional results for $n=\{100,1000\}$ for $p\in[7,14]$. 

\begin{table}[H]
    \centering
    \begin{tabular}{cS[table-format=1.3]S[table-format=1.3]S[table-format=1.3]|cS[table-format=1.3]S[table-format=1.3]}
\toprule
$p$ & 1 & 2 & 3 && 4 & 5 \\
$d$ &  &  &  &  &  \\
\midrule
3 & 0.3720 & 0.2080 & 0.1200 && 0.1467 & 0.1313 \\
4 & 0.2640 & 0.0900 & 0.0000 && \bftabnum -0.006 & \bftabnum -0.019 \\
5 & 0.2820 & 0.0800 & 0.0000 && \bftabnum -0.015 & \bftabnum -0.030 \\
6 & 0.2360 & 0.0511 & \bftabnum -0.040 & &\bftabnum -0.044 & \bftabnum -0.048 \\
\bottomrule
\end{tabular}
    \caption{HPR results for $n=100$ for $p=1,2,3$ on $d=3,4,5,6$ regular graphs. We sampled 10 graphs together with random initializations and report the average over the minimum $m_{\rm init}$ that reaches consensus in exactly $p$ steps from each run. For $p=4,5$ we select the minimal $m_{\rm init}$ over all runs of the HPR for $p=\{1,2,3\}$ when they were encountered during the runtime.}
    \label{tab:hpr-n=100}
\end{table}

\begin{table}[H]
    \centering
    \begin{tabular}{cS[table-format=1.5]S[table-format=1.5]S[table-format=1.5]|cS[table-format=1.5]S[table-format=1.5]}
\toprule
$p$ & 1 & 2 & 3 & &4 & 5 \\
$d$ &  &  &  &&  &  \\
\midrule
3 & 0.3840 & 0.2072 & 0.1200 & &0.1200 & 0.1195 \\
4 & 0.2622 & 0.0846 & \bftabnum -0.003 && 0.0027 & 0.0019 \\
5 & 0.2778 & 0.0892 & \bftabnum -0.005 && \bftabnum -0.008 & \bftabnum -0.009 \\
6 & 0.2292 & 0.0488 & \bftabnum -0.036 && \bftabnum -0.028 & \bftabnum -0.030 \\
\bottomrule
\end{tabular}
    \caption{HPR results for $n=1000$ for $p=1,2,3$ on $d=3,4,5,6$ regular graphs. We sampled 10 graphs together with random initializations and report the average over the minimum $m_{\rm init}$ from each run. For $p=4,5$ we select the minimal $m_{\rm init}$ over all runs of the HPR for $p=\{1,2,3\}$ when they were encountered during the runtime.}
    \label{tab:hpr-n=1000}
\end{table}
\vspace{-0.5cm}
\begin{table}[H]
    \centering
    \begin{tabular}{cS[table-format=1.3]S[table-format=1.3]|S[table-format=1.3]S[table-format=1.3]}
\toprule
& \multicolumn{4}{c}{$n$}\\
\cmidrule(lr){2-5}
& \multicolumn{2}{c}{$100$} & \multicolumn{2}{c}{$1000$} \\
\cmidrule(lr){2-3} \cmidrule(lr){3-5}
$p$ & 1 & 2 & 1 & 2 \\
$d$\\
\midrule
7 & 0.240 & 0.060 & 0.2428 & 0.1093  \\
8 & 0.220 & 0.024 & 0.2146 & 0.0284  \\
9 & 0.220 & 0.022 & 0.2166 & 0.0360  \\
10 & 0.200 & 0.022 & 0.2003 & 0.0170  \\
11 & 0.186 & 0.020  & 0.1960 & 0.0170 \\
12 & 0.180 & 0.002& 0.1847 & 0.0166 \\
13& 0.180 & 0.020  & 0.1871 & 0.0180 \\
14 & 0.172 & 0.006& 0.1740 & 0.0102  \\
\bottomrule
\end{tabular}
    \caption{HPR results for $n=100$ and $n=1000$ for $p=1,2$ on $d=\{7,8,9,10,11,12,13,14\}$ random regular graphs. We sampled 10 graphs together with random initializations and report the average over the minimum $m_{\rm init}$ from each run.}
    \label{tab:hpr-large-d}
\end{table}

\subsection{Parameter values for HPR on $d$-\rrgs with majority always-stay dynamics} \label{sec: appendix parameters HPR and lambda behavior}
The complete list of hyperparameters for the \hpr experiments is shown below in Table~\ref{tab:hyperparameters HPR}, used for generating the \hpr results in Tables~\ref{tab: all results},~\ref{tab:hpr-n=1000} and~\ref{tab:hpr-large-d}. For all experiments we set $\gamma=0.1$ and $T_{\rm max}=10^4$. 
Furthermore, the result files are available in the accompanying code.

\begin{centering}
\captionsetup{font=normalsize} 
\footnotesize
\begin{longtable}{llllll}
\toprule
 &  &  & bias $\pi$ & dampening $\varepsilon$ & temp $\lambda_{\mathrm{init}}$ \\
$d$ & $p$ & $n$ &  &  &  \\
\midrule
\multirow[t]{9}{*}{$3$} & \multirow[t]{3}{*}{$1$} & $100$ & $0.25$ & $0.20$ & $25$ \\
 &  & $1000$ & $0.25$ & $0.20$ & $25$ \\
 &  & $10000$ & $0.25$ & $0.20$ & $25$ \\
\cline{2-6}
 & \multirow[t]{3}{*}{$2$} & $100$ & $0.25$ & $0.20$ & $18$ \\
 &  & $1000$ & $0.25$ & $0.20$ & $18$ \\
 &  & $10000$ & $0.25$ & $0.20$ & $18$ \\
\cline{2-6}
 & \multirow[t]{3}{*}{$3$} & $100$ & $0.30$ & $0.20$ & $25$ \\
 &  & $1000$ & $0.30$ & $0.20$ & $25$ \\
 &  & $10000$ & $0.30$ & $0.20$ & $25$ \\
\cline{1-6} \cline{2-6}
\multirow[t]{9}{*}{$4$} & \multirow[t]{3}{*}{$1$} & $100$ & $0.30$ & $0.40$ & $25$ \\
 &  & $1000$ & $0.30$ & $0.40$ & $25$ \\
 &  & $10000$ & $0.30$ & $0.40$ & $25$ \\
\cline{2-6}
 & \multirow[t]{3}{*}{$2$} & $100$ & $0.20$ & $0.10$ & $25$ \\
 &  & $1000$ & $0.20$ & $0.10$ & $25$ \\
 &  & $10000$ & $0.20$ & $0.10$ & $25$ \\
\cline{2-6}
 & \multirow[t]{3}{*}{$3$} & $100$ & $0.30$ & $0.40$ & $20$ \\
 &  & $1000$ & $0.30$ & $0.40$ & $20$ \\
 &  & $10000$ & $0.30$ & $0.40$ & $20$ \\
\cline{1-6} \cline{2-6}
\multirow[t]{9}{*}{$5$} & \multirow[t]{3}{*}{$1$} & $100$ & $0.30$ & $0.40$ & $25$ \\
 &  & $1000$ & $0.30$ & $0.40$ & $25$ \\
 &  & $10000$ & $0.30$ & $0.40$ & $25$ \\
\cline{2-6}
 & \multirow[t]{3}{*}{$2$} & $100$ & $0.20$ & $0.10$ & $25$ \\
 &  & $1000$ & $0.20$ & $0.10$ & $25$ \\
 &  & $10000$ & $0.20$ & $0.10$ & $25$ \\
\cline{2-6}
 & \multirow[t]{3}{*}{$3$} & $100$ & $0.30$ & $0.15$ & $19$ \\
 &  & $1000$ & $0.30$ & $0.15$ & $17$ \\
 &  & $10000$ & $0.30$ & $0.15$ & $17$ \\
\cline{1-6} \cline{2-6}
\multirow[t]{9}{*}{$6$} & \multirow[t]{3}{*}{$1$} & $100$ & $0.20$ & $0.40$ & $25$ \\
 &  & $1000$ & $0.20$ & $0.40$ & $25$ \\
 &  & $10000$ & $0.20$ & $0.40$ & $25$ \\
\cline{2-6}
 & \multirow[t]{3}{*}{$2$} & $100$ & $0.20$ & $0.10$ & $20$ \\
 &  & $1000$ & $0.20$ & $0.10$ & $20$ \\
 &  & $10000$ & $0.20$ & $0.10$ & $20$ \\
\cline{2-6}
 & \multirow[t]{3}{*}{$3$} & $100$ & $0.20$ & $0.15$ & $15$ \\
 &  & $1000$ & $0.20$ & $0.15$ & $15$ \\
 &  & $10000$ & $0.20$ & $0.15$ & $15$ \\
\cline{1-6} \cline{2-6}
\multirow[t]{6}{*}{$7$} & \multirow[t]{3}{*}{$1$} & $100$ & $0.15$ & $0.05$ & $30$ \\
 &  & $1000$ & $0.15$ & $0.05$ & $30$ \\
 &  & $10000$ & $0.15$ & $0.05$ & $30$ \\
\cline{2-6}
 & \multirow[t]{3}{*}{$2$} & $100$ & $0.15$ & $0.05$ & $30$ \\
 &  & $1000$ & $0.15$ & $0.05$ & $30$ \\
 &  & $10000$ & $0.15$ & $0.05$ & $30$ \\
\cline{1-6} \cline{2-6}
\multirow[t]{6}{*}{$8$} & \multirow[t]{3}{*}{$1$} & $100$ & $0.20$ & $0.15$ & $15$ \\
 &  & $1000$ & $0.20$ & $0.15$ & $15$ \\
 &  & $10000$ & $0.20$ & $0.15$ & $15$ \\
\cline{2-6}
 & \multirow[t]{3}{*}{$2$} & $100$ & $0.20$ & $0.15$ & $15$ \\
 &  & $1000$ & $0.20$ & $0.15$ & $15$ \\
 &  & $10000$ & $0.20$ & $0.15$ & $15$ \\
\cline{1-6} \cline{2-6}
\multirow[t]{6}{*}{$9$} & \multirow[t]{3}{*}{$1$} & $100$ & $0.20$ & $0.15$ & $15$ \\
 &  & $1000$ & $0.20$ & $0.15$ & $15$ \\
 &  & $10000$ & $0.20$ & $0.15$ & $15$ \\
\cline{2-6}
 & \multirow[t]{3}{*}{$2$} & $100$ & $0.15$ & $0.05$ & $17$ \\
 &  & $1000$ & $0.15$ & $0.05$ & $17$ \\
 &  & $10000$ & $0.15$ & $0.05$ & $17$ \\
\cline{1-6} \cline{2-6}
\multirow[t]{4}{*}{$10$} & \multirow[t]{2}{*}{$1$} & $100$ & $0.20$ & $0.15$ & $10$ \\
 &  & $1000$ & $0.20$ & $0.15$ & $10$ \\
\cline{2-6}
 & \multirow[t]{2}{*}{$2$} & $100$ & $0.20$ & $0.15$ & $10$ \\
 &  & $1000$ & $0.20$ & $0.15$ & $10$ \\
\cline{1-6} \cline{2-6}
\multirow[t]{4}{*}{$11$} & \multirow[t]{2}{*}{$1$} & $100$ & $0.20$ & $0.15$ & $10$ \\
 &  & $1000$ & $0.20$ & $0.15$ & $10$ \\
\cline{2-6}
 & \multirow[t]{2}{*}{$2$} & $100$ & $0.15$ & $0.05$ & $25$ \\
 &  & $1000$ & $0.15$ & $0.05$ & $25$ \\
\cline{1-6} \cline{2-6}
\multirow[t]{4}{*}{$12$} & \multirow[t]{2}{*}{$1$} & $100$ & $0.20$ & $0.15$ & $10$ \\
 &  & $1000$ & $0.20$ & $0.15$ & $10$ \\
\cline{2-6}
 & \multirow[t]{2}{*}{$2$} & $100$ & $0.10$ & $0.10$ & $15$ \\
 &  & $1000$ & $0.10$ & $0.10$ & $15$ \\
\cline{1-6} \cline{2-6}
\multirow[t]{4}{*}{$13$} & \multirow[t]{2}{*}{$1$} & $100$ & $0.15$ & $0.05$ & $25$ \\
 &  & $1000$ & $0.15$ & $0.05$ & $25$ \\
\cline{2-6}
 & \multirow[t]{2}{*}{$2$} & $100$ & $0.10$ & $0.10$ & $15$ \\
 &  & $1000$ & $0.10$ & $0.10$ & $15$ \\
\cline{1-6} \cline{2-6}
\multirow[t]{4}{*}{$14$} & \multirow[t]{2}{*}{$1$} & $100$ & $0.20$ & $0.15$ & $10$ \\
 &  & $1000$ & $0.20$ & $0.15$ & $10$ \\
\cline{2-6}
 & \multirow[t]{2}{*}{$2$} & $100$ & $0.10$ & $0.10$ & $15$ \\
 &  & $1000$ & $0.10$ & $0.10$ & $15$ \\
\cline{1-6} \cline{2-6}
\bottomrule
\caption{\hpr hyperparameters for all the tested settings.}
\label{tab:hyperparameters HPR}
\end{longtable}
\end{centering}

\section{Simulated Annealing}
\label{sec: SA approach appendix}
In this section, we present the \sa~\cite{simulated_annealing} procedure for enhancing consensus in our setting. That is, we use \sa to obtain initializations of $d$-\rrgs leading to all-ones attractor via majority dynamics while minimizing $m_{\mathrm{init}}$. We compare the results of this widely used Markov Chain Monte Carlo (MCMC) optimization method~\cite{metropolis, hastings,simulated_annealing} with the \hpr algorithm in Section~\ref{sec:HPR}.

\noindent We define the energy function
\begin{equation}
    E(\bbar{\mathbf{x}}) = anm_{\mathrm{init}}(\bbar{\mathbf{x}}) - bn m_{\mathrm{attr}}(\bbar{\mathbf{x}})\,, \label{energy def} 
\end{equation}
which we aim to minimize. $n$ is the number of nodes and we introduce two parameters $a$ and $b$, which are interpretable as two inverse temperatures. The energy function~\eqref{energy def} aligns with our goal of finding initializations where minority becomes majority via majority dynamics: its minima are the states that minimize $m_{\mathrm{init}}$ and maximize $m_{\mathrm{attr}}$ (recall that homogeneous attractor corresponds to $m_{\mathrm{attr}}=1$). Note that the energy is a function only of the graph initialization because of the deterministic dynamics: $E(\bbar{\mathbf{x}}) = anm(\mathbf{x}^1) - bn m(\mathbf{x}^{p+1}) = anm(\mathbf{x}^1) - bn m(F^{\circ p}(\mathbf{x^1})) = E(\mathbf{x}^1)$. This is convenient because the combinatorial space to search is $S^n$ instead of $S^{n(p+1)}$. We construct the Boltzmann probability measure on graph configurations $\mathbf{x}^1=\mathbf{x}\in S^n$
\begin{equation}
    P(\mathbf{x}) = \frac{1}{Z} e^{-E(\mathbf{x})}\,. \label{prob SA}
\end{equation}
As we ``cool'' the system, i.e. we increase the inverse temperatures $a\sim b \rightarrow\infty$, the measure~\eqref{prob SA} concentrates on the ground-states of the energy~\eqref{energy def}. Hence, if we can sample from this probability distribution at lower and lower temperatures, we obtain (approximate) ground-states of the energy~\eqref{energy def}. The  \sa procedure achieves this by performing Metropolis-Hastings (MH) steps with increasing inverse temperatures. This MCMC process is defined by the transition rates based on the ratio $P(\mathbf{x}')/P(\mathbf{x}) = e^{-(E(\mathbf{x}') - E(\mathbf{x}))}$. If the probability of state $\mathbf{x}'$ is higher than the probability of $\mathbf{x}$ (hence $E(\mathbf{x}')<E(\mathbf{x})$) the MH step always accepts the new state $\mathbf{x}'$. If the state $\mathbf{x}'$ is less probable, MH accepts it only with probability $e^{-(E(\mathbf{x}') - E(\mathbf{x}))}$. Thus, the transition rates from state $\mathbf{x}$ to $\mathbf{x}'$ is
\begin{equation}
    p(\mathbf{x}'|\mathbf{x}) = \mathrm{min}\left(1, e^{-(E(\mathbf{x}') - E(\mathbf{x}))}\right)\,, \label{transition rates SA}
\end{equation}
which satisfy the detailed balance condition with the equilibrium distribution given by~\eqref{prob SA}. Moreover, by defining the transition rates by a ratio of probabilities~\eqref{prob SA} we get rid of the hard to compute partition function $Z$.

A slow cooling is essential for a good performance of the \sa procedure. We adopt the following cooling schedule. We start with inverse temperature values $a_i, b_i$ and increase them with the same rate $\alpha>1$, so that $a^{t+1}=\alpha a^t$ (and equivalently for $b^{t+1}$) until we reach the final inverse temperature values $a_f, b_f$. We chose the numerical values of $\{a_i,b_i, a_f, b_f, \alpha\}$ based on empirical observations of the performance of the \sa algorithm in our setting. The precise values used for the results in Table~\ref{tab: all results} are summarized in Table~\ref{tab: parameters for SA}. We initialize the binary node values uniformly at random and the trial state $\mathbf{x}'$ is created by choosing one node (uniformly at random) and flipping its sign. The entire  \sa procedure is summarized in Algorithm~\ref{alg:SA}. 

\begin{algorithm}
\caption{\textbf{S}imulated \textbf{A}nnealing $(G, S, F, E, \alpha, a_i, a_f, b_i, b_f, T_{\rm{max}}$)}\label{alg:SA}
\begin{algorithmic}
\item Initialize graph configuration $\mathbf{x}^{\mathrm{sol}}\in S^n$ uniformly at random and compute the energy $E(\mathbf{x}^{\mathrm{sol}})$, with $a=a_i$, $b=b_i$, according to~\eqref{energy def}. 
\State $t \gets 0$
\While{$\mathbf{x}^{\mathrm{sol}}$ is not a solution \textbf{and} $t\leq T_{\rm{max}}$}
\State Choose $i \in V$ uniformly at random and take $\mathbf{x}'_i = -\mathbf{x}^{\mathrm{sol}}_i$, $\mathbf{x}'_j = \mathbf{x}^{\mathrm{sol}}_j$ for all $j\in V\backslash i$.
\State $\Delta E \gets E(\mathbf{x}') - E(\mathbf{x}^{\mathrm{sol}})$.
\State $r \gets \rm{Random}(0,1)$
\If{$r\leq \rm{min}(1,e^{-\Delta E})$}
\State $\mathbf{x}^{\mathrm{sol}} \gets \mathbf{x}'$
\EndIf
\If{$a < a_f$}
\State $a \gets \alpha a$
\EndIf
\If{$b < b_f$}
\State $b \gets \alpha b$
\EndIf
\State $t \gets t+1$
\EndWhile
\end{algorithmic}
\end{algorithm}

\begin{table}[h!]
    \centering
    \large
    \begin{tabular}{c|l} 
    \toprule
      $d$ & $a_i, b_i, a_f, b_f$ \\ [1ex] 
    \midrule
     $3$ &  $0.015\,, 0.01\,, 4\,, 6$ \\ 
     $4$  & $0.015\,, 0.01\,, 4\,, 6$ \\ 
     $5$  & $0.015\,, 0.01\,, 4\,, 6$ \\ 
     $6$  &$0.015\,, 0.01\,, 4.5\,, 5$ \\
     \bottomrule
    \end{tabular}
    \caption{Parameter values used in the computation of the results presented in Table~\ref{tab: all results} for $p=3$. The cooling rate was set to $\alpha=1.0005$, and we used $T_{\mathrm{max}}=2n^3$ for the maximum \sa steps threshold.} \label{tab: parameters for SA}
\end{table}

\begin{figure}[h!]
    \centering
    \includegraphics[width=\linewidth]{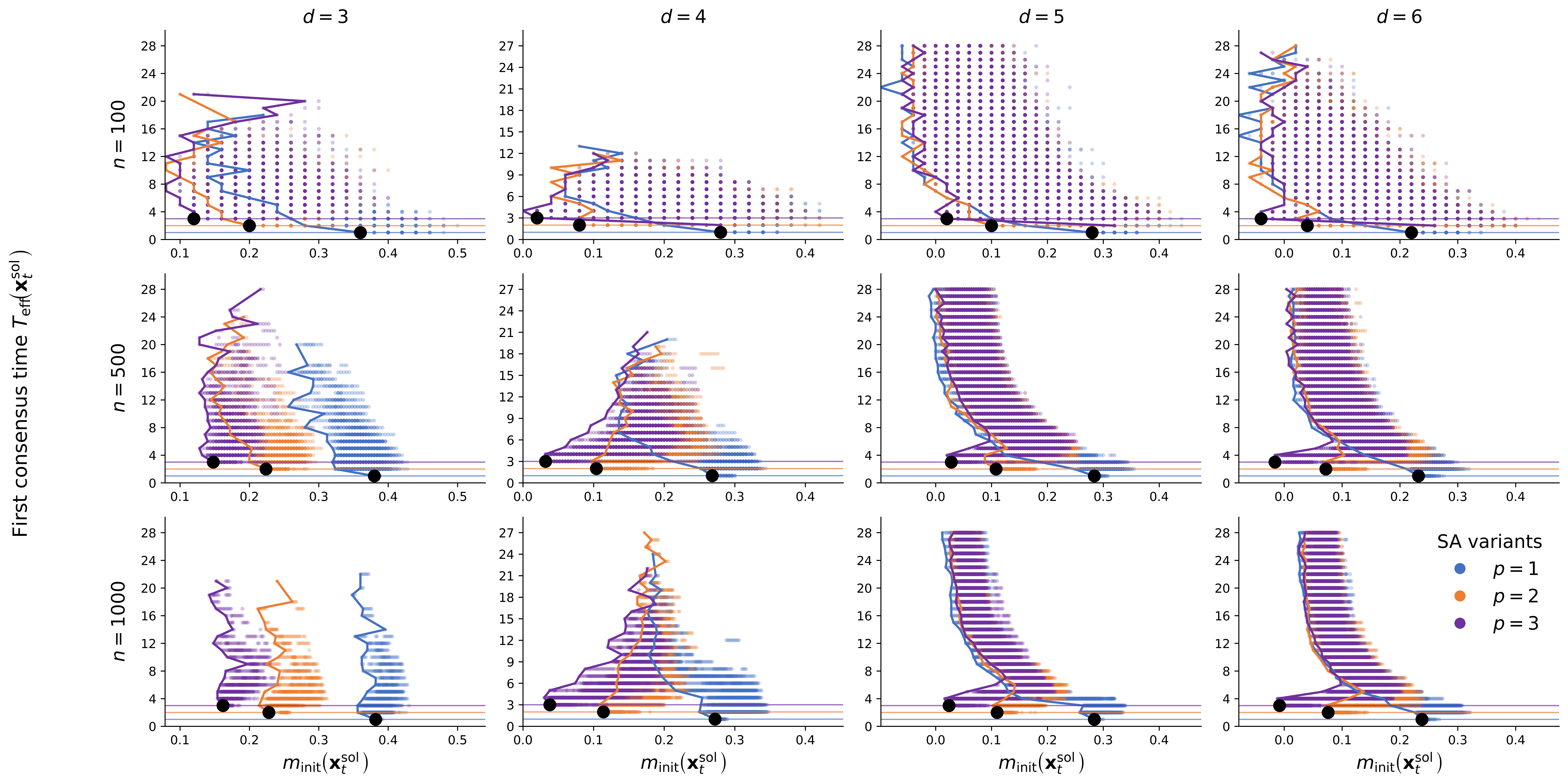}
    \caption{Values of $m_{\rm init}(\mathbf x^{\rm sol}_t)$ and the time to consensus $T_{\rm eff}(\mathbf x^{\rm sol}_t)$ under the majority rule, for all $\mathbf x^{\rm sol}_t$ from the trajectories of SA runs with $d=3,4,5,6$. We exclude every $\mathbf x^{\rm sol}_t$ that failed to reach consensus within $28$ time-steps. To avoid memory issues during the long runs of the SA procedure we plot the following combination of the SA runs. We ran the SA until we reach a solution ($T_{\rm eff}=p$) for 10 random graphs and plot every $10^4$th $(m_{\mathrm{init}}(\mathbf{x}_t^{\mathrm{sol}}), T_{\mathrm{eff}}(\mathbf{x}_t^{\mathrm{sol}}))$ pair. And we ran the SA for another 10 random graphs for $10^7$ time-steps and plot all $(m_{\mathrm{init}}(\mathbf{x}_t^{\mathrm{sol}}), T_{\mathrm{eff}}(\mathbf{x}_t^{\mathrm{sol}}))$ pairs without reaching the solution. This allows us to resolve the early-stage behavior in detail while also capturing the final magnetization values reached, where $m_{\rm init}(\mathbf{x}^{\rm sol}_t)$ fluctuates only weakly around its final value. The black lines are the lowest $m_{\rm init}$ for each $T_{\rm eff}>p$ with $p=1,2,3$ over all runs and black dots indicate the best $m_{\rm init}$  for a given $T_{\rm eff}=p$. Note that this figure reports different values than Table~\ref{tab: all results}, where we reported the average best initialization. The hyperparameter values are $a_i=0.015$, $b_i=0.01$, $a_f=4.5$ and $b_f=5$, except for the case $d=3$, $n=1000$ where $b_f=6$ was used.  The cooling rate was set to $\alpha=1.0005$, and for the maximum \sa steps threshold we used $T_{\mathrm{max}}=n^3$ and $T_{\mathrm{max}}=10^7$. }
    \label{fig:SA trajectories}
\end{figure}

We note that the behavior of the $m_{\mathrm{init}}(\mathbf{x}_t^{\mathrm{sol}})$ trajectories during the SA run depicted in Figure~\ref{fig:SA trajectories} depends on the chosen cooling mechanism. In addition, their behavior in finding solutions with $T_{\mathrm{eff}}(\mathbf{x}_t^{\mathrm{sol}})>p$ differs somewhat from that of the HPR trajectories shown in Figure~\ref{fig:time-ev-1}. For instance, the increase in the initial magnetization observed for $p<T_{\mathrm{eff}}(\mathbf{x}_t^{\mathrm{sol}})<12$ in the cases $p=3$, $d=5,6$, and $n=500,1000$ might be attributed to the fact that the temperature regime where the term in $a$ is better optimized than the term in $b$ does not persist long enough to generate these solutions. Furthermore, the solid lines corresponding to different values of $p$ tend to collapse onto a single curve at large $T_{\mathrm{eff}}(\mathbf{x}_t^{\mathrm{sol}})$. This behavior may stem from the structure of the energy function \eqref{energy def} and the chosen cooling parameters: the first term is identical for all $p$, while the second term exerts a similar influence on the magnetization at high $T_{\mathrm{eff}}(\mathbf{x}_t^{\mathrm{sol}})$ for $p=1, 2, 3$. We did not investigate these effects further, as this cooling regime empirically proved effective for finding solutions with $T_{\mathrm{eff}}(\mathbf{x}_t^{\mathrm{sol}})\leq p$, which was our primary focus.

\end{document}